\documentclass[pre,preprint,superscriptaddress,showpacs,preprintnumbers,eqsecnum,amsmath,amssymb]{revtex4}
\usepackage{graphicx}
\usepackage{dcolumn}
\usepackage{bm}
\usepackage{color}
\usepackage{epsfig}

\input xy 
\xyoption{all}

\begin{document}

\title{Statistical Scattering of Waves in Disordered Waveguides:
from Microscopic Potentials to Limiting Macroscopic Statistics}
\author{L.S. Froufe-P\'{e}rez}
\affiliation{Departamento de F\'{\i}sica de la Materia Condensada,
 and Instituto ``Nicol\'{a}s Cabrera'',
Universidad Aut\'{o}noma de Madrid,  E-28049 Madrid, Spain.}

\author{M. Y\'epez}
\affiliation{Instituto de F\'{\i}sica, Universidad Nacional Aut\'{o}noma de
M\'{e}xico, 01000 M\'{e}xico Distrito Federal, M\'{e}xico}

\author{P. A. Mello}
\affiliation{Instituto de F\'{\i}sica, Universidad Nacional Aut\'{o}noma de
M\'{e}xico, 01000 M\'{e}xico Distrito Federal, M\'{e}xico}

\author{J.J. S\'{a}enz}
\affiliation{Departamento de F\'{\i}sica de la Materia Condensada,
 and Instituto ``Nicol\'{a}s Cabrera'',
Universidad Aut\'{o}noma de Madrid,  E-28049 Madrid, Spain.}

\date{Fri., Oct. 20, 1pm. VERSION: froufe}

\begin{abstract}
We study the statistical properties of wave scattering in a disordered
waveguide. The statistical properties of a ``building block" of
length $\delta L$ are derived from a potential model and used to find the
evolution with length of the expectation value of physical quantities. In the
potential model the scattering units consist of thin potential slices,
idealized as delta slices, perpendicular to the longitudinal direction of the
waveguide; the variation of the potential in the transverse direction may be
arbitrary. The sets of parameters defining a given slice are taken to be
statistically independent from those of any other slice and identically distributed. In the
dense-weak-scattering limit, in which the potential slices are very weak and
their linear density is very large, so that the resulting mean free paths are
fixed, the corresponding statistical properties of the full waveguide depend only on the mean free paths and on no other property of the slice distribution. The {\em universality} that arises demonstrates the existence of a {\em generalized central-limit theorem}.

Our final result is a {\em diffusion equation} in the space of transfer
matrices of our system, which describes the evolution with the length $L$ of
the disordered waveguide of the transport properties of interest. In contrast
to earlier publications, in the present analysis the energy of
the incident particle is fully taken into account.

For one propagating mode, $N=1$, we have been able to solve the diffusion
equation for a number of particular observables, and the solution is in
excellent agreement with the results of microscopic calculations. In general,
we have not succeeded in finding a solution of the diffusion equation. We have
thus developed a numerical simulation, to be called ``random walk in the
transfer matrix space", in which the universal statistical properties of a
``building block"  are first
implemented numerically, and then the various building blocks are combined to find the statistical properties of the full waveguide. The
reported results thus obtained (in which use was made of a ``short-wavelength
approximation") are in very good agreement with those arising from truly
microscopic calculations, for both bulk and surface disorder.
\end{abstract}


\maketitle

\tableofcontents

\newpage

\section{Introduction}
\label{intro}

The statistical theory of certain complex wave interference phenomena, like the
statistical fluctuations of transmission and reflection of waves, is of
considerable interest in many fields of physics
\cite{Ishimaru78,Rytov89,Nieto-V90,Altshuler91,Sheng95,carlo97,Imrybook,Dattabook,Alhassid,mello-kumar,WRM}.
In the literature one has contemplated situations in which such a complexity
derives from the chaotic nature of the underlying classical dynamics, as in the
case of chaotic microwave cavities and quantum dots, or from the quenched
randomness of scattering potentials, as in the case of disordered conductors
or, more in general, disordered waveguides. It is the latter domain that will
interest us here.

In studies performed in such systems one has found remarkable {\em statistical regularities},
in the sense that the probability distribution for various macroscopic quantities involves a
rather small number of relevant physical parameters, or scaling parameters, while the rest of the
microscopic details serves as mere ``scaffolding".
In Ref. \cite{mello_shapiro}
it was shown that a limiting distribution of physical quantities indeed arises
in the so called {\em dense-weak-scattering limit}
(DWSL) and within a particular class of models:
the individual, microscopic, scattering units were defined through their transfer matrices and an ``isotropic" distribution of their phases was assumed.
The limiting distribution that was found constitutes a
generalized {\em central-limit theorem} (CLT).
Within this model only one relevant physical parameter occurs: the mean free path, which is the only property arising from the individual scattering units that survives in the DWSL.
This is consistent with the scaling hypothesis proposed by
E. Abrahams et al. \cite{gang_of_4}.
(When abandoning the DWSL, two parameters were needed in Ref. \cite{mello_rmp} to describe the conductance distribution.)
The result found in Ref. \cite{mello_shapiro} coincides with that of the maximum-entropy model that had been developed in Ref. \cite{mpk},
which gives rise to a diffusion equation known as the DMPK equation
(after Dorokhov \cite{dorokhov} and Mello, Pereyra and Kumar \cite{mpk}),
which can thus be interpreted as capturing the features arising from a CLT.

CLT's associated with products of matrices had been studied earlier, as, for instance,
in the well-known Oseledec theorem \cite{oseledec,carlo97};
the results of Refs. \cite{mello_shapiro,mpk} are consistent with this theorem in the
localized regime.

In spite of the successes of the diffusion equation of Ref. \cite{mpk} in the
study of the conductance distribution \cite{carlo97}, that equation fails to
give the proper description when the difference in behavior of the various
modes becomes relevant. A clear example was given in Ref. \cite{UAM}, where the
conductance distribution was studied in the crossover region $\langle G \rangle
\approx e^2 /h$. For waveguides with bulk disorder the description is
excellent, whereas for waveguides with surface disorder it is not satisfactory
 \cite{WRM,UAM2}.

A class of limiting distributions wider than that of Ref. \cite{mello_shapiro} was studied
by one of the present authors and S. Tomsovic in Ref. \cite{mello_tomsovic}
(to be referred to as MT), in which
the isotropy assumption of Ref. \cite{mello_shapiro} was relaxed to a large extent.
The DWSL played an essential role and the result was a more general CLT than that of
Ref. \cite{mello_shapiro},
expressed in terms of a generalized diffusion equation.
The scaling parameters that appear in MT are
the mean free paths (mfp's) for the various scattering processes that may occur in the
problem.
When the various mfp's can be represented by a single one, one encounters the same
diffusion equation that was studied in Ref. \cite{mpk} using a maximum-entropy model.
Thus the MT model appears as a possible candidate to study,
in the problem of waveguides with surface disorder,
the influence of the specific scattering properties of the relevant modes.

The ideas of MT are further developed in the ``Brownian-motion" model of Ref. \cite{houches}:
a waveguide of length $L$ is enlarged by adding a piece of thickness
$\delta L$ [to be called a ``building block" (BB)],
small on a macroscopic scale but still containing many scatterers, which
is likened to a Brownian particle which, in a time interval $\delta t$,
small on a macroscopic scale, still suffers many collisions from the molecules of the surrounding medium.
The transfer matrix for a BB is written as $M=I + \varepsilon$ and the independent parameters which $M$
depends upon are chosen so that $\varepsilon$ for the BB satisfies a number of properties,
reminiscent of those of a Brownian particle:
\begin{subequations}
\begin{eqnarray}
\langle\varepsilon\rangle_{\delta L} &=&0 + O(\delta L ^2)
\\
\langle\varepsilon \varepsilon\rangle_{\delta L}  &=& O(\delta L)   \; ,
\end{eqnarray}
\end{subequations}
while higher moments of $\varepsilon $ behave as higher powers of $\delta L$
(see Eqs. (3.73) and (3.74) of Ref. \cite{houches}).
The result of this analysis is the same diffusion equation as that of MT.

Though appealing the assumptions behind the Brownian-motion model of Ref. \cite{houches}
for the BB may be, they are, nevertheless, arbitrary.
Of course, they can be deduced from the MT model for the more
microscopic scattering units.
However, even these have a certain degree of arbitrariness.
It would be satisfactory if these models could be obtained in a unified way from a maximum-entropy ``ansatz":
this, however, is not known to the present authors at this time.

The motivation of the present paper is to {\em derive} from a potential model
the statistical properties of the BB and use them to find the ``evolution" with
length of the expectation value of physical quantities. Since the potential
model will be introduced at the level of the individual scattering units, the
approach to be presented here is, in a way, hybrid between the methods of MT
and Ref. \cite{houches}. We shall see that within the present model it is not
strictly true that the individual transfer matrices (resulting from the
individual potentials) for the various scatterers are identically distributed
[see Eq. (\ref{<epsr ^2>}) below], as was assumed in MT, and this fact will be
taken into account. The continuous limit is also treated here in a more
satisfactory way than in MT, and the energy appears explicitly in the following
presentation, in contrast to earlier publications.
We believe that the present model is physically more complete than that of Ref.
\cite{mpk}; it is also better founded than that of MT and Ref. \cite{houches},
in the sense that there is a lesser degree of arbitrariness in the assumptions,
although the resulting diffusion equation has a ``structure" similar to the one
obtained in MT and Ref. \cite{houches}. Our diffusion equation covers
situations not contemplated in these references: we shall find it suitable to
study wave-transport problems in which the physics of the various modes is
relevant; we shall also find a good description of the statistical properties
of quantities that involve phases, which were not described at all in previous
models. The reader is referred to Ref. \cite{mello_physicaA} for a preliminary
account of the results of the present paper.

The paper is organized as follows. In the next section we derive a
Fokker-Planck equation for the ``evolution" with the waveguide length $L$ of the
expectation value of the physical quantities of interest. That equation
represents the central result of the present paper and is given in Eq.
(\ref{diff eqn exp val sec2}), which we reproduce here for convenience:
\begin{eqnarray}
&&\frac{\partial \left\langle F(\bm{M})\right\rangle _L}{\partial L}
= \sum_{
\substack{
ijhl\lambda \mu
\\ abcd\alpha \beta
}
}
D_{ab,cd}^{ij,hl}(k,L) \left\langle M_{b \alpha }^{j \lambda } \;
M_{d \beta }^{l \mu} \; \frac{\partial ^2F(\bm{M})} {\partial M_{a
\alpha }^{i \lambda } \partial M_{c \beta }^{h \mu}} \right\rangle
_L .
\label{diff eqn exp val intro}
\end{eqnarray}
We notice that Eq. (\ref{diff eqn exp val intro}) contains no
drift term and is thus a diffusion equation. The physical observable
is denoted by $F(\bm{M})$, $\bm{M}$ being the transfer matrix of the
sample of length $L$. The quantities $D_{ab,cd}^{ij,hl}(k,L)$ play
the role of ``diffusion coefficients", which are defined in terms of
the second moments of $\varepsilon $ for the BB in Eq.
(\ref{<eps.eps>_D}) below 
(see the term linear in $\delta L$)
and are given explicitly in 
Eq. (\ref{D(k,L)})
in terms of the {\em mean free paths}. Notice that
the diffusion coefficients depend on the energy ($\sim k^2$) and
also on the length $L$ of the sample. 
The mean free paths depend only on the second
moments of the potential intensity of the individual impurities [see
Eq. (\ref{lab})], higher moments being irrelevant for the
diffusion equation: this is precisely what signals the existence of
a CLT. 
In order to derive the
diffusion equation (\ref{diff eqn exp val intro}) we need a
statistical model for the building block (BB): this is derived in
Sec. \ref{BB q1d} using a potential model for the random impurities.
Although the treatment of Sec. \ref{fokker-planck} is applicable to
the orthogonal as well as to the unitary symmetry classes
of Random-Matrix Theory
($\beta =1$ and $2$, respectively \cite{dyson}), the potential
model developed in Sec. \ref{BB q1d} assumes time-reversal
invariance, i.e., $\beta =1$. Some of the specific relations
derived there would have to be properly modified for the unitary
case, $\beta =2$. 
The results of Sec. \ref{BB q1d} which are needed for the
derivation of the diffusion equation (\ref{diff eqn exp val intro})
have an intrinsic interest as well, since they can be used to
describe the statistical scattering properties of thin slabs. The
diffusion equation (\ref{diff eqn exp val intro}) is first derived
for arbitrary energy, and only later the short-wavelength
approximation (SWLA) is contemplated; it is in this latter limit
that some of the results obtained earlier can be recovered. Needless
to say, we have no general way of finding either analytically or
numerically the solutions of the above diffusion equation. We thus
give in Sec. \ref{analytic} some simple examples in which the
analytic solution could be found; in Sec. \ref{numerical} we develop
a procedure to simulate numerically the diffusion process in
transfer-matrix space, and present some of the results that we have
been able to obtain so far. The conclusions of this work are given
in Sec. \ref{conclusions}. A number of appendices have been
included, in which some of the specific calculations are carried
out.


\section{Transport in q-1D disordered systems.
The combination law and the Smoluchowsky equation.
The diffusion equation}
\label{fokker-planck}

Consider a q-1D disordered system of uniform cross section,
connected, at both ends, to clean waveguides that support $N$
{\em open channels} each.
In the disordered region there is an underlying random potential to be specified later.
In some applications we shall be concerned with a 2D waveguide with a width to be denoted by $W$.

The scattering properties of the system will be described by means of its
transfer matrix $\bm{M}$, which can be written as
\begin{equation}
\bm{M}
=\left[
\begin{array}{ll}
M^{11} & M^{12} \\
M^{21} & M^{22}
\end{array}
\right]
\equiv
\left[
\begin{array}{ll}
\alpha &\beta \\
\gamma & \delta
\end{array}
\right] .
\label{M N 1}
\end{equation}
Each block $M^{ij}$ ($i=1,2$; $j=1,2$) in (\ref{M N 1}) is $N$-dimensional, so that $\bm{M}$ is $2N$-dimensional.
(The block $M^{12}$ will occasionally be denoted by $\beta $,
a symbol not to be confused with the index for universality classes in Random-Matrix Theory.)
One particular matrix element of the $ij$ block will be designated as
$M_{ab}^{ij}$, where $a, b \; (=1, \cdots ,N)$ denote the channels.
Some of the properties of the $\bm{M}$ matrix and its relation with the more conventional
reflection and transmission amplitudes, which are elements of the $\bm{S}$ matrix, are summarized in App. \ref{M-properties}.

The transfer matrix $\bm{M}$ will be considered to belong to one of the basic
symmetry classes introduced by Dyson in Quantum Mechanics \cite{dyson}.
Here we
shall be only concerned with scalar waves, so that, in applications to quantum
mechanics, we shall only have  ``spinless electrons". In the ``unitary'' case,
also denoted by $\beta =2$, the only restriction on $\bm{M}$ is flux
conservation, which is expressed by the pseudo-unitarity condition, Eq.
(\ref{pseudoun}). In the ``orthogonal'' case ($\beta =1$), time-reversal
invariance imposes the restriction given by Eq. (\ref{TRI N}). The
``symplectic" case ($\beta =4$) associated with half-integral spin will not
be considered here.

If the underlying potential has non-zero matrix elements between open and closed channels,
the $2N$-dimensional $\bm{M}$ matrix depends on an ``effective potential" which contains
information on closed channels, as explained in App. \ref{veff}.

Consider now two non-overlapping scatterers.
Their {\em extended transfer matrices}, $\tilde{\bm{M}}_1$ and  $\tilde{\bm{M}}_2$
(which include open {\em and} closed channels, are infinite-dimensional and
depend on the bare potential, as opposed to the effective one),
have the multiplicativity property
(see App. \ref{M_multiplicativity})
\begin{equation}
\tilde{\bm{M}}=\tilde{\bm{M}}_2 \tilde{\bm{M}}_1 .
\label{comb law ext M}
\end{equation}
In past publications by one of the authors (PAM)
(see, for instance, Ref. \cite{mello-kumar}),
closed channels have been neglected in the matrix multiplication of successive
scatterers.
In numerical simulations \cite{WRM,JJS_evanesc} one sees that for individual
configurations of the disordered system and in the calculation of the mean free
path, the inclusion of closed channels is important.
(In this paper, the expressions
``closed channels" and ``evanescent modes" will be taken as synonymous.)
On the other hand, for the
statistical fluctuations the conditions for neglecting the evanescent modes do
not appear to be very stringent. For a given mean free path, the statistical
properties of the different transport coefficients are found to be roughly
independent of the number of evanescent modes
(see also the discussion of the numerical simulations given in Sec. \ref{random_walk}).
In this article we shall thus follow the
earlier approximation and write the resulting transfer matrix
as the product of the individual open-channel transfer matrices.

\begin{figure}
\epsfig{file=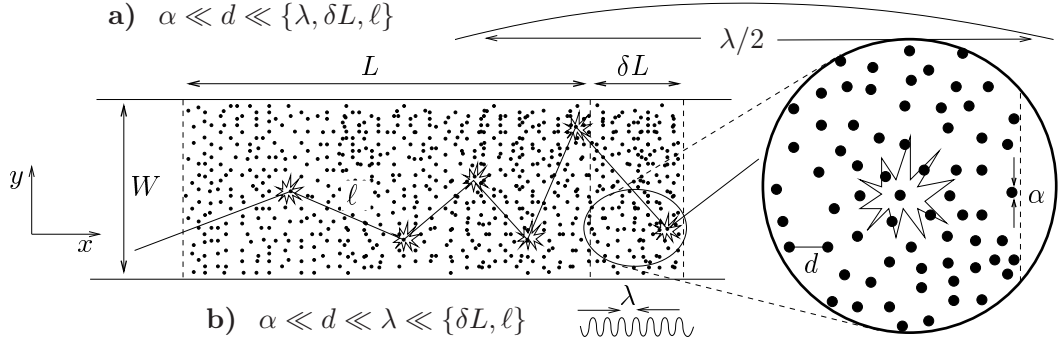,width=14cm,clip=}
\caption{ Schematic representation of a disordered wire and the building block
(BB); ({\em a}) and ({\em b}) correspond to the different regimes (see Section
\ref{BB q1d}) defined by the inequalities given in Eqs. (\ref{arb wl}) and
(\ref{swla regime}) ({\em short-wavelength approximation} (SWLA)),
respectively. }
\label{sketchBB}
\end{figure}

Suppose we start with a system containing $n$ scattering units
(to be defined at the
beginning of Sec. \ref{BB q1d})
and enlarge it
by adding, on its right-hand side, say, a slab, to be called a 
{\em building block} (BB), containing $m$ scattering units. 
Designating by $\bm{M}^{(L)}$ the transfer matrix of the original system and by $\bm{M}^{(\delta L)}$ that of
the BB, the resulting transfer matrix is
\begin{equation}
\bm{M}^{(L+\delta L)}=\bm{M}^{(\delta L)}\bm{M}^{(L)} .
\label{comb law M}
\end{equation}
We assume the BB to be of arbitrary thickness $\delta L$,
and to contain {\it many weak scatterers}
(see Fig. \ref{sketchBB}).
Its transfer matrix $\bm{M}^{(\delta L)}$ will be written as
\begin{equation}
\bm{M}^{(\delta L)}=\bm{I}_{2N}+\bm{\varepsilon} .
\label{M BB}
\end{equation}
The combination law for $\bm{M}$, Eq. (\ref{comb law M}), can be written as
\begin{subequations}
\begin{eqnarray}
\bm{M}^{(L+\delta L)} &=& \bm{M}^{(L)} + \delta \bm{M}  \\
&=& \bm{M}^{(L)} + \bm{\varepsilon M}^{(L)}.
\end{eqnarray}
\label{deltaM}
\end{subequations}

Consider now a function $F(\bm{M})$ of the $\bm{M}$ matrix, which we shall call
an ``observable'':
it could be, for instance, the transmission amplitude $t_{ab}$, or the conductance $G$,
which is proportional to the total transmission coefficient $T$.
We are interested in the expectation value $\left\langle F(\bm{M})\right\rangle _{n}$
of such an observable for a system containing $n$ impurities.
We first find below a recurrence relation with $n$ for that expectation value and then, in the continuous limit,
we shall find the equation
that governs the ``evolution'' of $\left\langle F(\bm{M})\right\rangle _{L}$
with increasing length $L$.

If we write a particular $\bm{M}$ matrix element as $M_{ab}^{ij}=\xi
_{ab}^{ij}+i\eta _{ab}^{ij}$, we may consider the observable $F(\bm{M})$ as a
function of all those $\xi _{ab}^{ij}$ and $\eta _{ab}^{ij}$ which are relevant
for the universality class $\beta $ in question. For instance, in the
orthogonal case ($\beta =1$), because of the TRI relation, Eq. (\ref{TRI 1}),
only the real and imaginary parts of the blocks $M_{ab}^{11}$ and $M_{ab}^{12}$
are relevant, while in the unitary case ($\beta =2$) the real and imaginary
parts of the four blocks $M_{ab}^{ij}$ ($i=1,2$; $j=1,2$) are needed.
Alternatively, we may consider $F(\bm{M})$ as a function  of all the relevant
$M_{ab}^{ij}$'s and their complex conjugates, a possibility which will be found
more convenient in what follows. For $\beta =1$, Eq. (\ref{TRI 1}) shows that
this is equivalent to expressing $F(\bm{M})$ as a function of {\em all} the
$M_{ab}^{ij}$ ($i=1,2$; $j=1,2$), enforcing Eq. (\ref{TRI 1}) at the end, while
for $\beta =2$ we need {\em all} the $M_{ab}^{ij}$ ($i=1,2$; $j=1,2$) {\em
and} their complex conjugates. In what follows we shall restrict
ourselves to the orthogonal case.


Writing the composition law for the $\bm{M}$ matrix as in Eq. (\ref{deltaM}), the expressions
for the observable $F(\bm{M})$ before and after adding the building block are related by the Taylor expansion
\begin{eqnarray}
&&F\left( \bm{M}^{(n+m)}\right) =F\left( \bm{M}^{(n)}+\delta
\bm{M}\right) 
=F\left( \bm{M}^{(n)}\right) 
+ \sum_{\substack{ i \lambda \\ a \alpha }} 
(\delta M_{a\alpha }^{i\lambda }) \; \frac{\partial F\left( \bm{M}\right)}
{\partial M_{a \alpha }^{i\lambda }}\Bigg|_{\bm{M}=\bm{M}^{(n)}}
\nonumber \\
&&\;\;\;\;\;\;\;+\frac 1{2!} \; \sum_{\substack{ i\lambda h\mu  \\
a\alpha c\beta }} (\delta M_{a \alpha }^{i \lambda }) \; (\delta
M_{c \beta }^{h \mu }) \; \frac{\partial ^2F\left(\bm{M}\right) }
{\partial M_{a \alpha }^{i \lambda }
\partial  M_{c \beta }^{h \mu }}
\Bigg|_{\bm{M}=\bm{M}^{(n)}}
+ \; \cdots \;,
\label{F(M+dM)}
\end{eqnarray}
where the lower indices $a,\alpha, \cdots$ on each $M$ indicate channels and run
over the values $1, \cdots, N$, while the upper indices $i,\lambda , \cdots$ identify
the block in Eq. (\ref{M N 1}) and take on the values $1,2$.

We take the expectation value of both sides of Eq. (\ref{F(M+dM)}) with respect to the enlarged
system containing $n+m$ scattering units and use Eq. (\ref{deltaM}),
considering the two pieces $n$ and $m$ to be {\em statistically independent}. We find
\begin{eqnarray}
&&\left\langle F(\bm{M})\right\rangle _{n+m} =\left\langle
F(\bm{M})\right\rangle _n + \sum_{\substack{ ij\lambda  \\
ab\alpha }} \left\langle \varepsilon _{ab}^{ij}\right\rangle_{m}
\left\langle M_{b \alpha}^{j \lambda } \; \frac{\partial F(\bm{M})}
{\partial M_{a \alpha }^{i \lambda }}\right\rangle _n \nonumber
\label{F(M+dM) 1}
\nonumber \\
&&\;\;\;\;\;\;\;+\; \frac 1{2!} \sum_{\substack{ ijhl\lambda \mu  \\
abcd\alpha \beta }} \left\langle \varepsilon _{ab}^{ij} \;
\varepsilon _{cd}^{hl} \right\rangle _{m} 
\left\langle M_{b \alpha }^{j \lambda } \; M_{d \beta }^{l \mu } \; \frac{\partial ^2F(\bm{M})} {\partial M_{a \alpha }^{i \lambda }\; 
\partial M_{c \beta }^{h \mu}} \right\rangle _n \; + \; \cdots \; .
\label{avF(M+dM)}
\end{eqnarray}
Here, $\left\langle \cdot \cdot \cdot \right\rangle _{n}$ denotes an
average evaluated with the probability density for the transfer matrix
of the original sample containing $n$ scattering units, i.e.,
\begin{equation}
\left\langle G(\bm{M}) \right\rangle _{n}
\equiv \left\langle G(\bm{M}^{(n)}) \right\rangle .
\label{<>n}
\end{equation}
A few comments are in order at this point.
We recall that the various matrix elements $M_{ab}^{ij}$ are not independent:
they are related by the pseudounitarity condition, Eq. (\ref{pseudoun}), arising from flux conservation, and by Eq. (\ref{TRI N}), associated with time-reversal invariance.
Of course, we can express the $M_{ab}^{ij}$'s in terms of independent parameters,
like those occurring in the ``polar representation" of the transfer matrix,
and perform a Taylor expansion (similar to the one above) with respect to such
independent parameters \cite{mello-kumar,mpk}.
Here, just as in Ref. \cite{mello_tomsovic}, we have found it simpler to
perform the expansion with respect to the matrix elements $M_{ab}^{ij}$
themselves, as in  Eq. (\ref{F(M+dM)}), in the understanding that in the resulting expression (\ref{F(M+dM)}) the
$M_{ab}^{ij}$'s and $\delta M_{ab}^{ij}$'s
have to be expressed in terms of independent parameters.
The average appearing in Eq. (\ref{avF(M+dM)}) is thus performed with a probability density for such independent parameters.
Since the latter depend on the underlying microscopic potentials,
in this paper we shall not propose a model for the
transfer matrix independent parameters, but rather for the microscopic potentials.
(See also Ref. \cite{mello_tomsovic}.)

The next step is to describe the problem in the
{\em dense-weak-scattering limit} (DWSL) briefly described in the Introduction
(and defined in Eqs. (\ref{dwsl}) below), so that
we can speak of the {\em continuous length} $L$ of the system and the length $\delta L$
of the BB.
Eq. (\ref{avF(M+dM)}) becomes
\begin{eqnarray}
&&\left\langle F(\bm{M})\right\rangle _{L + \delta L} =\left\langle
F(\bm{M})\right\rangle _L + \sum_{\substack{ ij\lambda \\
ab\alpha }} \left\langle \varepsilon _{ab}^{ij}\right\rangle_{L,
\delta L} \left\langle M_{b \alpha}^{j \lambda } \; \frac{\partial
F(\bm{M})} {\partial M_{a \alpha }^{i \lambda }}\right\rangle _L
\nonumber \label{F(M+dM) 1}
\nonumber \\
&&\;\;\;\;\;\;\;+\; \frac 1{2!} \sum_{\substack{ ijhl\lambda \mu  \\
abcd\alpha \beta }} \left\langle \varepsilon _{ab}^{ij} \;
\varepsilon _{cd}^{hl} \right\rangle _{L, \delta L} \left\langle
M_{b \alpha }^{j \lambda } \; M_{d \beta }^{l \mu } \;
\frac{\partial ^2F(\bm{M})} {\partial M_{a \alpha }^{i \lambda }\;
\partial M_{c \beta }^{h \mu}} \right\rangle _L \; + \; \cdots \; .
\label{avF(L+dL)}
\end{eqnarray}

To proceed, we need a statistical model for the BB.
For this purpose, as we mentioned in the previous paragraph,
a potential model is discussed in Sec. \ref{BB q1d}, in which the BB is
constructed as a collection of $m$ individual scattering units represented by
delta-potential slices. It is found that the first moment of $\bm{\varepsilon
}$ for the BB vanishes [see Eq. (\ref{av eps BB})], the second moments, in the
DWSL, admit an expansion in powers of $\delta L$ starting with $\delta L$
itself (see Eq. (\ref{<eps.eps>_D}), while higher moments behave as
higher powers thereof
(see the discussion following Eq. (\ref{Delta=(delta L)^q})). Also, the very important result emerges that {\em the dependence on
the cumulants of the potential higher than the second drops out in the DWSL}.
These results are reminiscent of the statistical behavior of the velocity
increment of a Brownian particle during a time interval $\delta t$ during which
many collisions from the surrounding medium have occurred \cite{chandra}.

When the moments of the BB, evaluated in the DWSL,
are substituted in Eq. (\ref{avF(L+dL)}),
we obtain, on the r.h.s of that equation, a power series in $\delta L$.
We also perform, on the l.h.s. of Eq. (\ref{avF(L+dL)}), a Taylor expansion of
$\left\langle F(\bm{M})\right\rangle _{L+\delta L}$
in powers of $\delta L$ around the ``initial" value
$\left\langle F(\bm{M})\right\rangle _{L}$.
We can then identify the coefficients of the various powers of $\delta L$
on the two sides of the equation.
In particular, the coefficients of $\delta L$ give
the {\em diffusion equation}
\begin{eqnarray}
&&\frac{\partial \left\langle F(\bm{M})\right\rangle _L}{\partial L}
= \sum_{\substack{ ijhl\lambda \mu  \\ abcd\alpha \beta }}
D_{ab,cd}^{ij,hl}(k,L) \left\langle M_{b \alpha }^{j \lambda } \;
M_{d \beta }^{l \mu} \; \frac{\partial ^2F(\bm{M})} {\partial M_{a
\alpha }^{i \lambda } \partial M_{c \beta }^{h \mu}} \right\rangle
_L .
\label{diff eqn exp val sec2}
\end{eqnarray}
The quantities
$D_{ab,cd}^{ij,hl}(k,L)$ play the role of ``diffusion coefficients": 
they are defined in Eq. (\ref{<eps.eps>_D}) below 
as proportional to the coefficient of the linear term in an expansion in powers of $\delta L$ of the second moment of $\varepsilon$ for the BB
and are given explicitly
in Eq. (\ref{D(k,L)}) in terms of the {\em mean free paths}.
The diffusion coefficients depend on the energy
($\sim k^2$) and also on the length $L$ of the sample.

We remark that, just as the coefficients of $\delta L$ in
Eq. (\ref{avF(L+dL)}) are expressible in terms of the mfp's,
the coefficients of higher-order terms in $\delta L$
have a similar property, because the contribution of higher moments becomes irrelevant in the DWSL.
Equating the coefficients of such higher-order terms on both sides of
Eq. (\ref{avF(L+dL)}) we obtain results which could be
derived from the diffusion equation (\ref{diff eqn exp val sec2}) by successive differentiations.
(See comment right after Eq. (\ref{diff eqn exp val sec3}).)

In the potential model discussed in the next section only the orthogonal case,
$\beta =1$, is contemplated. We expect a similar behavior for the unitary
class, $\beta =2$, although we do not have at the present moment the specific
expression for each diffusion coefficient in this case.

Eq. (\ref{diff eqn exp val sec2}) represents the central result of the present paper.
It depends only on the mean free paths which, in turn, depend only on the
{\em second moments of the individual delta-potential} strengths
[Eq. (\ref{lab})].
The fact that cumulants of the potential higher than the second
are irrelevant in the end signals the existence of a {\em generalized CLT}:
once the mfp's are specified, the limiting equation
(\ref{diff eqn exp val sec2})
is {\em universal}, i.e., independent of other details of the microscopic statistics.

\section{Statistical Properties of the Building Block}
\label{BB q1d}

In the present section we investigate the statistical scattering
properties of the BB which was used in
Sec. \ref{fokker-planck} to build a disordered system with a q1D geometry
(see Fig. \ref{sketchBB}).

Suppose that we model the scatterers constituting the BB by a sequence of
{\em thin slices}
(the scattering units referred to right above Eq. (\ref{comb law M}))
of cross section $W^{D-1}$
($D$ being the dimensionality of the waveguide).
From now on we denote the thickness of the slices by $2\alpha $ and their
separation by $d$.
(See Fig. \ref{sketchDeltaSlice} below.
Notice that in Fig. \ref{sketchBB} the same symbols refer to individual scatterers; here, a slice may contain
one or more of the individual scatterers shown in Fig. \ref{sketchBB}.)
The statistical properties of the potential slices will be
specified below (see Sec.  \ref{stat model}).
Inside $2\alpha $, the $r$-th scattering slice is described by the potential $V_r(x,\bm{y})$.
We denote by $x$ the coordinate along the waveguide and by $\bm{y}$
the coordinates in the transverse direction.
The distance $d$ between slices is taken to be much larger than $\alpha $,
but much smaller than the wavelength $\lambda$ of the incident wave and the
thickness $\delta L$ of the BB.
Initially we do not specify the ratio of the wavelength $\lambda$ to
$\delta L$ or the mean-free-path $\ell$ (to be defined later),
so we shall start out constructing the BB as a collection of $m$
thin slices satisfying the inequalities
\begin{subequations}
\begin{eqnarray}
\alpha  \ll  d  \ll \{ \lambda , \delta L, \ell \}.
\label{arb wl}
\end{eqnarray}
Later on, in section \ref{SWLA}, we shall find it advantageous to study a second
regime, in which $\delta L$ (and hence any final $L$) and $\ell$ contain many wavelengths, i.e.,
\begin{eqnarray}
\alpha  \ll  d  \ll \lambda  \ll  \{\delta L, \ell \}  \; ,
\label{swla regime}
\end{eqnarray}
\label{regime}
\end{subequations}
corresponding to what we shall call the
{\em short-wavelength approximation} (SWLA).

In principle we have no restriction on the dimensionality $D$ of the waveguide;
however, to be specific, we shall restrict the discussion to two-dimensional
waveguides with uniform width $W$. As we already indicated, in the potential
model to be presented below we shall be concerned with the orthogonal, or
$\beta =1$, symmetry class only.

\subsection{Properties of a single scattering slice}
\label{single delta slice}

Consider a single scattering slice
with potential $V(x,y)=\hbar^2 U(x,y)/(2m)$, centered at the origin of coordinates $x=0$, and
let $[U(x)]_{ab}$ be the matrix elements of $U(x,y)$ with respect to the ``transverse" states $ \chi _a(y)$ of the waveguide, i.e.,
\begin{equation}
[U(x)]_{ab} = \int_0 ^W \chi _a(y) U(x,y) \chi _b(y) dy ,
\label{Vab}
\end{equation}
with
\begin{equation}
\chi _a (y) = \sqrt{\frac2W} \sin \frac{\pi ay}{W},
\label{chi a}
\end{equation}
$a$ being an integer.
Under the conditions
\begin{subequations}
\begin{eqnarray}
k\alpha  &\ll& 1 \\
K_{ab}  \alpha   &\ll& 1  ,
\label{weak scatterer}
\end{eqnarray}
\end{subequations}
where $k=2\pi /\lambda = \sqrt{2mE}/\hbar$ and
\begin{equation}
K_{ab} ^2 =|U_{ab}| \equiv \big|[U(0)]_{ab} \big| ,
\end{equation}
we speak of a {\em thin} scatterer
(a thin barrier or well) and the dependence of the potential across the thickness $2\alpha $ is neglected.
On the other hand, the quantity
\begin{equation}
2\alpha \cdot U_{ab} \equiv u_{ab}
\label{u}
\end{equation}
(which has dimensions of $k$) is arbitrary.
Such a scatterer can be well approximated by the ``delta potential"
\begin{subequations}
\begin{eqnarray}
U(x,y) &=& u(y) \delta (x) ,
\label{delta slice 1}
\\
\left[U(x)\right]_{ab}&=&u_{ab} \; \delta (x)\;,
\label{delta slice 2}
\end{eqnarray}
\label{delta slice}
\end{subequations}
obtained formally taking the limits
\begin{subequations}
\begin{eqnarray}
|U_{ab}| &\rightarrow& \infty,
\label{Uab infty}
\\
\alpha  &\rightarrow& 0 ,
\label{a 0}
\end{eqnarray}
\label{lim delta pot}
\end{subequations}
in such a way that the quantity $u_{ab}$ of Eq. (\ref{u}) stays fixed.
From the inequalities (\ref{regime}) we see that the range $2\alpha $ of the potential
is the smallest length scale in the problem:
the limit (\ref{a 0}) is the extreme idealization of this situation.

Eqs. (\ref{delta slice}) define a delta-slice potential centered at the origin of coordinates.
The potential produced by the $r$-th delta slice, centered at $x=x_r$, is written as
\begin{subequations}
\begin{eqnarray}
U_r(x,y) &=& u_r(y)\delta (x-x_r)
\label{r delta slice (xy)} \\
\left[ U_r(x) \right]_{ab}
&=& (u_r)_{ab} \; \delta (x-x_r) .
\label{r delta slice mmn}
\end{eqnarray}
\label{r delta slice}
\end{subequations}
We remind the reader
that $U_r(x,y)$ has dimensions of $k^2$,
whereas $u_r(y)$ and $(u_r)_{ab}$ have dimensions of $k$.

A particle scattered by the potential of Eq. (\ref{delta slice})
inside the waveguide is described by the wave function
\begin{equation}
\psi (x,y)
=\sum_{a=1}^{\infty}\left[ \psi (x) \right]_a \chi _a (y) ,
\end{equation}
which satisfies Schr\"odinger's equation;
its components $\left[ \psi (x) \right]_a$ satisfy the coupled equations
\begin{subequations}
\begin{eqnarray}
\left( \frac{\partial^2}{\partial x^2} + k_a ^2  \right)\left[ \psi (x) \right]_a
&=& \sum_{b=1}^{\infty} \left[ \psi (x) \right]_b  {(u_r)}_{ab}\delta (x-x_r) \; ,
\hspace{1cm} 1\leq a\leq N
\label{coupled eqs open}\\
\left( \frac{\partial^2}{\partial x^2} - \kappa _a ^2  \right)\left[ \psi (x) \right]_a
&=& \sum_{b=1}^{\infty} \left[ \psi (x) \right]_b  {(u_r)}_{ab}\delta (x-x_r) \; ,
\hspace{1cm} a\geq N+1  .
\label{coupled eqs closed}
\end{eqnarray}
\label{coupled eqs}
\end{subequations}
Eq. (\ref{coupled eqs open}) refers to {\em open channels} and
Eq. (\ref{coupled eqs closed}) to {\em closed} ones.
The quantity $k_a$, defined by the relation
\begin{equation}
k_a^2 = k^2 -\left(\frac{\pi a}{W} \right)^2 ,
\label{ka}
\end{equation}
is the ``longitudinal" momentum for the open channel $a$, with the replacement
$k_a \Rightarrow i\kappa _a$ for closed channels \cite{mello-kumar}.
Notice that if
$N\pi < kW < (N+1)\pi$,
the problem admits precisely $N$ open channels.

The open-channel, $2N$-dimensional, transfer matrix $\bm{M}$
(that relates open-channel amplitudes on both sides of the potential) for the $r$-th slice,
to be designated by $\bm{M}_r$, will be written as
\begin{equation}
\bm{M}_r
=\left[
\begin{array}{cc}
M^{11}_r & M^{12}_r \\
\left[M^{12}_r\right]^* & \left[M^{11}_r\right]^*
\end{array}
\right]
\equiv\bm{I}_{2N}+\bm{\epsilon} _r  .
\label{Mr N 1}
\end{equation}
Since, eventually, we shall be interested in the limit of weak scatterers
in which $\bm{M}_r$ is close to the unit matrix, we have introduced
the difference $\bm{\epsilon} _r$ between $\bm{M}_r$ and the $2N$-dimensional unit matrix $\bm{I}_{2N}$.
In the above equation we have taken into account explicitly the fact that our system obeys time-reversal invariance (see App. \ref{M-properties}).
The $11$ and $12$ blocks of the matrix $\epsilon _r$ are
given by
\begin{subequations}
\begin{eqnarray}
(\epsilon _r)_{ab}^{11} &=& -i(\hat{v}_r)_{ab}e^{-i (k_a - k_b)x_r}
\equiv (\hat{v}_r)_{ab} (\vartheta _r)^{11} _{ab}
\label{eps11 N}
\\
(\epsilon _r)_{ab}^{12} &=& -i(\hat{v}_r)_{ab}e^{-i (k_a + k_b)x_r}
\equiv (\hat{v}_r)_{ab} (\vartheta _r)^{12} _{ab} \; ,
\label{eps 12 N}
\end{eqnarray}
\label{eps}
\end{subequations}
where $a$ and $b$ label the open channels and thus run from $1$ to $N$.
We have defined
\begin{equation}
(\vartheta _r)_{ab}^{jl}=(\vartheta (x_r))_{ab}^{jl} = i (-)^j
e^{i[(-)^j k_a + (-)^{l+1} k_b]x_r}
\label{theta}
\end{equation}
and we have introduced the {\em real} quantities
\begin{equation}
(\hat{v}_r)_{ab} =\frac{(\hat{u}_r)_{ab}}{2\sqrt{k_a k_b}}    \; ,
\label{vr N}
\end{equation}
where, as explained in App. \ref{veff}, $(\hat{u}_r)_{ab}$ is an ``effective" potential strength that takes into account
transitions to closed channels
[see also Ref. \cite{mello-kumar}, Eq. (3.134)].

In the above equations the strength of the various scatterers
is arbitrary.
As we already indicated, we shall be interested in
the situation of {\em weak scatterers}, defined by the inequality
\begin{equation}
|(\hat{u}_r)_{ab}| \ll \sqrt{k_a k_b} \; ,
\label{weak sc}
\end{equation}
which has to be added to the inequalities (\ref{regime}) in order
to complete the specification of the physical regime.

\subsection{Construction of the Building Block.
The regime (\ref{arb wl}).}
\label{construc of BB in general regime}

\subsubsection{The statistical model}
\label{stat model}

\begin{figure}
\epsfig{file=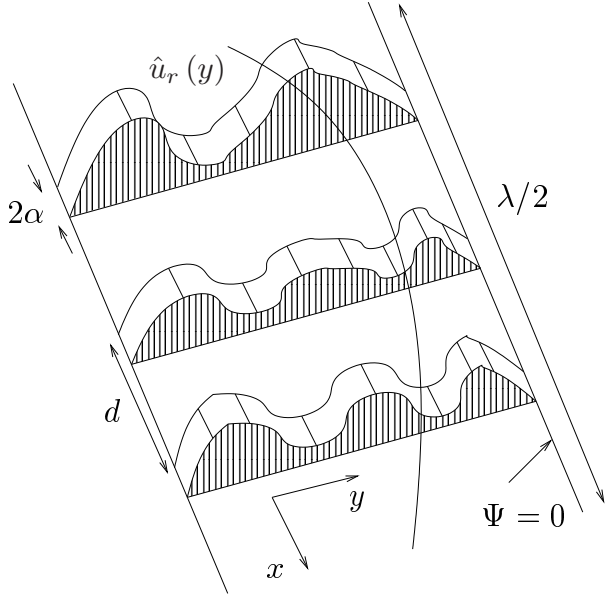,width=8cm,clip=}
\caption{Schematic representation of the construction of the building block
(BB) as a collection of ``thin potential slices".
}
\label{sketchDeltaSlice}
\end{figure}

The BB is assumed, for the time being, centered at $x=0$.
For the application to Eq. (\ref{avF(L+dL)}) the BB
will have to be translated to the interval $(L, L+\delta L)$; this will be done in
Sec. \ref{D and diff eqn}.
The BB is constructed
from $m$ delta slices located at the positions $x_r$
(see Fig.
\ref{sketchDeltaSlice}), i.e., assuming $m$ to be odd,
\begin{subequations}
\begin{eqnarray}
x_r &=&rd \\
r&=&-\frac{m-1}{2},\cdots,0,\cdots,\frac{m-1}{2}\\
\delta L &=& (m-1)d ,
\label{xr}
\end{eqnarray}
\end{subequations}
where $d$ denotes the distance between successive slices and
$\delta L$ the thickness of the BB.

The $m$ potentials $\hat{u}_r (y)$, $r=1,\cdots,m$, are assumed to be
{\em statistically independent} and
{\em identically distributed}.
We indicate the $p$-th moments of the individual $\hat{u}_r (y)$'s
and $\hat{v}_r (y)$ (which are related by the definition
(\ref{vr N})) as
\begin{subequations}
\begin{eqnarray}
\mu_p^{(u)}(a_1 b_1,a_2 b_2,\cdots, a_{p}b_{p})
=\Big\langle
(\hat{u}_r)_{a_1 b_1} (\hat{u}_r)_{a_2 b_2} \cdots (\hat{u}_r)_{a_{p}b_{p}}
\Big\rangle
\label{mu_p_u}
\\
\mu_p^{(v)}(a_1 b_1,a_2 b_2,\cdots, a_{p}b_{p})
=\Big\langle
(\hat{v}_r)_{a_1 b_1} (\hat{v}_r)_{a_2 b_2} \cdots (\hat{v}_r)_{a_{p}b_{p}}
\Big\rangle .
\label{mu_p_v}
\end{eqnarray}
\label{mu_p}
\end{subequations}
We assume, for simplicity, that all odd moments vanish, i.e.,
\begin{equation}
\mu_{2t+1}^{(u)}(a_1 b_1,a_2 b_2,\cdots, a_{2t+1}b_{2t+1}) = 0 .
\label{mu_(2t+1)}
\end{equation}
We thus have
\begin{subequations}
\begin{eqnarray}
\langle  (\hat{u}_r)_{ab} \rangle &=& \mu _1^{(u)}(ab) = 0
\label{<u>}
\\
\Big\langle  (\hat{u}_r)_{ab} (\hat{u}_s)_{cd}\Big\rangle
&=& \mu _2^{(u)}(ab,cd) \; \delta _{rs} \;
\label{<vv>}
\\
\cdots \;\;
\nonumber
\end{eqnarray}
\label{v,vv}
\end{subequations}
and similarly for the $\hat{v}_r$'s.
It is useful to introduce the correlation coefficient between
the matrix elements $(\hat{u}_r)_{ab}$ and $(\hat{u}_r)_{cd}$,
(which coincides with the  correlation coefficient between
$(\hat{v}_r)_{ab}$ and $(\hat{v}_r)_{cd}$) as
\begin{eqnarray}
C(ab,cd)
=\frac{\mu _2^{(u)}(ab,cd)}
{[\mu _2^{(u)}(ab) \mu _2^{(u)}(cd)]^{1/2}}
=\frac{\mu _2^{(v)}(ab,cd)}
{[\mu _2^{(v)}(ab) \mu _2^{(v)}(cd)]^{1/2}},
\label{corr u or v}
\end{eqnarray}
where $\mu _2^{(v)}(ab)\equiv \mu _2^{(v)}(ab,ab)$ denotes the variance
of $(\hat{v}_r)_{ab}$
(recall that $\mu _1^{(v)}(ab)=0$).
For even moments higher than the second we do not make, at this point, any special assumption; a particular scaling law will be assumed in  Eq. (\ref{mu_(2t)}) below.


From the statistics of the $(\hat{u}_r)_{ab}$'s
(and $(\hat{v}_r)_{ab}$'s) we can find the statistics of the
$({\epsilon }_r)_{ab}^{ij}$, using the relations
(\ref{eps}).
For instance, we find that the first moment of $({\epsilon }_r)_{ab}^{ij}$ vanishes, i.e.,
\begin{equation}
\Big\langle  (\epsilon _r)_{ab}^{ij} \Big\rangle = 0
\label{<eps r>}
\end{equation}
and that the second moments can be written as
\begin{equation}
\Big\langle  (\epsilon _r)_{ab}^{ij} (\epsilon _s)_{cd}^{hl}\Big\rangle
= \mu _2^{(v)}(ab,cd)
\left[ (\vartheta _r)_{ab}^{ij} (\vartheta _r)_{cd}^{hl}  \right] \delta _{rs} ,
\label{<epsr ^2>}
\end{equation}
where $(\vartheta _r)_{ab}^{ij}$ was defined in Eqs. (\ref{eps}) and
(\ref{theta}). The individual transfer matrices depend on the slice position
$x_r$ and, as a consequence, they are not identically distributed.

\subsubsection{The transfer matrix for the Building Block. Its first and second moments}
\label{first and second moments of BB}

The transfer matrix for the total sequence of $m$ delta slices is given by
\begin{subequations}
\begin{eqnarray}
\bm{M}^{(m)} &=&\bm{M}_m  \bm{M}_{m-1}\cdots \bm{M}_1
\label{M BB 1}\\
&=& \left(\bm{I}_{2N}+ \bm{\epsilon}_m \right)
\left(\bm{I}_{2N}+\bm{\epsilon}_{m-1} \right) \cdots
\left(\bm{I}_{2N}+\bm{\epsilon}_1 \right)
\label{M BB 2}\\
&=& \bm{I}_{2N}+\sum_r \bm{\epsilon}_r+ \sum_{r_1 > r_2}
\bm{\epsilon}_{r_1}  \bm{\epsilon}_{r_2}
+ \cdots
\nonumber
\\
&&  \cdots  + \sum_{r_1>...>r_{\mu }} \bm{\epsilon}_{r_1}  \cdots
\bm{\epsilon} _{r_{\mu }} + \cdots
\label{M BB 3}
\\
&\equiv& \bm{I}_{2N} + \bm{\varepsilon} .
\label{M BB 4}
\end{eqnarray}
\label{M BB 5}
\end{subequations}
The last line defines the matrix $\bm{\varepsilon}$
[that was already introduced in Eq. (\ref{M BB})] by which the total transfer matrix $\bm{M}$ of the BB differs from the unit matrix $\bm{I}_{2N}$; it is given by
\begin{subequations}
\begin{eqnarray}
\bm{\varepsilon}
&=&\sum_r \bm{\epsilon}_r
+ \sum_{r_1 > r_2} \bm{\epsilon}_{r_1}  \bm{\epsilon}_{r_2}
+ \cdots +\sum_{r_1>...>r_{\mu }} \bm{\epsilon}_{r_1}  \cdots \bm{\epsilon} _{r_{\mu }} + \cdots .
\label{eps BB 5}
\\
&\equiv&\sum_{\mu =1}^{m} \bm{\varepsilon}^{(\mu )},
\label{eps BB 6}
\end{eqnarray}
\label{eps BB 7}
\end{subequations}
where the last line defines the contribution to $\bm{\varepsilon}$ of order $\mu $ in the individual $\epsilon _r$'s.
Our aim is to find the statistical properties
--in particular the moments--
of the matrix $\bm{\varepsilon}$.
In the future we shall use the notation $\left\langle \cdots\right\rangle _{\delta L}$ to indicate an average associated with the BB, i.e.,
\begin{equation}
\left\langle G(\bm{M}) \right\rangle _{\delta L}
\equiv \left\langle G(\bm{M}^{(m)}) \right\rangle ,
\label{<>dL}
\end{equation}
just as in Eq. (\ref{<>n}).
For the average of $\bm{M}$ we trivially find, from Eqs. (\ref{M BB 1}), (\ref{M BB 2})
and the fact the various $\epsilon _r$' are statistically independent and average to zero [Eq. (\ref{<eps r>})],
\begin{equation}
\left\langle \bm{M}\right\rangle_{\delta L}
= \left\langle
\bm{M}_m\right\rangle \cdots \left\langle \bm{M}_1 \right\rangle =I_{2N} .
\label{av M}
\end{equation}
Thus Eq. (\ref{M BB 4}) implies that the {\em first moment} of $\bm{\varepsilon} $ vanishes, i.e.,
\begin{equation}
\left\langle \bm{\varepsilon} \right\rangle_{\delta L} =0 ,
\label{av eps BB}
\end{equation}
as could also have been obtained by averaging Eq. (\ref{eps BB 7}) directly:
\begin{subequations}
\begin{eqnarray}
\left\langle \bm{\varepsilon} \right\rangle_{\delta L}
&=&\sum_{\mu =1}^{m}
\left\langle \bm{\varepsilon}^{(\mu )} \right\rangle_{\delta L}
\\
&=& \sum_r \left\langle\bm{\epsilon}_r \right\rangle
+ \sum_{r_1 > r_2} \left\langle\bm{\epsilon}_{r_1}  \bm{\epsilon}_{r_2} \right\rangle
+ \cdots
+ \sum_{r_1>...>r_{\mu }}\left\langle \bm{\epsilon}_{r_1}  \cdots \bm{\epsilon} _{r_{\mu }} \right\rangle + \cdots
\nonumber \\
&=& 0.
\label{av eps BB 2}
\end{eqnarray}
\end{subequations}

For the {\em second moments} of $\varepsilon $ we have, from
Eq. (\ref{eps BB 6})
\begin{subequations}
\begin{eqnarray}
\left\langle  \varepsilon_{a b}^{i j}
\varepsilon_{c d}^{h l} \right\rangle _{\delta L}
&=& \sum_{\mu, \mu ' =1}^{m}
\left\langle  \left[ \varepsilon ^{(\mu )} \right]_{a b}^{i j}
\left[ \varepsilon ^{(\mu ')} \right]_{c d}^{h l} \right\rangle _{\delta L}
\label{vareps^2}
\nonumber \\
&=&
\left\langle  \left[ \varepsilon ^{(1)} \right]_{a b}^{i j}
\left[ \varepsilon ^{(1)} \right]_{c d}^{h l} \right\rangle _{\delta L}
\label{vareps^2 2}
\\
&+&\left\langle  \left[ \varepsilon ^{(1)} \right]_{a b}^{i j}
\left[ \varepsilon ^{(2)} \right]_{c d}^{h l} \right\rangle _{L'}
+\left\langle  \left[ \varepsilon ^{(2)} \right]_{a b}^{i j}
\left[ \varepsilon ^{(1)} \right]_{c d}^{h l} \right\rangle _{\delta L}
\label{vareps^2 3}
\\
&+&\left\langle  \left[ \varepsilon ^{(2)} \right]_{a b}^{i j}
\left[ \varepsilon ^{(2)} \right]_{c d}^{h l} \right\rangle _{\delta L}
+\left\langle  \left[ \varepsilon ^{(3)} \right]_{a b}^{i j}
\left[ \varepsilon ^{(1)} \right]_{c d}^{h l} \right\rangle _{\delta L}
+\left\langle  \left[ \varepsilon ^{(1)} \right]_{a b}^{i j}
\left[ \varepsilon ^{(3)} \right]_{c d}^{h l} \right\rangle _{\delta L}
\label{vareps^2 4}
\\
&+& \cdots . \nonumber
\end{eqnarray}
\label{vareps^2 series}
\end{subequations}
The second line, Eq. (\ref{vareps^2 2}), is {\em second order} in the individual
$[\epsilon _r]_{a b}^{i j}$ and hence in the potentials $(\hat{v}_r)_{a b}$,
and the successive lines are higher order in these quantities.

\paragraph{The second-order term in the second-moment expansion,
Eq. (\ref{vareps^2 2})}
\label{sec_order sec_mom}

The second-order term, Eq. (\ref{vareps^2 2}), in the second moment expansion
can be written using Eqs. (\ref{eps BB 7}) and (\ref{<epsr ^2>}) as
\begin{subequations}
\begin{eqnarray}
\left\langle  \left[ \varepsilon ^{(1)} \right]_{a b}^{i j}
\left[ \varepsilon ^{(1)} \right]_{c d}^{h l} \right\rangle _{\delta L}
&=&
\sum _{r,s}
\left\langle (\epsilon _{r})_{ab} ^{ij}
(\epsilon _{s})_{cd}^{hl}  \right\rangle
\label{var-eps^2 1b}
\\
&=&\sum _{r,s}
\langle (\hat{v}_r)_{ab} (\hat{v}_s)_{cd}  \rangle
\left[ (\vartheta _r)_{ab}^{ij} (\vartheta _s)_{cd}^{hl} \right]
\label{var-eps^2 2b}
\\
&=&
\frac{\mu _2^{(v)}(ab,cd)}{d}
\sum _{r}\left[ (\vartheta _r)_{ab}^{ij} (\vartheta _r)_{cd}^{hl} \right] \cdot d .
\label{var-eps^2 2c}
\end{eqnarray}
\label{var-eps^2 2}
\end{subequations}

From the definition of the correlation coefficient between pairs of matrix elements, Eq.  (\ref{corr u or v}),
we can write the fraction in Eq. (\ref{var-eps^2 2c}) as
\begin{eqnarray}
\frac{\mu _2^{(v)}(ab,cd)}{d}
= C(ab,cd)
\left[
\frac{\mu _2^{(v)}(ab)}{d}
\; \frac{\mu _2^{(v)}(cd)}{d}
\right ]^{1/2}
\equiv \frac{C(ab,cd)}{\sqrt{\ell_{ab}(k) \; \ell_{cd}(k)}} \; .
\label{<vv>/d}
\end{eqnarray}
Here we have used the standard definition of the
{\em mean free path} (mfp)
$\ell_{ab}$
associated with the incoherent sum of reflections from channel $b$ to $a$
from a sequence of $\nu =1/d$ scatterers per unit length, i.e.,
\begin{equation}
\frac{1}{\ell_{ab}(k)}
=\nu \Big\langle \left|(r_1 (k))_{ab}\right| ^2\Big\rangle ,
\label{lab 1}
\end{equation}
together with the fact that the average reflection coefficient
for a delta slice is $r$-independent and approximately given,
in the weak-scattering regime, Eq. (\ref{weak sc}), by
[see Eqs. (\ref{r,t(M)beta1}), (\ref{Mr N 1}) and (\ref{eps})]
\begin{equation}
\Big\langle|(r_1)_{ab}|^2 \Big\rangle
\approx\Big\langle[(\hat{v}_1)_{ab}]^2\Big\rangle .
\label{<R>}
\end{equation}
We can write the following equivalent expressions for the inverse mfp:
\begin{equation}
\frac{1}{\ell_{ab}(k)}
=\nu \mu _2^{(v)}(ab)
=\frac{\mu _2^{(v)}(ab)}{d}
=\frac{\mu _2^{(u)}(ab)}{4k_a k_b d}
\equiv \frac{\tilde{\mu}_2^{(u)} (ab)}{4k_a k_b} .
\label{lab}
\end{equation}
where the energy dependence of the mfp is exhibited explicitly.
In the last member of the above equation we have defined the quantity $\tilde{\mu}_2^{(u)} (ab)$ as
\begin{equation}
\tilde{\mu}_2^{(u)} (ab) = \frac{\mu _2^{(u)}(ab)}{d}  \; .
\label{mu_2_tilde}
\end{equation}
Since our delta slice is spatially symmetric in the $x$ direction, we have the
same result for the mfp for the transmission, out of the incident flux, from
channel $b$ to channel $a$. Within the present model there is thus no
distinction between the so called {\em transport} and {\em scattering mfp}'s
 \cite{Ziman}.

We now  turn to the summation in Eq. (\ref{var-eps^2 2c}).
We shall evaluate it in the {\em dense-weak-scattering limit} (DWSL) which we now define
(see Eqs. (\ref{dwsl}) below).
This limit was already referred to in Secs. \ref{intro} and \ref{fokker-planck}.
Within the regime defined by the inequalities (\ref{arb wl}) we have already
considered $\alpha $ as the smallest length scale occurring in the problem and simplified the
situation by literally taking the limit $\alpha \to 0$ [Eq. (\ref{a 0})].
With regards to the next length scale in our regime, i.e., the distance $d$ between successive
scattering slices, we shall again be interested in a simplifying
limit.
For a {\em fixed energy (and hence fixed $\lambda $), fixed $\delta L$ and mfp's},
it will be convenient to take the {\em continuous limit}
\begin{subequations}
\begin{eqnarray}
d &\to& 0
\label{d to 0}
\\
m &\to& \infty,
\label{CL}
\end{eqnarray}
in such a way that
\begin{equation}
md = \delta L
\end{equation}
remains fixed.
In order to keep the mfp's $\ell_{ab}$ of Eq. (\ref{lab}) finite,
we carry to an extreme the weak-scattering condition (\ref{weak sc})
and, {\em for a fixed energy}, take the limit in which the individual scattering units become infinitely weak, i.e.,
\begin{equation}
\mu _2^{(u)}(ab) \to 0 ,
\end{equation}
\label{dwsl}
\end{subequations}
in such a way that, together with $d\to 0$, Eq. (\ref{d to 0}), the ratio $\tilde{\mu}_{2}^{(u)}(ab)$
of Eq. (\ref{mu_2_tilde}), and hence the $\ell_{ab}(k)$ of Eq. (\ref{lab}), remain fixed.
This limit has to be considered as the extreme idealization of the inequality (\ref{weak sc}) and
of the inequality
$d \ll \{\lambda, \delta L, \ell \}$ of (\ref{arb wl}) for fixed energy,
$\delta L$ and mfp's.

We have already assumed in Eq. (\ref{mu_(2t+1)}) that all the odd moments of $\hat{u}$ and $\hat{v}$ vanish. We shall now assume that the even moments,
Eq. (\ref{mu_p}) with $p=2t$, scale with $d$ as
\begin{equation}
\mu_{2t}^{(u)}(a_1b_1, \cdots, a_{2t}b_{2t})
=d^{t}\; \tilde{\mu}_{2t}^{(u)}(a_1b_1, \cdots, a_{2t}b_{2t}),
\label{mu_(2t)}
\end{equation}
$\tilde{\mu}_{2t}^{(u)}(a_1b_1, \cdots, a_{2t}b_{2t})$ being independent of $d$, with a similar expression for
$\mu_{2t}^{(v)}(a_1b_1, \cdots, a_{2t}b_{2t})$.
Eq. (\ref{mu_2_tilde}) is the particular case of this last equation for $t=1$.

In the DWSL, the $\sum_r$ appearing in Eq. (\ref{var-eps^2 2c})
tends to an integral, which we denote by
\begin{eqnarray}
\Delta_{ab,cd}^{ij,hl}(k, \delta L)
&\equiv& \lim_{DWS}
\sum _{r}\left[ (\vartheta _r)_{ab}^{ij} (\vartheta _r)_{cd}^{hl} \right] \cdot d
\nonumber \\
&=& \int_{- \delta L /2}^{\delta L /2}
\vartheta _{ab}^{ij}(x) \vartheta _{cd}^{hl}(x) dx \; ,
\label{Delta(k)}
\end{eqnarray}
where $\vartheta _{ab}^{ij}(x)$ is given by Eq. (\ref{theta}) with $x_r$ replaced by $x$.
We find explicitly
\begin{equation}
\Delta _{ab,cd}^{ij,hl}(k, \delta L)
= (-)^{i+h+1}\frac{\sin\frac{K_{ab,cd}^{ij,hl} \; \delta L}{2}}
{\frac{K_{ab,cd}^{ij,hl}}{2}} \; ,
\label{Delta}
\end{equation}
a quantity with dimensions of length, $K_{ab,cd}^{ij,hl}$ being given by
\begin{equation}
K_{ab,cd}^{ij,hl} = (-1)^i k_a + (-1)^{j+1} k_b + (-1)^h k_c +
(-1)^{l+1} k_d . 
\label{K}
\end{equation}
From Eq. (\ref{K}), and using the notation of Eq. (\ref{bar}), we readily find the symmetry relations
\begin{equation}
K_{ab,cd}^{ij,hl}=K_{cd,ab}^{hl,ij}
=-K_{ab,cd}^{\bar{i}\bar{j},\bar{h}\bar{l}}   \;,
\label{symm_K}
\end{equation}
so that
\begin{equation}
\Delta _{ab,cd}^{ij,hl}(k, \delta L)
=\Delta _{cd,ab}^{hl,ij}(k, \delta L)
=\Delta _{ab,cd}^{\bar{i}\bar{j},\bar{h}\bar{l}}(k, \delta L)   \;.
\label{symm_Delta}
\end{equation}

We thus have, for the expression (\ref{var-eps^2 2}) in the DWSL:
\begin{equation}
\lim_{DWS}
\left\langle
\left[ \varepsilon ^{(1)} \right]_{a b}^{i j}
\left[ \varepsilon ^{(1)} \right]_{c d}^{k l}
\right\rangle _{\delta L}
=\frac{C(ab,cd)}
{\sqrt{\ell_{ab}(k) \ell_{cd}(k)}}
\Delta_{ab,cd}^{ij,hl}(k, \delta L)  \; ,
\label{<vareps^2>}
\end{equation}
{\em a result valid for arbitrary $k$ and $\delta L$}.

For the application to Eq. (\ref{avF(L+dL)})
we shall need the expansion of the moments of $\varepsilon $ in powers of 
$\delta L$, with
the BB translated to the interval $(L, L+\delta L)$; 
this will be done in Sec. \ref{D and diff eqn} below.
For the time being we perform that expansion, for simplicity, with the BB centered at the origin.
We see from Eq. (\ref{Delta}) that the leading term of
$\Delta _{ab,cd}^{ij,hl}(k, \delta L)$
in an expansion in powers of $\delta L$ is linear in $\delta L$
[as is obvious from the integral definition itself, Eq. (\ref{Delta(k)})], i.e.,
\begin{equation}
\Delta _{ab,cd}^{ij,hl}(k, \delta L)
= (-)^{i+h+1}\delta L + O(\delta L)^2.
\label{Delta_leading}
\end{equation}
As a result, Eq. (\ref{<vareps^2>}) shows that {\em the leading term in an expansion
in powers of $\delta L$
of the second-order contribution to the second moments of $\varepsilon$
for the BB behaves, in the DWSL}, as
\begin{equation}
\lim_{DWS}
\left\langle
\left[ \varepsilon ^{(1)} \right]_{a b}^{i j}
\left[ \varepsilon ^{(1)} \right]_{c d}^{h l}
\right\rangle _{\delta L}
= (-)^{i+h+1} \frac{C(ab,cd)}{\sqrt{\ell_{ab}(k) \ell_{cd}(k)}} \delta L
+ O(\delta L)^2 \;.
\label{<vareps^2>(1,1)leading}
\end{equation}

\paragraph{The fourth-order term in the second-moment expansion, Eq. (\ref{vareps^2 4})}
\label{second moment BB propto v2}

A similar analysis is performed in App. \ref{4th_order in second_moment},
Eq. (\ref{vareps(2)vareps(2) DWSL}),
for the fourth-order contribution to the second moments of
$\varepsilon $, Eq. (\ref{vareps^2 4}):
it is shown that the leading term of such a quantity, in an expansion in powers of $\delta L$, behaves, in the DWSL, as
$\left(\delta L / \ell \right)^2$,
where $\ell$ denotes a typical mfp [see Eq. (\ref{<vareps^2>leading 1})].
From this result and Eq. (\ref{<vareps^2>(1,1)leading}) we thus have
\begin{equation}
\lim_{DWS}
\left\langle
\varepsilon _{a b}^{i j}
\varepsilon _{c d}^{h l}
\right\rangle _{\delta L}
= (-)^{i+h+1}
\frac{C(ab,cd)}
{\sqrt{\ell_{ab}(k)\ell_{cd}(k)}} \delta L
+ O(\delta L)^2 \;.
\label{<vareps^2>leading}
\end{equation}

The analysis of the two above particular cases is generalized to
arbitrary moments in App. \ref{pth moment varepsilon}.
For an even moment ($p=2t$) in the DWSL,
the lowest-order term in Eq. (\ref{vareps^p})
(this term is of order $2t$ in the $\hat{v}_r$'s)
has a leading term in an expansion in powers of $\delta L$
which behaves as $(\delta L / \ell)^t$.
Higher-order terms in (\ref{vareps^p}) are higher order in $\delta L$.
Also, {\em the dependence on the cumulants of the potential higher than the second drops out in the DWSL}.
The contribution to the second moments obtained above,
Eq. (\ref{<vareps^2>leading}),
represents, for $t=1$, a particular case of this general result.
For an odd moment ($p=2t+1$), the corresponding term behaves as
$(\delta L / \ell)^{t+1}$.

In conclusion, this proves the behavior of the moments of $\varepsilon$
that was mentioned in Sec. \ref{fokker-planck}, right after Eq. (\ref{avF(L+dL)}).

\subsection{The Diffusion Coefficients and the Diffusion Equation}
\label{D and diff eqn}

We now generalize the above analysis to the situation in which the BB lies in the interval $(L, L+\delta L)$.
The integral in
Eq. (\ref{Delta(k)}) has to be performed in that interval
[the notation $\langle \cdots \rangle _{L, \delta L}$ in
Eq. (\ref{avF(L+dL)}) and in some of the following equations indicates this fact] and
Eq. (\ref{<vareps^2>}) becomes
\begin{equation}
\lim_{DWS}
\left\langle
\left[ \varepsilon ^{(1)} \right]_{a b}^{i j}
\left[ \varepsilon ^{(1)} \right]_{c d}^{h l}
\right\rangle _{L, \delta L}
=\frac{C(ab,cd)}
{\sqrt{\ell_{ab}(k) \ell_{cd}(k)}}
\Delta_{ab,cd}^{ij,hl}(k, \delta L)
e^{iK_{ab,cd}^{ij,hl}(L+\frac{\delta L}{2})} ,
\label{<vareps^2> L,dL}
\end{equation}
while the expansion in Eq. (\ref{<vareps^2>leading})
[taking into account Eq. (\ref{<vareps^2>leading 2})] is now
\begin{eqnarray}
\lim_{DWS}
\left\langle \varepsilon ^{ij}_{ab}
\varepsilon ^{hl}_{cd}
\right\rangle _{L, \delta L}
&=& 2D^{ij,hl}_{ab,cd}(k,L)\delta L
\nonumber \\
&& + \left[
iK_{ab,cd}^{ij,hl}D_{ab,cd}^{ij,hl}(k,L)
+2\sum_{\alpha '\beta ', \lambda ' \mu '}
D_{a \alpha ',c \beta '}^{i\lambda ',h \mu '}(k,L)
D_{\alpha ' b,\beta 'd}^{\lambda 'j,\mu 'l}(k,L)
\right](\delta L)^2
\nonumber \\
&& +O(\delta L)^3 \; ,
\label{<eps.eps>_D}
\end{eqnarray}
where $K_{ab,cd}^{ij,hl}$ was defined in Eq. (\ref{K}).
In Eq. (\ref{<eps.eps>_D}) we have defined the ``diffusion coefficients''
$D^{ij,hl}_{ab,cd}(k,L)$
\begin{eqnarray}
D^{ij,hl}_{ab,cd}(k,L)
&=&(-)^{i+ h +1}
\frac{C(ab,cd)}{2\sqrt{\ell_{ab}(k)\ell_{cd}(k)}}
e^{iK_{ab,cd}^{ij,hl}L} \; ,
\label{D(k,L)}
\end{eqnarray}
which {\em depend on the energy}
(through the energy dependence of the mfp's and through $K_{ab,cd}^{ij,hl}$)
and also {\em on the length} $L$.
Notice that the diffusion coefficients are, in general, complex numbers;
this, however, should not worry the reader, because the evolution of real observables will always turn out to be real
[see, for instance, Eq. (\ref{e-alfa/alfa*}) below].

From the relations (\ref{symm_K}) we readily find for the diffusion coefficients the symmetry properties
\begin{equation}
D^{ij,hl}_{ab,cd}(k,L)
=D^{hl,ij}_{cd,ab}(k,L)
=[D^{\bar{i} \bar{j},\bar{h}\bar{l}}_{ab,cd}(k,L)]^{\ast}.
\label{symm_D}
\end{equation}

We introduce the expansion (\ref{<eps.eps>_D}) and a similar one for higher
moments of $\varepsilon $ on the r.h.s. of Eq. (\ref{avF(L+dL)}),
thus obtaining a power series in $\delta L$:
\begin{eqnarray}
\left\langle F(\bm{M})\right\rangle _{L + \delta L}
&=&\left\langle F(\bm{M})\right\rangle _L
+\sum_{\substack{ ijhl, \lambda \mu  \\ abcd, \alpha \beta }}
\Bigg\{
D_{ab,cd}^{ij,hl}(k,L) \delta L
\nonumber \\
&&
+ \left[
\frac12 iK_{ab,cd}^{ij,hl}  D_{ab,cd}^{ij,hl}(k,L)
+
\sum_{\alpha '\beta ', \lambda ' \mu '}
D_{a \alpha ',c \beta '}^{i\lambda ',h \mu '}(k,L)
D_{\alpha ' b,\beta 'd}^{\lambda 'j,\mu 'l}(k,L)
\right]
(\delta L)^2
\Bigg\}
\nonumber \\
&&\hspace{1cm}\times
\left\langle
M_{b \alpha }^{j \lambda } \; M_{d \beta }^{l \mu } \;
\frac{\partial ^2F(\bm{M})} {\partial M_{a \alpha }^{i \lambda }\;
\partial M_{c \beta }^{h \mu}} \right\rangle _L
\nonumber \\
&&+ O(\delta L)^2 \; .
\label{avF(L+dL) a}
\end{eqnarray}
The curly bracket in this last equation corresponds to the BB second moment of Eq. (\ref{<eps.eps>_D});
the contribution [which starts with $(\delta L)^2$]
 of the third and higher moments is just indicated in the last line.
We also perform on the l.h.s. of Eq. (\ref{avF(L+dL)}) a Taylor expansion of
$\langle F(\bm{M})\rangle_{L+\delta L}$ in powers of $\delta L$ around the ``initial" value $\langle F(\bm{M})\rangle_L$, i.e.,
\begin{equation}
\left\langle F(\bm{M})\right\rangle _{L+\delta L}
=\left\langle F(\bm{M})\right\rangle _{L}
+\frac{\partial \left\langle F(\bm{M})\right\rangle _{L}}{\partial L} \;\delta L
+\frac{1}{2!}
\frac{\partial^2 \left\langle F(\bm{M})\right\rangle _{L}}
{\partial L^2} \;(\delta L)^2 + \cdots.
\label{Taylor <F>(L+dL) k}
\end{equation}
We then identify the coefficients of the various powers of $\delta L$
in Eqs. (\ref{avF(L+dL) a}) and (\ref{Taylor <F>(L+dL) k}).
In particular, the coefficients of $\delta L$ give the diffusion equation,
Eq. (\ref{diff eqn exp val sec2}), derived in Sec. \ref{fokker-planck},
which we reproduce here:
\begin{eqnarray}
&&\frac{\partial \left\langle F(\bm{M})\right\rangle _L}{\partial L}
= \sum_{\substack{ ijhl\lambda \mu  \\ abcd\alpha \beta }}
D_{ab,cd}^{ij,hl}(k,L) \left\langle M_{b \alpha }^{j \lambda } \;
M_{d \beta }^{l \mu} \; \frac{\partial ^2F(\bm{M})} {\partial M_{a
\alpha }^{i \lambda } \partial M_{c \beta }^{h \mu}} \right\rangle
_L .
\label{diff eqn exp val sec3}
\end{eqnarray}
Equating the coefficients of higher powers of $\delta L$ in
Eqs. (\ref{avF(L+dL) a}) and (\ref{Taylor <F>(L+dL) k}) we obtain results which could be derived from the diffusion equation (\ref{diff eqn exp val sec3})
by successive differentiations.
We have verified this statement explicitly for the coefficients of $(\delta L)^2$
in the specific one-channel case treated in Sec. \ref{analytic} below.

The diffusion equation (\ref{diff eqn exp val sec3})
governs the evolution with length of the expectation value of physical observables.
The expectation values appearing in Eq. (\ref{diff eqn exp val sec3}) must fulfill, for $L=0$, the ``initial condition"
\begin{equation}
\left\langle F(\bm{M})\right\rangle _{L=0} = F(\bm{I}) ,
\label{in_condition}
\end{equation}
obtained by setting $\bm{M}=\bm{I}$ in the expression $F(\bm{M})$ for the observable, since for $L=0$ the scattering system is absent.
More general initial
conditions are discussed in Ref. \cite{carlo97}.

As was indicated earlier, the cumulants of the potential higher than the second are irrelevant in the end;
this signals the existence of a
{\em generalized central-limit theorem} (CLT):
once the mfp's are specified, the limiting equation
(\ref{diff eqn exp val sec3})
is {\em universal}, i.e., independent of other details of the microscopic statistics.

Since the structure of the present diffusion equation is essentially the same as the structure
of the one derived in MT (Ref. \cite{mello_tomsovic}, Eq. (3.18)),
it is worthwhile, for the sake of comparison, to summarize, at this point, the MT model.
In MT the statistical assumptions are made at the level of the {\em individual scattering units},
just as in the present paper
(the same units that were also contemplated in Ref. \cite{mello_shapiro});
however, the assumptions are not made for the potentials, but rather for
the corresponding transfer matrices.
In MT, the transfer matrix for each scattering unit is close to the unit matrix and is written as
$\bm{M}_r=\bm{I} + \bm{\epsilon }_r$, just as in our Eq. (\ref{Mr N 1}) above;
it is further expressed in terms of independent parameters
(in the Pereyra representation \cite{pereyra}), for which various statistical assumptions are made:

i) The first moment and some of the second moments of the independent parameters are chosen so that
the resulting $\langle\epsilon _r\rangle =0$
(see Eqs. (3.15) and (3.16) of MT;
with this feature, there is no drift term in the resulting Fokker-Planck equation),
while the remaining second moments of the independent parameters are kept arbitrary.

ii) The individual scattering units are statistically independent and identically distributed.

iii) The energy does not appear explicitly, but only as the energy at which the resulting mfp's have
to be evaluated.

iv) In order to obtain explicit expressions for the diffusion coefficients,
in the analysis that follows from Eq. (3.18) of Ref. \cite{mello_tomsovic}
a more explicit model was postulated for the second moments
mentioned in i) above.

In the present paper, assumption i) is a consequence of the vanishing of the first moment of the individual potentials,
Eq. (\ref{<u>}), thus giving Eq. (\ref{<eps r>}).
Assumption ii) has to be contrasted with Eq. (\ref{<epsr ^2>}) above, which shows that, here, the transfer matrices for the individual scattering units are not identically distributed.
As it has already been stressed, in contrast to assumption iii)
the energy appears now explicitly.
Finally, the additional assumptions mentioned in iv) are, to some extent, arbitrary;
they are compared below with those arising from the short-wavelength approximation of the present model.

\subsection{The Short-Wavelength Approximation.
The regime (\ref{swla regime}).}
\label{SWLA}

In the DWSL the above expressions are exact for all energies.
We now turn to a different regime, to be called the
{\em short-wavelength approximation} (SWLA),
defined by the inequalities (\ref{swla regime}).
The regime to be studied is analogous to the geometrical optics limit
studied in optics \cite{born_wolf}.
Essentially, we shall assume that we can fit many wavelengths inside a BB, i.e.,
\begin{equation}
\lambda \ll \delta L \;, \;\;\;\;\; {\rm or}  \;\;\;\;\;\;  k \delta L \gg 1 \;  ,
\label{kdL>1}
\end{equation}
so that in this regime
{\em only lengths much larger than the wavelength actually enter
the description}.

To this end we go back to Eq. (\ref{avF(L+dL)}) which,
after setting
$\left\langle \varepsilon _{ac}^{ik}\right\rangle_{L,\delta L}=0$
because of (\ref{av eps BB}), we rewrite here for convenience:
\begin{eqnarray}
\left\langle F(\bm{M})\right\rangle _{L+\delta L , k}
&=&\left\langle F(\bm{M})\right\rangle _{L,k} +  \frac{1}{2!}
\sum_{\substack{ ijhl  \\ abcd }} \left\langle \varepsilon
_{ab}^{ij} \; \varepsilon _{cd}^{hl} \right\rangle _{L,\delta L; k}
\sum_{\substack{ \lambda \mu  \\ \alpha \beta }} \left\langle
M_{b\alpha }^{j \lambda } \; M_{d \beta }^{l\mu } \; \frac{\partial
^2F(\bm{M})} {\partial M_{a \alpha}^{i \lambda }\;
\partial M_{c \beta }^{h \mu}}
\right\rangle _{L,k}
\nonumber \\
&&\hspace{1cm} + \cdots .
\label{avF(L+dL) 1}
\end{eqnarray}
We have indicated explicitly the $k$ dependence of the various expectation values.
We first analyze below the BB factors appearing on the r.h.s. of the above equation, and then the remaining expectation values.

1) The BB factor
$\left\langle
\varepsilon _{ab}^{ij} \;
\varepsilon _{cd}^{hl}
\right\rangle _{L,\delta L ; k}$ can be written, from
Eq. (\ref{vareps^2 series}), as
\begin{equation}
\left\langle
\varepsilon _{ab}^{ij} \;
\varepsilon _{cd}^{hl}
\right\rangle _{L,\delta L; k}
= \left\langle
[\varepsilon ^{(1)}]_{ab}^{ij} \;
[\varepsilon  ^{(1)}]_{cd}^{hl}
\right\rangle _{L,\delta L; k}
+\left\langle
[\varepsilon  ^{(2)}]_{ab}^{ij} \;
[\varepsilon  ^{(2)}]_{cd}^{hl}
\right\rangle _{L,\delta L; k}
+\cdots .
\label{ee=e1e1+e2e2}
\end{equation}

The first term on the r.h.s. of this last equation is given by
Eq. (\ref{<vareps^2> L,dL}), and its contribution to
(\ref{avF(L+dL) 1}) is given by

\begin{eqnarray}
&&\frac12 \sum_{\substack{ ijhl  \\ abcd }} \left\langle \left[
\varepsilon ^{(1)} \right]_{ab}^{ij} \left[ \varepsilon ^{(1)}
\right]_{cd}^{hl} \right\rangle _{L,\delta L; k} \sum_{\substack{
\lambda \mu  \\ \alpha \beta }}
\langle \left( \cdot \cdot \cdot \right) ^{ijhl\lambda \mu } \rangle_{L,k}
\nonumber \\
&&\hspace{1cm}
=\frac12
\sum_{\substack{ ijhl  \\ abcd  \\ \left(K=0\right) }} \frac{C(ab,cd)}{\sqrt{\ell_{ab}(k)\ell_{cd}(k)}}
\Delta _{ab,cd}^{ij,hl}(k, \delta L)
e^{iK\left(L+\frac{\delta L}{2}\right)}
\sum_{\substack{ \lambda \mu  \\ \alpha \beta }}
\langle \left( \cdot \cdot \cdot \right) ^{ijhl\lambda \mu }
_{abcd \alpha \beta }  \rangle _{L,k}
\nonumber \\
&&\hspace{1cm} +\frac12
\sum_{\substack{ ijhl  \\ abcd \\ (K \neq 0) }} \frac{C(ab,cd)}{\sqrt{\ell_{ab}(k)\ell_{cd}(k)}}
\Delta _{ab,cd}^{ij,hl}(k, \delta L)
e^{iK\left(L+\frac{\delta L}{2}\right)}
\sum_{\substack{ \lambda \mu  \\ \alpha \beta }}
\langle  \left( \cdot \cdot \cdot \right) ^{ijhl\lambda \mu }
_{abcd \alpha \beta }\rangle _{L,k}
\; .
\nonumber \\
&&\hspace{1cm}
\label{<vareps^2> L 1}
\end{eqnarray}
In this equation, $K$ is an abbreviation for $K_{ab,cd}^{ij,hl}$ which was defined in Eq. (\ref{K}), and $\Delta_{ab,cd}^{ij,hl}(k, \delta L)$ was given in Eq. (\ref{Delta}).
We have also used the notation
\begin{equation}
\left\langle\left( \cdot \cdot \cdot \right) ^{ijhl\lambda \mu }
_{abcd \alpha \beta }\right\rangle _{L,k}
\equiv
\left\langle M^{j\lambda }_{b \alpha }M^{l\mu }_{d \beta } 
\frac{\partial ^{2} F\left(M\right) } 
{\partial M^{i\lambda}_{a \alpha }\partial M^{h\mu}_{c \beta }}
\right\rangle_{L,k} ,
\label{<...> 2}
\end{equation}
as an abbreviation for the last factor appearing on the r.h.s. of Eq. (\ref{avF(L+dL) 1}).
In the one but last line in Eq. (\ref{<vareps^2> L 1}) the sum is over the combinations of  indices that make $K_{ab,cd}^{ij,hl}=0$, while in the
last line it is over those combinations that make
$K_{ab,cd}^{ij,hl} \neq 0$.

The second term on the r.h.s. of Eq. (\ref{ee=e1e1+e2e2}) is given in
Eq. (\ref{vareps(2)vareps(2) DWSL 1}) and, using a similar convention as in the last equation, its contribution to (\ref{avF(L+dL) 1}) can be written as
\begin{eqnarray}
&&\frac12 \sum_{\substack{ ijhl  \\ abcd }} \left\langle \left[
\varepsilon ^{(2)} \right]_{ab}^{ij} \left[ \varepsilon ^{(2)}
\right]_{cd}^{hl} \right\rangle _{L,\delta L; k} \sum_{\substack{
\lambda \mu  \\ \alpha \beta }}
\langle \left( \cdot \cdot \cdot \right) ^{ijhl\lambda \mu }
_{abcd \alpha \beta } \rangle_{L,k}
\nonumber \\
&&\hspace{1cm}= \frac12
\sum_{\substack{ ijhl\lambda ' \mu '  \\ abcd\alpha ' \beta ' \\
(K_1=K_2=0) }}
\frac{C(a \alpha ', c \beta ')}
{\sqrt{\ell_{a \alpha'}(k)\ell_{c \beta '}(k)}}
\frac{C(\alpha ' b, \beta ' d)}
{\sqrt{\ell_{\alpha ' b}(k)\ell_{\beta ' d}(k)}}
\nonumber \\
&&\hspace{2cm}
\times\Delta_{a \alpha ',\alpha ' b, c \beta ',\beta 'd  }
^{i \lambda ',\lambda ' j, h \mu ', \mu ' l} (k, {\cal R}(\delta L)) e^{i(K_1+K_2)(L+\frac{\delta L}{2})}
\sum_{\substack{ \lambda \mu  \\ \alpha \beta }}
\langle \left( \cdot \cdot \cdot \right) ^{ijhl\lambda \mu }
_{abcd \alpha \beta }\rangle_{L,k}
\nonumber \\
&&\hspace{1cm}+
\frac12
\Bigg\{
\sum_{\substack{ ijhl\lambda ' \mu '  \\
abcd\alpha ' \beta ' \\
(K_1\neq 0, K_2\neq 0) }} + \sum_{\substack{ ijhl\lambda ' \mu '  \\
abcd\alpha ' \beta ' \\
(K_1=0, K_2 \neq 0) }} + \sum_{\substack{ ijhl\lambda ' \mu '  \\
abcd\alpha ' \beta ' \\
(K_1 \neq 0, K_2=0) }} \Bigg\} \frac{C(a \alpha ', c \beta)}
{\sqrt{\ell_{a \alpha '}(k)\ell_{c \beta}(k)}} \frac{C(\alpha ' b,
\beta ' d)} {\sqrt{\ell_{\alpha ' b}(k)\ell_{\beta ' d}(k)}}
\nonumber \\ \nonumber \\
&&\hspace{2cm}
\times\Delta_{a \alpha ',\alpha ' b, c \beta ',\beta 'd  }
^{i \lambda ',\lambda ' j, h \mu ', \mu ' l} (k, {\cal R}(\delta L)) e^{i(K_1+K_2)(L+\frac{\delta L}{2})}
\sum_{\substack{\lambda \mu  \\ \alpha \beta }}
\langle \left( \cdot \cdot \cdot \right) ^{ijhl\lambda \mu }
_{abcd \alpha \beta } \rangle_{L,k} \; .
\label{<vareps^2> L 2}
\end{eqnarray}
We recall that $K_1$ and $K_2$ are defined in Eq. (\ref{K1,K2}).

Higher-order contributions occurring on the r.h.s of
Eq. (\ref{ee=e1e1+e2e2}) can be obtained from the analysis of
App. \ref{pth moment varepsilon}.

We now analyze the consequences of the inequality (\ref{kdL>1})
for the above expressions
(\ref{<vareps^2> L 1}) and (\ref{<vareps^2> L 2}),
which so far are exact.
It will be convenient to take the wavenumber $k$ as
\begin{equation}
k=\frac{(N+1/2)\pi }{W},
\label{k in middle}
\end{equation}
i.e., halfway between the threshold for the last open channel and that for the first closed one,
so that the longitudinal momenta are given by
$k_a = k\sqrt{1-[a/(N+1/2)]^2}$.
From Eq. (\ref{K}) we see that when $K_{ab,cd}^{ij,hl} \neq 0$, $K_{ab,cd}^{ij,hl}$ is proportional to $k$
(the coefficients only depending on channel indices), so that
$K_{ab,cd}^{ij,hl} \delta L \gg 1$.
As a result:

i) in Eq. (\ref{<vareps^2> L 1}) the sum with $K=0$ gives the largest contribution
(proportional to $\delta L$, as we now analyze in detail),
while the sum with $K\neq 0$,
which contains $K$ in the denominator of
$\Delta _{ab,cd}^{ij,hl}(k, \delta L)$, will be neglected.

Let us be more specific about the combination of indices $ab,cd$ and
$ij,hl$ that give
rise to $K=0$ in Eq. (\ref{<vareps^2> L 1}).
Take, for instance, $i=j=h=l=1$. Since $k_a, k_b, k_c, k_d$ are incommensurate,
$K_{ab,cd}^{11,11}=k_b - k_a + k_d - k_c$
(see Eq. (\ref{K})) can only vanish if $a=b$ and $c=d$,
or $a=d$ and $b=c$. On the other hand,
$K_{ab,cd}^{12,12}=-(k_a + k_b + k_c + k_d)$ never vanishes.
We thus have, for
$\Delta _{ab,cd}^{ij,hl}(k, \delta L)$, defined for arbitrary $k$ and $\delta L$ in Eq. (\ref{Delta}), the approximate result:
\begin{subequations}
\begin{eqnarray}
\Delta _{ab,cd}^{ij,hl}(k, \delta L) \approx (-1)^{i+h+1}
\delta _{K0}\cdot \delta L,
\label{Delta HEL gen}
\end{eqnarray}
(here, $\delta _{K0}$ is Kronecker's delta which takes on the value 1 when $K=0$ and vanishes otherwise)
or, more explicitly:
\begin{eqnarray}
\Delta _{ab,cd}^{11,11}(k, \delta L)
&\approx&
-\frac{\delta _{ab} \delta _{cd} + \delta _{ad} \delta _{bc}}
{1+\delta _{ac}}
\;\delta L
\\
\Delta _{ab,cd}^{11,22}(k, \delta L)
&\approx&
\frac{\delta _{ab} \delta _{cd} + \delta _{ac} \delta _{bd}}
{1 + \delta _{ad}}
\;\delta L
\\
\Delta _{ab,cd}^{12,21}(k, \delta L)
&\approx&
\frac{\delta _{ac} \delta _{bd} + \delta _{ad} \delta _{bc}}
{1+ \delta _{ab}}
\;\delta L
\\
\Delta _{ab,cd}^{11,12}(k, \delta L)
&\approx& \Delta _{ab,cd}^{11,21}(k, \delta L)
\approx \Delta _{ab,cd}^{12,12} (k, \delta L)
\approx 0 \; .
\end{eqnarray}
\label{Delta HEL}
\end{subequations}
We can thus write
$\left\langle  \left[ \varepsilon ^{(1)} \right]_{a b}^{i j}
\left[ \varepsilon ^{(1)} \right]_{c d}^{h l} \right\rangle _{\delta L}$
in the DWSL, followed by the SWLA, as
\begin{subequations}
\begin{eqnarray}
\lim_{DWS}\left\langle  \left[ \varepsilon ^{(1)} \right]_{a b}^{i j}
\left[ \varepsilon ^{(1)} \right]_{c d}^{k l} \right\rangle _{\delta L}
\approx
\frac{C(ab,cd)}{\sqrt{\ell_{ab}(k)\ell_{cd}(k)}}
(-1)^{i+h+1}
\delta _{K0}\cdot \delta L ,
\label{<vareps2> order2 HEL DWSL}
\end{eqnarray}
One finds explicitly in the various cases
($C(a,c)$ being an abbreviation for $C(aa,cc)$):
\begin{eqnarray}
\lim_{DWS}
\left\langle  \left[ \varepsilon ^{(1)} \right]_{a b}^{11}
\left[ \varepsilon ^{(1)} \right]_{c d}^{11} \right\rangle _{\delta L}
&\approx&
-\frac{1}{1+\delta _{ac}}
\Big[
C(a,c) \frac{\delta L}{\sqrt{\ell_{aa}(k)\ell_{cc}(k)}}
\; \delta _{ab} \delta _{cd}
+ \frac{\delta L}{\ell_{ab}} \delta _{ad}\delta _{bc}
\Big ]
\\  \nonumber \\
\lim_{DWS}
\left\langle  \left[ \varepsilon ^{(1)} \right]_{a b}^{11}
\left[ \varepsilon ^{(1)} \right]_{c d}^{22} \right\rangle _{\delta L}
&\approx&
\frac{1}{1+\delta _{ad}}
\Big[
C(a,c) \frac{\delta L}{\sqrt{\ell_{aa}(k)\ell_{cc}(k)}}
\;\delta _{ab} \delta _{cd}
+ \frac{\delta L}{\ell_{ab}} \delta _{ac}\delta _{bd}
\Big ]
\\ \nonumber \\
\lim_{DWS}
\left\langle  \left[ \varepsilon ^{(1)} \right]_{a b}^{12}
\left[ \varepsilon ^{(1)} \right]_{c d}^{21} \right\rangle _{\delta L}
&\approx& \frac
{\delta _{ac}\delta _{bd} + \delta _{ad}\delta _{bc}}
{1+ \delta _{ab}}
\frac{\delta L}{\ell_{ab}(k)}
\\ \nonumber \\
\lim_{DWS}\left\langle
\left[ \varepsilon ^{(1)} \right]_{ab}^{11}
\left[ \varepsilon ^{(1)} \right]_{cd}^{12}\right\rangle_{\delta L}
&\approx&\lim_{DWS}\left\langle
\left[ \varepsilon ^{(1)} \right]_{ab}^{11}
\left[ \varepsilon ^{(1)} \right]_{cd}^{21}\right\rangle_{\delta L}
\approx \lim_{DWS}\left\langle \left[ \varepsilon ^{(1)} \right]_{ab}^{12}
\left[ \varepsilon ^{(1)} \right]_{cd}^{12}\right\rangle_{\delta L}
\approx  0 .
\nonumber \\
\end{eqnarray}
\label{<vareps vareps DWSL HEL>}
\end{subequations}
Other combinations can be found from TRI, Eqs. (\ref{TRI eps 1}).
The result is that
{\em in the DWSL, followed by the SWLA, the second-order contribution
to a second moment of $\varepsilon $ for the BB is either negligible or
behaves as $\delta L / \ell$, $\ell$ denoting a typical mfp}.

One can write Eqs. (\ref{<vareps vareps DWSL HEL>}) as
\begin{equation}
\lim_{DWS}\left\langle  \left[ \varepsilon ^{(1)} \right]_{a b}^{i j}
\left[ \varepsilon ^{(1)} \right]_{c d}^{k l} \right\rangle _{\delta L}
\approx 2 \tilde{D}_{ab,cd}^{ij,hl}(k)\cdot \delta L,
\label{<vareps vareps DWSL HEL> D}
\end{equation}
where we have defined the diffusion coefficients in the SWLA as
\begin{equation}
\tilde{D}_{ab,cd}^{ij,hl}
=(-1)^{i+h+1}\frac{C(ab,cd)}{2\sqrt{\ell_{ab}(k)\ell_{cd}(k)}}
\delta _{K0}
\end{equation}
which, from Eq. (\ref{<vareps vareps DWSL HEL>}), take the
explicit form
\begin{subequations}
\begin{eqnarray}
\tilde{D}_{ab,cd}^{11,11}(k)&=&
-\frac{1}{1+\delta _{ac}}
\Big[
\frac{C(a,c)}{2\sqrt{\ell_{aa}(k)\ell_{cc}(k)}} \delta _{ab} \delta _{cd}
+ \frac{1}{2\ell_{ab}} \delta _{ad}\delta _{bc}
\Big ]
\\  \nonumber \\
\tilde{D}_{ab,cd}^{11,22}(k)&=&
\frac{1}{1+\delta _{ad}}
\Big[
\frac{C(a,c)}{2\sqrt{\ell_{aa}(k)\ell_{cc}(k)}} \delta _{ab} \delta _{cd}
+ \frac{1}{2\ell_{ab}} \delta _{ac}\delta _{bd}
\Big ]
\\ \nonumber \\
\tilde{D}_{ab,cd}^{12,21}(k)&=& \frac
{\delta _{ac}\delta _{bd} + \delta _{ad}\delta _{bc}}
{1+ \delta _{ab}}
\frac{1}{2\ell_{ab}(k)}
\\ \nonumber \\
\tilde{D}_{ab,cd}^{11,12}(k)
&=&\tilde{D}_{ab,cd}^{11,21}(k)=\tilde{D}_{ab,cd}^{12,12}(k) = 0 .
\end{eqnarray}
\label{<D HEL>}
\end{subequations}
These diffusion coefficients depend on the energy through the mfp's
{\em only}.

ii) Eq. (\ref{Delta(k) 1}) shows that
{\em in the DWSL, followed by the SWLA, the fourth-order contribution}
(\ref{vareps(2)vareps(2) DWSL}) {\em to a second moment of $\varepsilon $ for the BB is either negligible or
behaves as $(\delta L / \ell)^2$, $\ell$ denoting a typical mfp}.
Thus in Eq. (\ref{<vareps^2> L 2}) we keep only the sum for $K_1=K_2=0$ and
neglect the other summations, the result being thus proportional to
$(\delta L / \ell)^2$.

We finally obtain, for the BB second moments of Eq. (\ref{ee=e1e1+e2e2}) in the SWLA:
\begin{eqnarray}
\lim_{DWS}
\left\langle \varepsilon ^{ij}_{ab}
\varepsilon ^{hl}_{cd}
\right\rangle _{L, \delta L ; k}
&\approx& 2\tilde{D}^{ij,hl}_{ab,cd}(k)\cdot\delta L
+ 2\sum_{\alpha '\beta ', \lambda ' \mu '}
\tilde{D}_{a \alpha ',c \beta '}^{i\lambda ',h \mu '}(k)
\;\tilde{D}_{\alpha ' b,\beta 'd}^{\lambda 'j,\mu 'l}(k)
\cdot(\delta L)^2
\nonumber \\
&& +O(\delta L)^3 \; .
\label{<eps.eps>_D SWLA}
\end{eqnarray}

2) Similar arguments applied to the analysis of
App. \ref{pth moment varepsilon} lead to the result that
a $(2t)$-th moment of $\varepsilon $ for the BB
can either be neglected because it contains $k_a$'s in the
denominator, or it gives a contribution to Eq. (\ref{avF(L+dL) 1})
which is porportional to $(\delta L/\ell)^{t}$,
whereas a $(2t+1)$-th moment contributes as $(\delta L/\ell)^{t+1}$.

3) We need some knowledge about the behavior of the averages
$\langle \cdots \rangle_{L,k}$ appearing in
Eq. (\ref{avF(L+dL) 1}) in the SWLA.
We shall assume that, for large enough $k$, we can approximate
\begin{equation}
\langle \cdots \rangle_{L,k} \approx \langle \cdots \rangle_{L}^{(0)} \; ,
\label{<>0}
\end{equation}
where the r.h.s. represents a function {\em smooth to all scales} of $L$
and whose energy dependence only appears through the mfp's $\ell_{ab}(k)$.
This {\em ansatz}, which seems merely reasonable at this point, is verified in a particular case in Sec. \ref{analytic} below.
In the analysis that follows we shall assume that {\em the energy is kept fixed}, so that {\em the mfp's will be taken as fixed parameters} and will be written as $\ell_{ab}$.
Likewise, we shall write $\tilde{D}^{ij,hl}_{ab,cd}$
for the diffusion coefficients.

We now make use of the results
1i), 1ii) and 2) above, as well as the assumption (\ref{<>0}), to write Eq. (\ref{avF(L+dL) 1}) in the SWLA as
\begin{eqnarray}
\left\langle F(\bm{M})\right\rangle _{L + \delta L}^{(0)}
&\approx&\left\langle F(\bm{M})\right\rangle _L ^{(0)}
\nonumber \\
&&
+\sum_{\substack{ ijhl, \lambda \mu  \\ abcd, \alpha \beta }}
\left[
\tilde{D}_{ab,cd}^{ij,hl} \cdot \delta L
+ \sum_{\alpha '\beta ', \lambda ' \mu '}
\tilde{D}_{a \alpha ',c \beta '}^{i\lambda ',h \mu '}
\; \tilde{D}_{\alpha ' b,\beta 'd}^{\lambda 'j,\mu 'l}
\cdot(\delta L)^2
\right]
\nonumber \\
&&\hspace{3cm}\times
\left\langle
M_{b \alpha }^{j \lambda } \; M_{d \beta }^{l \mu } \;
\frac{\partial ^2F(\bm{M})} {\partial M_{a \alpha }^{i \lambda }\;
\partial M_{c \beta }^{h \mu}} \right\rangle _L ^{(0)}
\nonumber \\
&&+ O(\delta L)^2 \; .
\label{avF(L+dL) SWLA a}
\end{eqnarray}
The square bracket in this last equation corresponds to the BB second moment appearing in Eq. (\ref{avF(L+dL) 1});
the contribution of the third and higher moments is just indicated in the last line, in accordance with 2) above.

We now assume that the quantity
$\left\langle F(\bm{M})\right\rangle _{L+\delta L}^{(0)}$ appearing
on the l.h.s. of Eq. (\ref{avF(L+dL) SWLA a}) can be expanded in a Taylor series around the value $L$,
and that $\delta L$ is smaller than the radius of convergence $R$ of the expansion, i.e.,
$\lambda \ll \delta L < R$, so that:
\begin{equation}
\left\langle F(\bm{M})\right\rangle _{L+\delta L}^{(0)}
=\left\langle F(\bm{M})\right\rangle _{L}^{(0)}
+\frac{\partial \left\langle F(\bm{M})\right\rangle _{L}^{(0)}}{\partial L} \;\delta L
+\frac{1}{2!}
\frac{\partial^2 \left\langle F(\bm{M})\right\rangle _{L}^{(0)}}
{\partial L^2} \;(\delta L)^2 + \cdots.
\label{Taylor <F>(L+dL)}
\end{equation}
Comparing the coefficients of $\delta L$ in
Eqs. (\ref{avF(L+dL) SWLA a}) and
(\ref{Taylor <F>(L+dL)}) we finally find:
\begin{eqnarray}
&&\frac{\partial \left\langle
F(\bm{M})\right\rangle_L^{(0)}}{\partial L}
= \sum_{\substack{ ijhl\lambda  \mu   \\
abcd\alpha \beta }}  \tilde{D}_{ab,\; cd}^{ij , \; hl} \left\langle
M_{b\alpha}^{j\lambda} \; M_{d\beta}^{l\mu} \; \frac{\partial
^2F(\bm{M})} {\partial M_{a\alpha}^{i\lambda} M_{c\beta}^{h\mu}}
\right\rangle _L ^{(0)} \; . \label{diff eqn exp val swla}
\end{eqnarray}
In the SWLA we have thus ended up with an evolution equation for the ``smooth" quantities defined in Eq. (\ref{<>0}).

We need to fix the initial conditions appropriate to
Eq. (\ref{diff eqn exp val swla}). If we require
Eq. (\ref{in_condition}) for the exact expectation values, i.e.,
$\langle F(\bm{M}) \rangle_{L=0,k} = F(\bm{I})$, and $F(\bm{I})$ is $k$-independent, then Eq. (\ref{<>0}) implies
\begin{equation}
\langle F(\bm{M}) \rangle_{L=0}^{(0)} = F(\bm{I}).
\label{<>0_swla}
\end{equation}

More detailed assumptions than (\ref{<>0}) on the structure of the expectation
value $\langle \cdots \rangle_{L,k}$ appearing in Eq. (\ref{avF(L+dL) 1}) in
the SWLA are presented in App. \ref{swla from details of <>} for the
one-channel case, $N=1$. There, a rederivation of Eq. (\ref{diff eqn exp val
swla}) using such assumptions is also discussed.

The derivations given above of both diffusion equations, Eq. (\ref{diff eqn exp
val sec3}), valid for arbitrary energies, and Eq. (\ref{diff eqn exp val
swla}), valid in the SWLA, use, as a starting point, Eq. (\ref{avF(L+dL) 1}),
which describes the result of adding a BB to an already existing waveguide of
length $L$. This is also the starting point of the derivation given in App.
\ref{swla from BB}. We believe that it would be very instructive to rederive
the diffusion equation in the SWLA, Eq. (\ref{diff eqn exp val swla}), starting
directly from the more general one, Eq. (\ref{diff eqn exp val sec3}), since
such a derivation would shed more light on the nature of the various
approximations involved. However, we have succeeded in fulfilling this goal
only in the one-channel case, $N=1$; the derivation is presented in App.
\ref{swla from exact diff eqn}.

\section{Applications of the diffusion equation}
\label{applications}

\subsection{Analytic examples}
\label{analytic}

In this section we study a simple example in which the diffusion equation
(\ref{diff eqn exp val sec3}) can be solved exactly.
We restrict the analysis to a one-channel geometry ($N=1$) and consider,
as examples of the observable $F(\bm{M})$, the quantities
\begin{subequations}
\begin{eqnarray}
M^{11}M^{22} &=& \alpha \alpha ^{\ast} =\frac{1}{tt^{\ast}} \equiv \frac{1}{T}
\\
M^{11}M^{12} &=& \alpha \beta = - \left(\frac{r}{t^2}\right)^{\ast },
\end{eqnarray}
\end{subequations}
where we have used Eq. (\ref{r,t(M)beta1}) to establish the connection
with reflection and transmission amplitudes.
We shall give only the main results of the calculation, some of
the details being presented in App. \ref{derivation-analytic}.

For the one-channel case, the diffusion equation
(\ref{diff eqn exp val sec3}) can be written as
\begin{eqnarray}
&&\frac{\partial \left\langle F(\bm{M})\right\rangle _L}{\partial L}
= \sum_{\substack{ ijhl \lambda  \mu }} D^{ij,hl}(k,L) \left\langle
M^{j \lambda } \; M^{l \mu} \; \frac{\partial ^2F(\bm{M})} {\partial
M^{i \lambda } \partial M^{h \mu}} \right\rangle _L , \label{diff
eqn N1}
\end{eqnarray}
where the diffusion coefficient $D^{ij,hl}(k,L)$
is given explicitly in Eq. (\ref{K,D N=1}).
For simplicity, we have suppressed all channel indeces, which would take the value 1.
We emphasize that in the DWSL this equation is exact, in the sense that it is valid for all energies.

The mfp is energy dependent. However, in the present calculation we keep the energy fixed and so the mfp is taken as a fixed parameter and will be written as $\ell$.
One can write all the evolution equations in terms of the ratio of the
length $L$ to the mfp $\ell$
\begin{equation}
s=L/\ell \; ,
\label{s}
\end{equation}
and essentially the ratio of the mfp to the wavelength $\lambda $
\begin{equation}
x_0=2k\ell.
\label{x0}
\end{equation}

Using the diffusion coefficients of Eq. (\ref{K,D N=1}) one finds the pair of
coupled equations
\begin{subequations}
\begin{eqnarray}
\frac{\partial \left\langle \alpha \alpha ^{\ast }\right\rangle_s}
{\partial s}
&=&\left\langle \alpha \beta \right\rangle_s
e^{ix_{0}s}+\left( 2\left\langle \alpha \alpha ^{\ast}\right\rangle_s -1\right)
+\left\langle \alpha ^{\ast}\beta ^{\ast }\right\rangle_s e^{-ix_{0}s},
\label{e-alfa/alfa*} \\
\frac{\partial \left\langle \alpha \beta \right\rangle_s }
{\partial s}
&=&-\left\langle \alpha \beta \right\rangle_s
-\left(2\left\langle \alpha \alpha ^{\ast }\right\rangle_s -1\right)
e^{-ix_{0}s}-\left\langle \alpha ^{\ast }\beta ^{\ast}\right\rangle_s e^{-2ix_{0}s},
\label{e-alfa/beta}
\end{eqnarray}
\label{e-alfa/alfa*,alfa/beta}
\end{subequations}
which have to be solved with the initial conditions at $s=0$:
\begin{subequations}
\begin{eqnarray}
\left\langle \alpha \alpha ^{\ast }\right\rangle _{s=0}&=&1
\\
\left\langle
\alpha \beta \right\rangle _{s=0}&=&0 .
\end{eqnarray}
\label{in_conditions}
\end{subequations}

The second derivatives of the observable $F(\bm{M})$ appearing on the r.h.s of
the diffusion equation (\ref{diff eqn N1}) produce, in general,
quantities which are different from the observable $F(\bm{M})$ itself,
whose average we wish to study.
One then needs to compute the evolution of these other quantities and this,
in turn, generates still new ones.
In the example considered here,
Eq. (\ref{e-alfa/alfa*,alfa/beta}) shows that
the evolution of
$\left\langle \alpha \alpha ^{\ast }\right\rangle$
involves $\left\langle \alpha \alpha ^{\ast }\right\rangle$ {\em and}
$\left\langle \alpha \beta \right\rangle$,
and similarly for the evolution of
$\left\langle \alpha \beta \right\rangle$:
we thus find a pair of coupled equations which ``close'',
in the sense that the quantities occurring on the r.h.s. are the same as on the l.h.s.

The evolution equations (\ref{e-alfa/alfa*,alfa/beta}) for the real quantity
$\left\langle \alpha \alpha ^{\ast }\right\rangle_s$
and the complex quantity
$\left\langle\alpha \beta \right\rangle_s$ can be written as the
triplet of coupled equations (\ref{e-1,2,3}), which can be solved using the method of Laplace transforms,
using the initial conditions (\ref{in_conditions}), with the result
\begin{subequations}
\begin{eqnarray}
\left\langle \alpha \alpha ^{\ast }\right\rangle_s
&=& \frac 12
+ \frac 12 \left[
\frac{p_{1}^{2}+2p_{1}+x_{0}^{2}}
{\left( p_{1}-p_{2}\right) \left( p_{1}-p_{3}\right)}e^{p_{1}s}
+\frac{p_{2}^{2}+2p_{2}+x_{0}^{2}}
{\left(p_{2}-p_{1}\right) \left( p_{2}-p_{3} \right)}e^{p_{2}s}
+\frac{p_{3}^{2}+2p_{3}+x_{0}^{2}}
{\left( p_{3}-p_{1}\right) \left( p_{3}-p_{2}\right) }e^{p_{3}s}
\right],
\label{s-1}
\nonumber \\ \\
\left\langle \alpha \beta \right\rangle_s
&=&-\left[ \frac{p_{1} + i x_0}
{\left(p_{1}-p_{2}\right) \left( p_{1}-p_{3}\right)}e^{(p_{1}-ix_0)s}
+\frac{p_{2}+ i x_0}{\left( p_{2}-p_{1}\right)
\left(p_{2}-p_{3}\right) }e^{(p_{2}-ix_0)s} \right.
\nonumber \\ \nonumber \\
&&\hspace{7cm}\left.+\frac{p_{3} + i x_0}
{\left(p_{3}-p_{1}\right) \left( p_{3}-p_{2}\right) }e^{(p_{3}-ix_0)s}
\right] .
\label{s-2}
\end{eqnarray}
\label{s_1,2,3}
\end{subequations}
In this equation, $p_{1}$, $p_{2}$ and $p_{3}$ are the roots of the third degree polynomial
$P\left( p\right) =p^{3}+x_{0}^{2}p-2x_{0}^{2}$, with
$p_{1}\;\epsilon \;\mathbb{R}$,
$p_{2}, p_{3}\;\epsilon\; \mathbb{C}$ and $p_{3}=p_{2}^{\ast }$.

The solutions (\ref{s_1,2,3}) are {\em exact}, being valid for {\em
arbitrary length $L$, mfp $\ell$ and wavenumber $k$}.
Moreover, as shown below,  the solutions of
the diffusion equation are in full quantitative agreement with the
statistical averages obtained from numerical solutions of the
one-dimensional wave equation.

\begin{figure}[h]
\epsfig{file=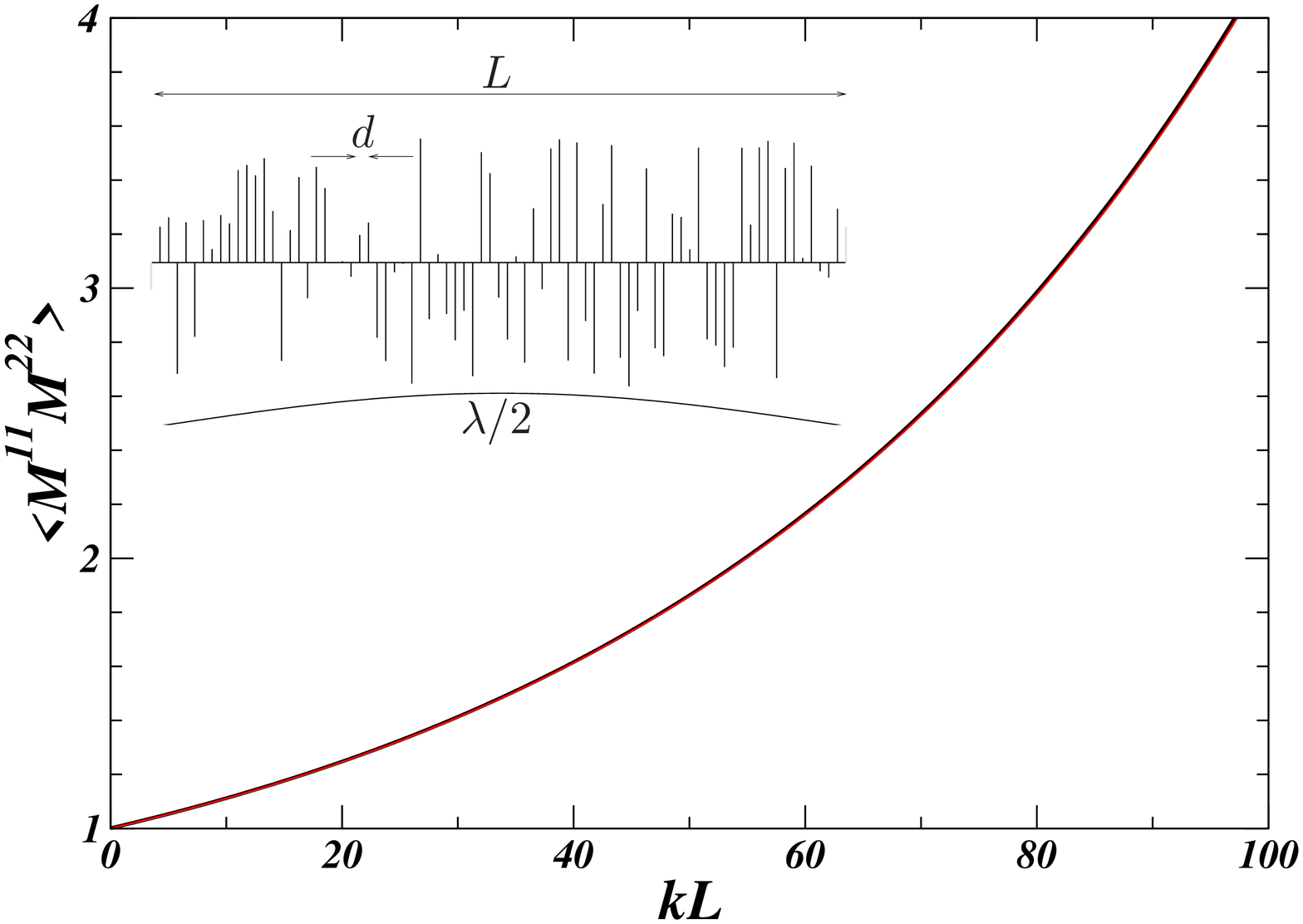,width=10cm,clip=,keepaspectratio}
\caption{$\langle M^{11}M^{22}\rangle = \langle \alpha \alpha^*
\rangle$ versus $kL$. Numerical results (bold line) from the
one-dimensional model sketched in the inset are indistinguishable
from the analytical results, Eq. (\ref{s-1}). The results correspond
to $x_0=2k \ell=200$ and $d/\ell = 10^{-3}$.}
\label{N1_1_num_sim}
\end{figure}

\begin{figure}[h]
\epsfig{file=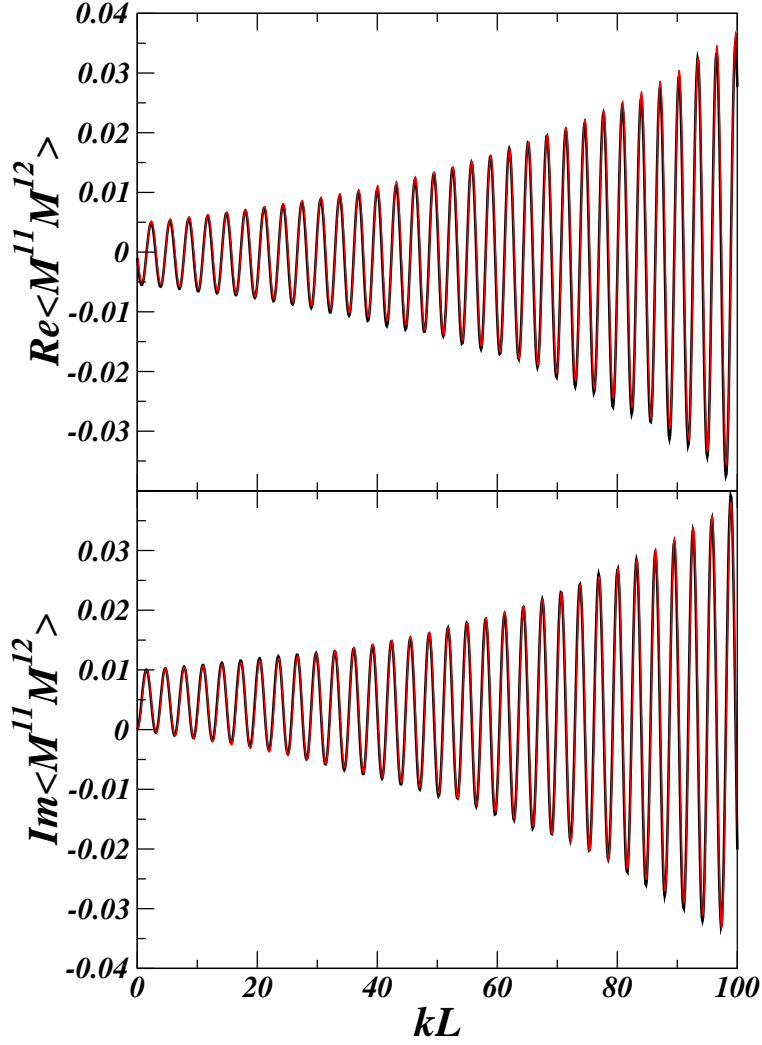,width=10cm,clip=,keepaspectratio} \caption{Real (top) and
imaginary (bottom) parts of  $\langle M^{11}M^{12}\rangle = \langle \alpha
\beta \rangle$ as a function of $kL$. Numerical results (bold line)  are
indistinguishable from the analytical results. The parameters are the same as
in Fig. \ref{N1_1_num_sim}.} \label{N1_2_num_sim}
\end{figure}

A one-dimensional version of the delta-slice model discussed in
Sec. \ref{stat model} is sketched in the inset of Fig.
\ref{N1_1_num_sim}.
Notice that in a 1D problem there are no evanescent modes.
The system of length $L$ is constructed from
``delta potentials'', $U_r(x)=u_{r}\delta(x-x_{r})$
(recall that $U_r(x)$ and $u_{r}$ have dimensions of $k^2$ and $k$, respectively), located at
the positions $x_r=rd$ ($r=0,1,2 \dots$);
$u_{r}$ is assumed to be
uniformly distributed over the interval $[-u_{max},+u_{max}]$. The
mean free path,  obtained from  Eq. (\ref{lab}), is simply given by:
\begin{equation}
\frac{d}{\ell} = \frac{1}{3} \left( \frac{u_{max}}{2k} \right)^2.
\end{equation}
The results of the numerical calculations for $\langle \alpha
\alpha^* \rangle$ and $\langle \alpha \beta \rangle$ versus $L$ are
shown in Figs. \ref{N1_1_num_sim} and \ref{N1_2_num_sim},
respectively. Averages were obtained from $10^7$ different
microscopic realizations. Numerical results
(bold line)  are indistinguishable from the analytical solution of
the diffusion equation (Eqs. (\ref{s-1}) and (\ref{s-2})).

It will be interesting to see what these results reduce to in the SWLA
discussed in Sec. \ref{SWLA} above.
In preparation for this, we first consider a {\em fixed} value of $s=L/\ell$
and take $x_0 = 2 k\ell \gg 1$.
From Eq. (\ref{e10}) one can expand the functions
$\left\langle \alpha \alpha ^{\ast }\right\rangle_s$,
$\left\langle \alpha \beta \right\rangle_s$
in powers of $1/x_{0}$;
in terms of the original variables $k$, $\ell$ and $L$ they take the form
\begin{subequations}
\begin{eqnarray}
\left\langle \alpha \alpha ^{\ast }\right\rangle_s
&=& \frac 12 (1 +  e^{2\frac{L}{\ell}})
+\frac{2}{(2k\ell)^2}
\left[
-\left(1+2\frac{L}{\ell}\right)e^{2\frac{L}{\ell}}
+e^{-\frac{L}{\ell}}\frac{e^{2ikL}+e^{-2ikL}}{2}
\right]
+ O\left(\frac{1}{k\ell}\right)^3
\nonumber \\ \\
\left\langle \alpha \beta \right\rangle_s
&=&\frac{i}{2k\ell}\left(
e^{-\frac{L}{\ell}} - e^{2\frac{L}{\ell}}e^{-2ikL}
\right)
+\frac{2}{(2k\ell)^2}
\left[
\frac{5-3\frac{L}{\ell}}{4}e^{-\frac{L}{\ell}}
-\left(
e^{2\frac{L}{\ell}}e^{-2ikL}
+\frac14 e^{-\frac{L}{\ell}}e^{-4ikL}
\right)
\right]
\nonumber \\
&&\hspace{6cm}+ O\left(\frac{1}{k\ell}\right)^3  \; .
\end{eqnarray}
\label{<aa*> <ab> expansion}
\end{subequations}
The solutions (\ref{<aa*> <ab> expansion}) satisfy the differential equations (\ref{e-alfa/alfa*,alfa/beta}) together with the initial conditions
(\ref{in_conditions}) to every order in the expansion in powers of
$1/2k\ell$.

Notice that the {\em ansatz} made in Eq. (\ref{<>0}) is verified explicitly in this example, with the result
\begin{subequations}
\begin{eqnarray}
\left\langle \alpha \alpha ^{\ast }\right\rangle _s ^{(0)}
&=&
\frac{1}{2} (1+e^{2\frac{L}{\ell}}) ,
\label{<aa*>0} \\
\left\langle \alpha \beta \right\rangle _s ^{(0)} &=& 0,
\label{<ab>0}
\end{eqnarray}
\end{subequations}
which represents, in this particular case, the SWLA discussed in Sec. \ref{SWLA}.
The result (\ref{<aa*>0}) agrees with what had been obtained earlier as a solution of the diffusion equation of
Ref. \cite{mpk} for $N=1$, also known as Melnikov's equation.
Notice that
\begin{equation}
\left\langle \beta \beta ^{\ast }\right\rangle _s ^{(0)}
= \left\langle \alpha \alpha ^{\ast }\right\rangle _s ^{(0)}-1
= \left\langle \frac{R}{T}  \right\rangle _s ^{(0)}
=\frac{1}{2} (e^{2\frac{L}{\ell}}-1)
\label{landauer}
\end{equation}
represents the well known exponential increase of Landauer's resistance
\cite{landauer}.

If, in Eq. (\ref{<aa*> <ab> expansion}), we further expand the exponentials
$e^{2L/\ell}$,  $e^{-L/\ell}$, in powers of $1/\ell$, we end up with an expansion of $\left\langle \alpha \alpha ^{\ast }\right\rangle _s$,
$\left\langle \alpha \beta \right\rangle _s$
in powers of $1/\ell$.
The result found in Eq. (\ref{<vareps^2> L,dL}), setting $L=0$ and interpreting $\delta L$ as $L$,
is precisely the term proportional to
$1/\ell$ in such an expansion;
we have verified the consistency of the two results
up to $O(1/\ell)$.

\subsection{Random walk in the transfer matrix space: Numerical simulations.}
\label{numerical}

As we have shown, the diffusion equation, Eq. (\ref{diff eqn exp val sec3}),
determines the statistical properties of transport for any physical observable
and it only depends on the mean free paths $\ell_{ab}$. Once the various
$\ell_{ab}$ are specified, the statistical distributions are universal, i.e.,
independent of other details of the microscopic statistics. However, in order
to know the exact shape of the distribution of a given observable we have to
solve the diffusion equation. This is a challenging problem even in the
isotropic case \cite{mpk} (where all the mfp's are equivalent, $\ell_{ab} =
\ell$). Here, instead of a direct solution of the multidimensional diffusion
equation we have followed an alternative way that can be seen as a
generalization of a random walk in the transfer matrix space (Ref.
\cite{luis_thesis}). The method, based on our previous
theoretical description, can be summarized as follows:

1) We first obtain a set of mean free paths
from a given microscopic potential model for the building block or,
eventually, from specific experiments on very thin slabs.

2) We generate an ensemble of transfer matrices having their first
and second moments equal to those corresponding to a BB of a certain
length $\delta L$.

3) The transfer matrix for a system of length $L=P\delta L$ is
obtained by combining $P$ building block matrices randomly chosen from the
ensemble. This procedure can be repeated again and again in order to
obtain the statistical distribution of any physical quantity. As
predicted by the CLT associated with the composition of BB's explained in
App. \ref{clt_bb},
higher order moments of the BB matrix elements play no
role in the final statistics.

The statistical distributions of different physical quantities will
be shown to be in full agreement with the results of  exact
microscopic numerical calculations for a model system. This shows
that validity of the diffusion equation given in
Eq. (\ref{diff eqn exp val sec3}) goes beyond  the various formal limits discussed in Sec. \ref{BB q1d}.

\subsubsection{Microscopic potential model and mean free paths}

\begin{figure}[h]
\epsfig{file=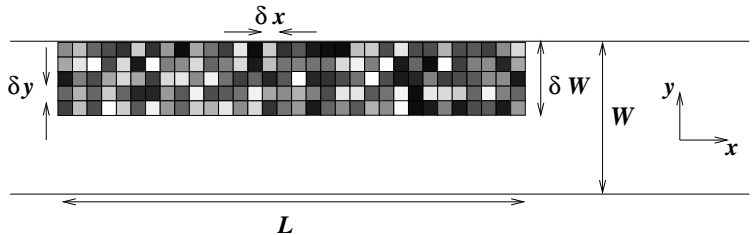,width=10cm,clip=} \caption{ Schematic representation of
the microscopic model based on random potentials. Each square of the plot, or
``cell", represents a region of constant random potential.}
\label{sketch_microscopic}
\end{figure}

Let us consider the potential model sketched in Fig.
\ref{sketch_microscopic}.
In this model, a 2D waveguide with
perfectly reflecting walls has a region of length $L$ which is
divided into small ``cells" of dimensions $\delta x\times\delta y$.
The working wavelength is chosen to be such that $\delta x,\delta
y \ll \lambda$. In the language of Sec. \ref{single delta slice},
the potential in the $r$-th slice, Eq. (\ref{r delta slice}),
is replaced here, for finite $\delta x$, by
\begin{equation}
U_r(x,y) = u_r (y) \frac{\theta _{\delta x}(x-x_r)}{\delta x} ,
\label{Ur finite dx}
\end{equation}
where $\theta _{\delta x}(x-x_r)$ takes the value 1 inside the interval
$(x_r - \delta x /2, \; x_r + \delta x /2)$ and 0 outside.
Should $\delta x$ tend to zero, the expression
in (\ref{Ur finite dx}) would tend to that of Eq. (\ref{r delta slice (xy)}).
Inside the $r$-th slice, the potential is taken to be constant within each cell, i.e.,
\begin{equation}
u_r (y) = \sum _s u_s \theta _{\delta y}(y-y_r),
\label{u_(u) for cells}
\end{equation}
so that
\begin{equation}
U_r(x,y) = \sum _s U_s \theta _{\delta x}(x-x_r) \theta _{\delta y}(y-y_r),
\label{Ur finite dx 1}
\end{equation}
with $U_s = u_s / \delta x$.
The constant values $U_s$ of the potential inside each cell located in the region $W - \delta W < y <  W$
is sampled from a uniform distribution within the interval
$\left[-U_0,U_0\right]$.
Outside the region defined by $W - \delta W < y <  W$, the potential is taken to be zero.

In order to get the mfp's corresponding to our model system, we
follow the same steps leading to Eq. (\ref{lab}) in Sec. \ref{BB q1d}
above.
In the limit $\delta x, \delta y \ll \delta W$,
and neglecting the coupling to evanescent modes, i.e., using the ``bare" potential $u$ instead of the ``effective" one $\hat{u}$
(see text following Eq. (\ref{vr N}) and App. \ref{veff}), we obtain
\begin{eqnarray}
\frac{1}{\ell_{ab}}
= \frac{\langle \left[ v_{ab} \right]^2\rangle}{\delta x}
= \frac{U_0^2}{3} \frac{\delta x \delta y}{4 k_a k_b}
\int_{W-\delta W}^W \chi_a^2(y) \chi_b^2(y) dy ,
\end{eqnarray}
where $\chi_a(y)$ are the transverse eigenfunctions of the clean
waveguide (Eq. (\ref{chi a})). The mfp's for bulk disordered
systems, i.e., when the disordered potential covers the whole
section of the waveguide ($\delta W = W$) are simply given by
\begin{eqnarray}
\left.\frac{1}{\ell_{ab}}\right|_{\rm{bulk}} = \frac{U_0^2}{3}
\frac{\delta x \delta y}{4 k_a k_b} \frac{2 + \delta_{ab}}{2W}.
\label{bulk mfp}
\end{eqnarray}
In order to analyze a surface disordered waveguide, we shall also
consider the limit $\delta W \ll W$,
\begin{eqnarray}
\left.\frac{1}{\ell_{ab}}\right|_{\rm{surface}}
= \frac{U_0^2}{3}
\frac{\delta x \delta y}{4 k_a k_b} \left( \frac{4 \pi^4}{W^2} a^2
b^2 \delta W \right)  .
\label{surface mfp}
\end{eqnarray}

\subsubsection{Random transfer matrices for a building block}

In order to generate an ensemble of random transfer matrices whose
first and second moments are given, it is useful to describe the
transfer matrix elements of the BB as a function of the
$2N^{2}+N$ independent parameters of the Pereyra representation
[see App. \ref{M-properties}, Eq. (\ref{RepPereyra})]. The matrix
$\varepsilon $ of Eq. (\ref{M BB 4}) can be expressed (in that
representation) as
\begin{subequations}
\begin{eqnarray}
\varepsilon ^{11} &=&e^{ih}\sqrt{1+\eta \eta ^{\ast }}-1,  \label{CF1a} \\
\varepsilon ^{12} &=&e^{ih}\eta ,  
\label{CF1b}
\end{eqnarray}
\label{CF1}
\end{subequations}
where $h$ is an arbitrary $N\times N$ Hermitian matrix
(thus contributing $N^{2}$ parameters) and $\eta $ is an
arbitrary $N\times N$\ complex symmetric matrix (thus contributing
$N^{2}+N$ parameters).

Applying successive approximations to Eqs. (\ref{CF1}) it is
possible to invert them to express the matrices $h$\ and $\eta $
as functions of the blocks $\varepsilon ^{ij}$, i.e.,
\begin{subequations}
\begin{eqnarray}
h &=&-i\varepsilon ^{11}+\frac{i}{2}\left( \varepsilon
^{12}\varepsilon ^{21}+\varepsilon ^{11}\varepsilon ^{11}\right)
+O\left( \varepsilon
^{3}\right) ,  \label{CF2a} \\
\eta &=&\varepsilon ^{12}-\varepsilon ^{11}\varepsilon
^{12}+O\left( \varepsilon ^{3}\right) .  \label{CF2b}
\end{eqnarray}
\label{CF2}
\end{subequations}

The aim is to derive the statistical properties of the matrices
$h$ and $ \eta $ in terms of those of the blocks $\varepsilon
^{ij}$ which we derived in the previous section; we shall do this
in the SWLA (see Sec. \ref{SWLA}). We can use Eqs.
(\ref{CF2}), (\ref{<D HEL>}) and (\ref{<eps.eps>_D SWLA}) to
obtain (in powers of $\delta L$) the first and second moments of
the matrix elements $\eta _{ab}$, $h_{ab}$. For the first moments
we obtain
\begin{subequations}
\begin{eqnarray}
\left\langle h_{ab}\right\rangle_{\delta L} &=&O\left(\delta L\right)^{2}, \label{4.20a} \\
\left\langle \eta _{ab}\right\rangle _{\delta L} &=&O\left( \delta
L\right) ^{2},  \label{4.20b}
\end{eqnarray}
\label{4.20}
\end{subequations}
and for the second moments
\begin{subequations}
\begin{eqnarray}
\left\langle h_{ab}h_{cd}\right\rangle _{\delta L}
&=&-\left\langle\varepsilon_{ab}^{11}\varepsilon_{cd}^{11}\right\rangle_{\delta L}+
\cdots =\frac{\delta L}{1+\delta _{ac}}
\left[\delta_{ab}\delta_{cd}\frac{C\left(aa,cc\right)}
{\sqrt{\ell_{aa}\ell_{cc}}}+\frac{\delta_{ad}\delta _{bc}}{\ell _{ab}}\right]\label{4.21a} \\
&&+O\left( \delta L\right) ^{2},  \notag \\
\left\langle h_{ab}h_{cd}^{\ast }\right\rangle _{\delta L}
&=&-\left\langle \varepsilon _{ab}^{11}\varepsilon
_{cd}^{22}\right\rangle _{\delta L}+\cdots =\frac{\delta
L}{1+\delta_{ad}}\left[\delta_{ab}\delta_{cd}\frac{C\left(aa,cc\right)
}{\sqrt{\ell _{aa}\ell_{cc}}}+\frac{\delta_{ac}\delta _{bd}}{\ell _{ab}}\right]   \label{4.21b} \\
&&+O\left( \delta L\right) ^{2},  \notag \\
\left\langle h_{ab}\eta _{cd}\right\rangle _{\delta L}
&=&-i\left\langle \varepsilon _{ab}^{11}\varepsilon
_{cd}^{12}\right\rangle _{\delta L}+\cdots
=O\left( \delta L\right) ^{2},  \label{4.21c} \\
\left\langle h_{ab}\eta _{cd}^{\ast }\right\rangle _{\delta L}
&=&-i\left\langle \varepsilon _{ab}^{11}\varepsilon
_{cd}^{21}\right\rangle
_{\delta L}+\cdots =O\left( \delta L\right) ^{2},  \label{4.21d} \\
\left\langle \eta _{ab}\eta _{cd}\right\rangle _{\delta L}
&=&\left\langle \varepsilon _{ab}^{12}\varepsilon
_{cd}^{12}\right\rangle _{\delta L}+\cdots
=O\left( \delta L\right) ^{2},  \label{4.21e} \\
\left\langle \eta _{ab}\eta _{cd}^{\ast }\right\rangle _{\delta L}
&=&\left\langle \varepsilon _{ab}^{12}\varepsilon
_{cd}^{21}\right\rangle _{\delta L}+\cdots =\frac{\delta
_{ac}\delta _{bd}+\delta _{ad}\delta _{bc}}{1+\delta
_{ab}}\frac{\delta L}{\ell _{ab}}+O\left( \delta L\right) ^{2}.
\label{4.21f}
\end{eqnarray}
\label{4.21}
\end{subequations}

To generate the ensemble of random transfer matrices for $\eta $
and $h$ in the SWLA, we need to know the statistical properties of
the real and imaginary parts of the matrix elements $\eta _{ab}$
and $h_{ab}$, to be denoted as
\begin{subequations}
\begin{eqnarray}
\eta _{ab}^{R} &\equiv &\textrm{Re}\eta_{ab}=\frac{1}{2}\left(
\eta _{ab}+\eta _{ab}^{\ast }\right) ,\qquad \eta _{ab}^{I}\equiv
\textrm{Im}\eta _{ab}=\frac{1}{2i}\left( \eta _{ab}-\eta
_{ab}^{\ast }\right) , \label{R_Ia} \\
h_{ab}^{R} &\equiv
&\textrm{Re}h_{ab}=\frac{1}{2}\left(h_{ab}+h_{ab}^{\ast }\right)
,\qquad h_{ab}^{I}\equiv
\textrm{Im}h_{ab}=\frac{1}{2i}\left(h_{ab}-h_{ab}^{\ast }\right) .
\label{R_Ib}
\end{eqnarray}
\label{R_I}
\end{subequations}
Using Eqs. (\ref{4.20}) and (\ref{4.21}) we find
\begin{subequations}
\begin{eqnarray}
\left\langle \left( \eta _{ab}^{R}\right) ^{2}\right\rangle
_{\delta L} &=&\left\langle \left( \eta _{ab}^{I}\right)
^{2}\right\rangle _{\delta L}=\frac{\delta L}{2\ell
_{ab}}+O\left(\delta L\right) ^{2},\;
\hspace{5mm}\forall a,b, \label{h_eta_a} \\
\left\langle \left( h_{ab}^{R}\right) ^{2}\right\rangle _{\delta
L} &=&\left\langle \left( h_{ab}^{I}\right) ^{2}\right\rangle
_{\delta L}=\frac{\delta L}{2\ell _{ab}}+O\left( \delta L\right)
^{2},
\hspace{5mm} a\neq b,
\label{h_eta_b} 
\\
\left\langle \left( h_{aa}\right) ^{2}\right\rangle _{\delta L}
&=&\frac{\delta L}{\ell _{aa}}+O\left( \delta L\right) ^{2},  \label{h_eta_c} 
\\
\left\langle h_{aa}h_{bb}\right\rangle _{\delta L}
&=&\frac{C\left( aa,bb\right) }{\sqrt{\ell _{aa}\ell _{bb}}}\delta
L+O\left( \delta L\right)
^{2},  \label{h_eta_d} 
\\
\left\langle h_{ab}^{R}h_{cd}^{R}\right\rangle _{\delta L}
&=&\left\langle h_{ab}^{I}h_{cd}^{I}\right\rangle _{\delta
L}=\left\langle h_{ab}^{R}h_{cd}^{I}\right\rangle _{\delta
L}=O\left( \delta L\right)^{2} ,
\hspace{5mm}a\neq b\neq c\neq d,  
\label{h_eta_e} 
\\
\left\langle h_{ab}^{R}h_{ad}^{R}\right\rangle _{\delta L}
&=&\left\langle h_{ab}^{I}h_{ad}^{I}\right\rangle _{\delta
L}=\left\langle h_{ab}^{R}h_{ad}^{I}\right\rangle _{\delta
L}=O\left( \delta L\right)^{2} ,
\hspace{5mm} a \neq b\neq d, 
\\
\left\langle h_{ab}^{R}\eta _{cd}^{R}\right\rangle _{\delta L}
&=&\left\langle h_{ab}^{I}\eta _{cd}^{I}\right\rangle _{\delta
L}=\left\langle h_{ab}^{R}\eta _{cd}^{I}\right\rangle _{\delta
L}=\left\langle h_{ab}^{I}\eta _{cd}^{R}\right\rangle _{\delta
L}=O\left(\delta L\right) ^{2},  \label{h_eta_f} 
\\
\left\langle \eta _{ab}^{R}\eta _{cd}^{I}\right\rangle _{\delta L}
&=&\left\langle h_{ab}^{R}h_{cd}^{I}\right\rangle_{\delta
L}=O\left( \delta L\right) ^{2}  \label{h_eta_g} 
\\
\left\langle \eta _{ab}^{R}\eta _{cd}^{R}\right\rangle _{\delta L}
&=&\left\langle \eta _{ab}^{I}\eta _{cd}^{I}\right\rangle_{\delta L}
=O \left(\delta L\right) ^{2},
\hspace{5mm} a\neq c,\;b\neq d,  \label{h_eta_h} 
\\
\left\langle \eta _{ab}^{R}\eta _{ad}^{R}\right\rangle _{\delta L}
&=&\left\langle \eta _{ab}^{I}\eta _{ad}^{I}\right\rangle _{\delta
L}=\left\langle \eta _{ab}^{R}\eta _{ad}^{I}\right\rangle _{\delta
L}=O\left( \delta L\right) ^{2},
\hspace{5mm} a \neq b\neq d.
\end{eqnarray}
\label{h_eta}
\end{subequations}
We recall that the diagonal elements $h_{aa}$ are real
since $h$ is a Hermitian matrix.

From now on, to generate the ensemble we shall consider a
potential which is delta correlated in the transverse direction;
in that case we have
\begin{equation}
\frac{C\left( aa,bb\right) }{\sqrt{\ell _{aa}\ell
_{bb}}}=\frac{1}{\ell _{ab}},  \label{4.22}
\end{equation}
which allows rewriting Eqs.
(\ref{h_eta_c})-(\ref{h_eta_d}) as one equation:
\begin{equation}
\left\langle h_{aa}h_{bb}\right\rangle _{\delta L}=\frac{\delta
L}{\ell _{ab}}+O\left( \delta L\right) ^{2}.  \label{h_eta_cd}
\end{equation}

Therefore, in the SWLA, real and imaginary parts of the matrix
elements of $\eta $\ and off-diagonal matrix elements of $h$ are, 
to order $\delta L$,
uncorrelated, with zero mean, Eq.
(\ref{4.20}), and with variance $\delta L/2\ell_{ab}$, Eqs.
(\ref{h_eta_a}) and (\ref{h_eta_b}). For these elements we have
used two different distributions giving the same variances:
\begin{subequations}
\begin{eqnarray}
P_{1}\left( x\right)  &=&\frac{1}{4\sqrt{3}\sigma }\left[ \theta
\left( x+\sqrt{3}\sigma \right) -\theta \left( x-\sqrt{3}\sigma
\right) \right] ,
\label{D_Pa} \\
P_{2}\left( x\right)  &=&\frac{1}{2}\delta \left( x-\sigma \right)
+\frac{1}{2}\delta \left( x+\sigma \right) ,  \label{D_Pb}
\end{eqnarray}
\label{D_P}
\end{subequations}
$\theta \left( x\right) $ being the usual step function
and $\sigma ^{2}=\mathrm{var}\left( x\right) $. As we can see from
the CLT of App. \ref{clt_bb}, the final results only depend on the
coefficients proportional to $\delta L$, while the rest of the
details of the distributions do not play any role.

In contrast, the diagonal elements of the $h$ matrices are
correlated, Eq. (\ref{h_eta_cd}). In order to generate numerically
a set of uncorrelated variables from the diagonal
elements of the $h$ matrices we have performed an orthogonal
transformation on the diagonal terms $h_{aa}$,
\begin{equation}
h_{aa}'=\sum_b O_{ab}h_{bb} ,
\label{nn.230}
\end{equation}
in such a way that the
covariance matrix $C_{ab}\equiv\left\langle h_{aa}h_{bb}\right\rangle =\delta
L/\ell_{ab}$ is diagonalized, to obtain
\begin{equation} \left\langle h'_{aa}h'_{bb}\right\rangle
=\delta_{ab}\sigma_{a}^{2} .
\label{nn.240}
\end{equation}
Hence we can numerically generate a set of $N$ uncorrelated variables $h'_{aa}$
with zero mean and a variance given by the eigenvalues of the
$C_{ab}=\delta L/\ell_{ab}$ matrix and, after that, obtain,
by the change of coordinates (\ref{nn.230}), the $h_{aa}$
variables  which are properly correlated.

\subsubsection{Random walk in the transfer-matrix space: Statistical conductance distributions}
\label{random_walk}

Once we have numerically generated an ensemble of transfer matrices
with its first and second moments correct up to order $\delta L$, we
can obtain a transfer matrix corresponding to a system of length
$L=P\delta L$ by multiplying $P$ transfer matrices of the ensemble
of BB's taken at random. Numerically this procedure is unstable
because the pseudounitary group, to which the transfer matrices belong,
is non-compact \cite{mello-kumar}. This property leads to numerical
instabilities as the norm of the transfer matrix elements can grow
without limit. Instead of using the product of transfer matrices, we
obtain the scattering matrix associated with each transfer matrix
(Eqs. (\ref{r,t(M)beta1})), and then combine different scattering
matrices to obtain the scattering matrix for the system of length
$L$ (Eqs. (\ref{SM_combine})).

For a given set of mean free paths $\ell_{ab}$ we choose the length $\delta L$
of the BB in such a way that $\delta L/\ell_{ab} \ll 1$ for all channels. With
this, we generate random transfer matrices as explained above and, for each
one, we obtain the corresponding scattering matrix.
Applying Eq. (\ref{SM_combine}) $P$ times
we obtain the scattering matrix corresponding to a system of length $L=P\delta L$.
This procedure can be repeated as many times as needed to obtain the desired statistical properties.

A detailed numerical analysis of the statistical properties is
beyond the scope of the present work and will be discussed
elsewhere.  Here we just focus on the statistical distribution of
the conductance and the intriguing discrepancies between surface and
bulk disordered systems \cite{UAM,UAM2}.
\begin{figure}[h]
\epsfig{file=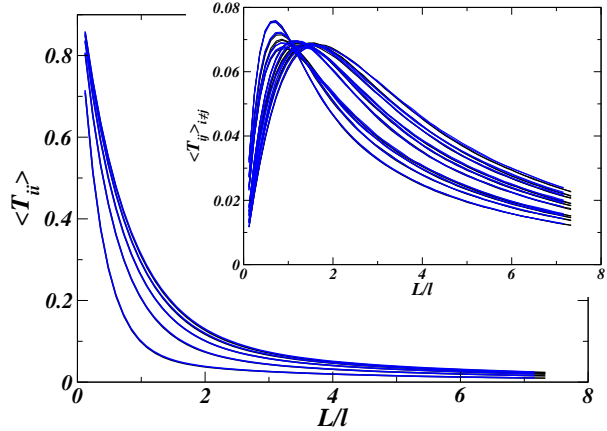,width=8cm,clip=} \caption{{\bf Bulk disordered
waveguides:} Average transmittances $\left\langle T_{ii}\right\rangle $
(channel in = channel out), as a function of $L/\ell$. The inset shows the
equivalent results for $\left\langle T_{ij}\right\rangle $ with $i \ne j$. The results based on the numerical solution of the Schr\'odinger equation 
(microscopic calculation)
and the random walk simulation of the diffusion equation in the SWLA overlap.}
\label{plot_T_ii_bulk}
\end{figure}

\paragraph{Bulk Disorder}
The behavior of the average transmittances
$\left\langle T_{ii}\right\rangle $ (channel in = channel out), for  bulk
disordered wires, is plotted in Fig. \ref{plot_T_ii_bulk} as a function of
$L/\ell$, $\ell$ being the averaged transport mean free path,
\begin{equation}
\frac{1}{\ell} \equiv \frac{1}{N} \sum_{ab} \frac{1}{\ell_{ab}}.
\end{equation}
The inset shows the equivalent results for $\left\langle T_{ij}\right\rangle $
with $i \ne j$.  The random-walk simulation was performed in the SWLA. We have also solved numerically the Schr\"odinger equation for the same model system (sketched in Fig. \ref{sketch_microscopic}). We followed an implementation of the so-called \emph{generalized scattering matrix} (GSM) method (see for example Ref. \cite{JJS_evanesc}). 
The first step consists in the calculation of the set of
transverse eigenfunctions and the scattering matrix for each slice of length
$\delta x$. The combination of two consecutive slices is done by mode matching
at the interface. After that we combine scattering matrices to obtain the
scattering matrix of the whole system.  It is important to mention that this
calculation is performed using both propagating and evanescent modes and hence, this method can be considered as exact.  The statistical properties of any transport parameter obtained from $10^5$ different realizations were found to converge for 3 evanescent modes. The calculations have been done starting from the set of mean free paths $\ell_{ab}$ given by Eq. (\ref{bulk mfp}) for 
$kW=5.5\pi$ (corresponding to 5 propagating modes), $U_0 W^{2}=100$ and $\delta x/W=\delta y/W=1/50$. The exact numerical results for the average transmission coefficients are indistinguishable from the random-walk simulations.

\begin{figure}[h]
\epsfig{file=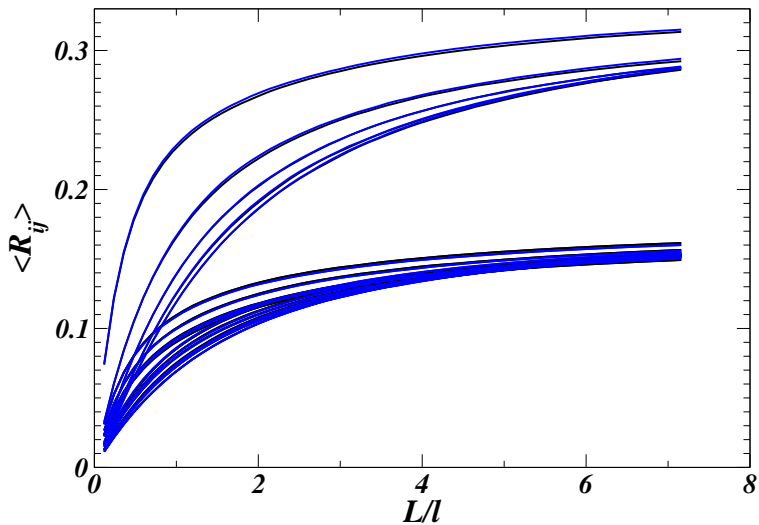,height=10cm,angle=-90,clip=} \caption{{\bf Bulk
disordered waveguides:} Average reflectance $\left\langle R_{ij}\right\rangle
$  as a function of $L/\ell$.
The upper curves correspond to $i=j$ and the lower curves to $i \neq j$.
The results based on the numerical solution of
the Schr\'odinger equation (microscopic calculation) and the random walk simulation of the diffusion equation in
the SWLA overlap.}
\label{plot_R_ij_bulk}
\end{figure}

The random-walk results for the average reflection coefficients $\left\langle R_{ij}\right\rangle $
for bulk disorder (shown in Fig. \ref{plot_R_ij_bulk}) are also in good
agreement with our numerical results as well as with previous numerical work
\cite{PGM} (using a two-dimensional Tight-Binding model with Anderson
disorder). The set of
reflection coefficients corresponding to
backscattering ($\left\langle R_{ii}\right\rangle$) are consistent with an
enhanced backscattering factor $\left\langle R_{ii}\right\rangle/\left\langle
R_{ij}\right\rangle \approx 2$, as expected from the DMPK equation.

\begin{figure}[h]
\epsfig{file=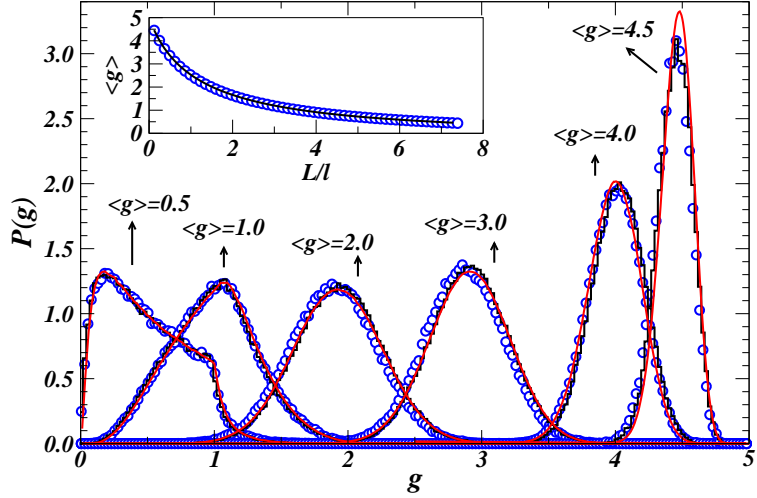,height=10cm,angle=-90,clip=} \caption{{\bf Bulk
disordered waveguides:} Distribution of the dimensionless conductance, $P(g)$,
for different conductance averages, $\langle g \rangle$. The three different
curves based on different approaches overlap. Circles correspond to the random
walk simulation of the diffusion equation in the SWLA. The continuous line
represents the results of the Montecarlo simulation of Ref. \cite{UAM}. The
histogram lines are the results based on the numerical solution of the 
Schr\"odinger equation (microscopic calculation). The inset shows the average conductance as a function of $L/\ell$.}
\label{plot_PG_bulk}
\end{figure}

The distribution of the dimensionless conductance, $ P(g)$ 
[with $g=\text{trace}(tt^{\dagger})$], 
for bulk disordered wires is plotted in Fig.
\ref{plot_PG_bulk} for different conductance averages, $\langle
g \rangle$. The inset shows the average conductance as a function of $L/\ell$.
The exact numerical results for the conductance distribution (histogram lines
in Fig. \ref{plot_PG_bulk}) are indistinguishable from the
random walk
simulations (open circles). 
For comparison we also plot (continuous line) the exact result of
the diffusion equation of Ref. \cite{mpk} (DMPK equation) obtained from a
Montecarlo simulation \cite{UAM}. 
Despite the slight channel anisotropy of transport,
the results are compatible with those of the DMPK equation.

\paragraph{Surface Disorder}

\begin{figure}[h]
\epsfig{file=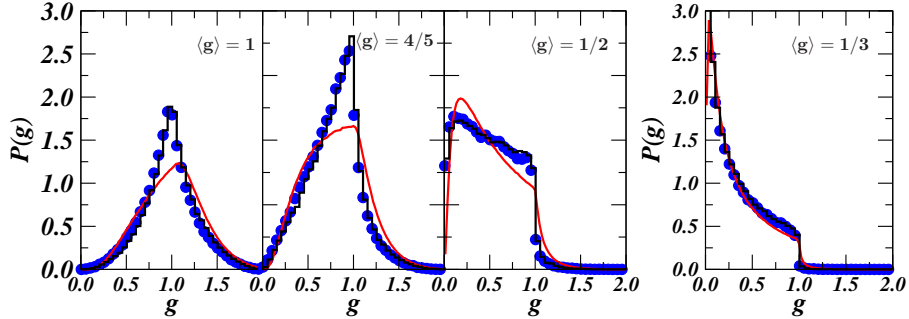,width=12cm,clip=} \caption{{\bf Surface disordered
waveguides:} Distribution of the dimensionless conductance, $P(g)$, for
different conductance averages, $\langle g \rangle$.  
Circles correspond to the random walk simulation of the diffusion equation in the SWLA. 
The histogram lines are the results based on the numerical solution of the Schr\"odinger equation (microscopic calculation). 
The equivalent results for bulk disorder (continuous line, DMPK) are also shown for comparison.} 
\label{plot_PG_surf}
\end{figure}

In the case of surface disorder, the mean free paths are very different from
those obtained for a uniform (bulk) distribution of scatterers. In particular, the dependence of $\ell_{ab}$ on $a^2b^2$ 
[see Eq. (\ref{surface mfp})]
reflects the strong channel anisotropy of transport in surface disordered waveguides
\cite{WRM,AGM_apl,freilikher,izrailev_1,izrailev_2,rendon}. 
This could be the origin of the differences between bulk and surface distributions.  
Previous numerical calculations for surface disordered waveguides, showed that close to
the onset of localization, the conductance distributions presented an
unexpected sharp cusp-like shape \cite{UAM2}. The distribution of the
dimensionless conductance for surface disordered wires obtained from the random
walk simulation in the SWLA is plotted in Fig.
\ref{plot_PG_surf} (open circles) for different conductance averages. 
The exact solution of the Schr\"odinger equation (microscopic calculation; histograms) is again in full agreement with the
diffusion equation and with previous numerical work \cite{WRM,UAM2}. The
calculations have been done starting from the set of mean free paths
$\ell_{ab}$ given by Eq. (\ref{surface mfp}) for $\delta W=0.1W$,
$U_{0}=100/W^{2}$, $kW=5.5\pi$ ($N=5$), $\delta x=10\delta y=W/50$.

It is worth noticing that when the disordered region is confined
close to the surface, the mean free paths can be extremely large (for
example, for the present calculation, $\ell \approx
1.50\times10^{4}W$). The exact numerical solution of the wave
equation is then extremely expensive in terms of computation time
compared to the random walk simulations based on the statistical
properties of the BB.

\begin{figure}[h]
\epsfig{file=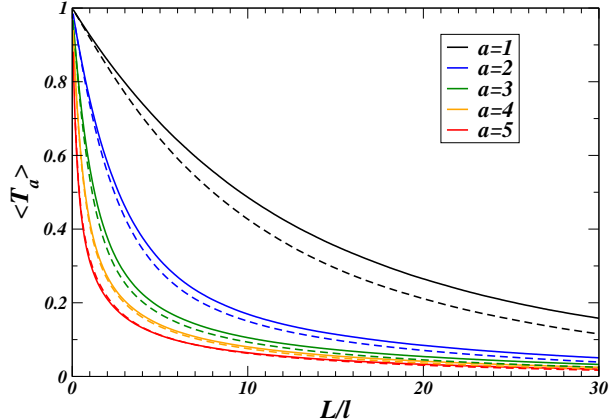,height=8cm,angle=-90,clip=}
\caption{{\bf Surface disordered waveguides:}
Average transmittances $\left\langle
T_{a}\right\rangle $  as a function of $L/\ell$.  The agreement between results
based on the numerical solution of the 
Schr\"odinger equation (microscopic calculation; continuous lines) and the
random walk simulation of the diffusion equation in the SWLA (dashed lines) is
not as good as for the conductance distributions.} \label{plot_T_i_surf}
\end{figure}

Although the random walk in the SWLA  accurately reproduces the
exact conductance distributions, it is not in full agreement with the
statistical properties of the different transmittances. As an example, Fig.
\ref{plot_T_i_surf} shows the behavior of $\langle T_a \rangle = \sum_b \langle
T_{ba} \rangle$ versus $L/\ell$ for both the exact numerical results
(continuous lines) and the random walk (dashed lines). The disagreement could
be associated to the use of an approximate expression (Eq. (\ref{surface mfp}))
for the mean free paths. For small lengths compared to the mean free paths, the
average reflectance is given by (see also Eq. (\ref{lab 1}))
\begin{equation}
\left\langle R_{ab}\right\rangle =\frac{L}{\ell_{ab}} + \cdots .
\label{nn.360}
\end{equation}
We could then have obtained the different mean free paths $\ell_{ab}$ for all
modes by performing a linear fitting of the numerical results to Eq.
(\ref{nn.360}). However, as long as the energy is not very close to the onset
of new propagating channels, we found that  the numerical mfp's are well
described by Eqs. (\ref{bulk mfp}) and (\ref{surface mfp}) within the numerical
accuracy. The discrepancy
could then be associated to the limitations of the
SWLA. The generalization of the random walk method beyond the SWLA is in
progress.

In summary, we have implemented a numerical method to obtain the
statistical properties of the transport coefficients using the
diffusion equation derived
in this work. We have extensive
numerical evidence of the suitability of our model to describe the
statistics of wave transport in disordered waveguides. It is worth
noticing that our model exactly reproduces the conductance
distributions obtained from the microscopic model even though this
one contains as many evanescent modes as needed to perform the
calculation in an exact manner. The only parameters needed to obtain
the statistics of any transport coefficient are the mean free paths
$\ell_{ab}$, as it is implied by the diffusion equation, all
the statistical properties being fixed at any length once all $\ell_{ab}$
parameters are fixed.


\section{Conclusions and Discussion}
\label{conclusions}

The central result of the present paper is the Fokker-Planck equation,
Eq. (\ref{diff eqn exp val sec2}), which describes the evolution with the length
$L$ of a disordered waveguide of transport properties which can be expressed in terms of the transfer matrix $\bm{M}$ of the system.

Our starting point is a potential model in which the scattering units consist of thin potential slices (taken as delta slices for convenience) perpendicular to the longitudinal direction of the waveguide, the variation of the potential in the transverse direction being arbitrary.
A statistical law for the potential slices is specified, as detailed in
Sec. \ref{stat model}:
in particular, the parameters of a given slice are taken to be statistically independent from those of any other slice, so that we are dealing here with the situation of uncorrelated (at least in the longitudinal direction) disorder.
Our result is obtained in the so called {\em dense-weak-scattering limit}, denoted by DWSL in the text, in which each potential slice is very weak and
the linear density of slices is very large, so that the resulting mean free paths (mfp's)
are fixed (see Eq. (\ref{dwsl})).
The statistical properties of a building block (denoted by BB) of length $\delta L$, say,
are first derived; the BB is then added to a waveguide of length $L$ to obtain a composition law, from which the diffusion equation is eventually derived.
In the DWSL, the statistical properties of the BB, and hence of the full system,
depend only on the mfp's which, in turn, depend only on the second moments of the individual delta-potential strengths. Cumulants of the potential higher than the second are irrelevant in the limit, signalling  the existence of a {\em generalized central-limit theorem} (CLT): once the mfp's are specified, the limiting equation
(\ref{diff eqn exp val sec2}) is {\em universal}, i.e., independent of other details of the microscopic statistics.

One important characteristic of the present analysis, compared with previous ones, is that the energy of the incident particle is fully taken into account,
a consequence being that the generalized diffusion coefficients appearing in the diffusion equation (\ref{diff eqn exp val sec2}) depend on the wavenumber $k$ of the incident wave and on the length $L$.

The diffusion equation (\ref{diff eqn exp val sec2}) for expectation values is very difficult to solve, the main reason being explained in the text, right below
Eq. (\ref{in_conditions}).
The original DMPK equation \cite{mpk} for the probability distribution of certain parameters of the transfer matrix was solved exactly for the unitary symmetry class only
\cite{beenakker_rejaei}, whereas for the evolution of expectation values arising from that same equation for a large number of open channels, $N \gg 1$,
an iterative procedure was developed to find the result as an expansion in powers of $1/N$ \cite{mello-kumar}.
In the present case, in Sec. \ref{analytic} we have been able to solve
Eq. (\ref{diff eqn exp val sec2}) exactly for $N=1$, but only for a few particular observables: the solution is in excellent agreement with the results of a microscopic calculation.
However, not even for $N \gg 1$ have we been able to develop an analytic iterative procedure like the one we mentioned above;
even numerically we have not succeeded in developing a method to solve
Eq. (\ref{diff eqn exp val sec2}).
We have thus tackled the problem of extracting information from the analysis of the present paper from a different point of view, based on the study of the BB itself,
which was shown to have universal statistical properties.
First, we should remark that the BB is useful not only as an intermediate step to obtain the diffusion equation;
it is interesting as a physical system in itself, i.e., a slab.
In the paper we obtained its statistical properties up to order $\delta L$ only, with some extension to order $(\delta L)^2$. In principle, although it represents a tedious task, the procedure could be carried on to at least a few more powers of $\delta L$.
A similar expansion was performed in an earlier publication \cite{mello_tomsovic}.
Second, the BB was used in Sec. \ref{numerical} to develop the method that we called
``random walk in the transfer matrix space",
which was essential for the numerical analysis based on the results of the present work.
The results reported in that section showed excellent agreement with the corresponding microscopic calculations.
Efforts towards an analytical and/or numerical treatment of the diffusion equation (\ref{diff eqn exp val sec2}) itself would be very important.

In Sec. \ref{SWLA} we develop the short-wavelength approximation, denoted as SWLA in the text, which bears resemblance to the geometrical optics limit studied in optics.
The results of this approximation allow making a connection with some of our previous
work, in which the energy did not appear explicitly in the analysis.
We should remark that the numerical results of the random walk in the transfer matrix space reported in Sec. \ref{numerical} were performed within this approximation.

In the analysis presented in this paper, the presence of evanescent modes
for a single slice appears in the effective potential
$(\hat{u}_r)_{ab}$ that occurs in Eq. (\ref{vr N}) and is used to construct the open-channel transfer matrix;
the effective potential takes into account transitions to evanescent modes.
Our statistical law is thus postulated for the matrix elements of the effective potential.
However, as we mentioned in Sec. \ref{fokker-planck} around
Eq. (\ref{comb law ext M})
and in Sec. \ref{random_walk}, the transfer matrix for a sequence of scatterers
was constructed multiplying open-channel transfer matrices, i.e., ignoring the presence of evanescent modes in the combination law.
Nonetheless, the final agreement with microscopic calculations is very good.
An important question for future investigation is thus to understand the effect of evanescent modes when combining subsystems to form the whole waveguide.

In the potential model developed here the property of time-reversal invariance is satisfied and the treatment is also restricted to scalar waves. In the language of random-matrix theory, we are dealing with the orthogonal symmetry class, or
$\beta =1$.
For possible applications to electronic systems, it would be interesting to extend the analysis to the unitary and symplectic cases, $\beta =2$ and $\beta =4$, respectively.

As explained in the Introduction, in earlier publications
(like Refs. \cite{mpk,mello-kumar})
the notion of maximum entropy in conjunction with a
number of physical constraints played an important role in selecting the
distribution of the BB: in a way, that selection captured the features arising from a CLT.
We think that it would be very interesting to investigate the question whether the results presented here can be obtained within such a framework.

Finally, since the results of our model have been compared successfully only with microscopic computer simulations, we think that it would be very challenging to measure these same quantities in the laboratory, in order to make comparisons with real-life experiments.

\acknowledgments

The authors would like to thank A. Garc\'{\i}a-Mart\'{\i}n, 
P. Garc\'{\i}a-Mochales, N. Kumar and P. A. Serena  for interesting discussions. This work was supported by the
Spanish MCyT (Ref. No. BFM2003-01167)  and the EU Integrated Project ``Molecular Imaging" (Contract No: LSHG-CT-2003-503259).
P.A.M. aknowledges Conacyt support through contract No. 42655.
M.Y also thanks Conacyt for its upport through Scholarship No. 179710.
We are also grateful to the Max Planck Institut f\"ur Physik Komplexer Systeme in Dresden, for supporting a long-term visit of P.A.M. and a short one of L.S.F.P and J.J.S., during which important progress on this paper was achieved.
\appendix

\section{Some Properties of the Transfer Matrix}
\label{M-properties}

\subsection{Transfer and Scattering Matrices}
\label{M_S matrices}

The transfer matrix $\bm{M}$ is closely related to a perhaps more
familiar object: the scattering or $\bm{S}$ matrix, that relates
incoming to outgoing waves:
\begin{equation}
\bm{S}=\left(\begin{array}{cc}
r & t'\\
t & r'\end{array}\right).
\end{equation}
The $N$-dimensional blocks of the $\bm{M}$ matrix (Eq. (\ref{M N 1})),
\begin{equation}
\bm{M} =\left[
\begin{array}{ll}
M^{11} & M^{12} \\
M^{21} & M^{22}
\end{array}
\right] \equiv \left[
\begin{array}{ll}
\alpha &\beta \\
\gamma & \delta
\end{array}
\right],
\end{equation}
are related to the reflection and transmission matrices $r,t$ for left
incidence and $r',t'$ for right incidence as
\begin{subequations}
\begin{eqnarray}
r&=&-\delta ^{-1}\gamma, \hspace{1cm} t'= \delta ^{-1}
\\
t &=& (\alpha ^{\dagger})^{-1} , \hspace{1cm}  r' = \beta \delta ^{-1} .
\end{eqnarray}
\label{r,t(M)beta2}
\end{subequations}
The physical property of flux conservation (FC) requires the
$\bm{S}$ matrix to be unitary ($\bm{S}\bm{S}^\dagger = \bm{1}$)
while the $\bm{M}$ matrix must satisfy the {\em pseudounitarity}
condition
\begin{equation}
\bm{M}^{\dagger }\bm{\Sigma} _z\bm{M}=\bm{\Sigma} _z .
\label{pseudoun}
\end{equation}
This is the only condition that $\bm{M}$ satisfies in the unitary, or
$\beta =2$, case. If, in addition, the system is time-reversal invariant (TRI), i.e.,
in the orthogonal case $\beta =1$, we have the extra condition
\begin{equation}
\bm{M}^{*}=\bm{\Sigma} _x\bm{M} \bm{\Sigma} _x,
\label{TRI N}
\end{equation}
where $\bm{\Sigma}_z$ and $\bm{\Sigma}_x$ have the structure of
Pauli matrices:
\begin{equation}
\bm{\Sigma}_z = \left(\begin{array}{cc}
1 & 0\\
0 & -1 \end{array}\right) \quad , \quad \bm{\Sigma}_x =
\left(\begin{array}{cc}
0 & 1\\
1 & 0 \end{array}\right).
\end{equation}
Eq. (\ref{TRI N}) implies
\begin{equation}
M^{22}=\left( M^{11}\right) ^{*},\qquad M^{21}=\left( M^{12}\right)
^{*}, \label{TRI 1}
\end{equation}
so that in Eq. (\ref{M N 1}) only the two blocks $M^{11}$ and
$M^{12}$, or $\alpha $ and $\beta $, need be considered. The
relation with the reflection and transmission matrices in the TRI case is now
\begin{subequations}
\begin{eqnarray}
r&=&-(\alpha ^{\ast}) ^{-1}\beta ^{\ast}, \hspace{1cm} t'= (\alpha ^{\ast}) ^{-1}
\\
t &=& (\alpha ^{\dagger})^{-1} , \hspace{1cm}  \;\;\;\;\;\; r' = \beta (\alpha ^{\ast}) ^{-1} .
\end{eqnarray}
\label{r,t(M)beta1}
\end{subequations}
In this TRI case, the matrix $\bm{\varepsilon}$ for the BB, defined in Eq. (\ref{M BB}), must satisfy the relations
\begin{equation}
\bm{\varepsilon} ^{*}=\bm{\Sigma} _x\bm{\varepsilon} \Sigma _x,  \label{TRI eps}
\end{equation}
so that
\begin{subequations}
\begin{eqnarray}
\varepsilon ^{21} &=&(\varepsilon ^{12})^{*}   \\
\varepsilon ^{22} &=&(\varepsilon ^{11})^{*}  .
\label{TRI eps}
\end{eqnarray}
\label{TRI eps 1}
\end{subequations}
Introducing the notation
\begin{equation}
\overline{1}=2,\qquad \overline{2}=1,
\label{bar}
\end{equation}
the TRI relations (\ref{TRI eps 1}) can be written as
\begin{equation}
\epsilon _{a b}^{\bar{i} \; \bar{j}}=(\epsilon _{a b}^{i j})^{*} .
\label{TRI_eps_bar}
\end{equation}

\subsection{The transfer matrix in terms of independent parameters}
\label{M_independent parameters}

FC and TRI symmetry imply that the actual number of independent
parameters of a transfer matrix is $2N^2+N$. This fact can be taken
explicity into account
by writing $\bm{M}$ in a {\em polar representation} as
\cite{mello-kumar}
\begin{equation}
\bm{M}=\left[\begin{array}{cc}
\bm{u} & 0\\
 0 & \bm{u}^* \end{array}\right]\left[\begin{array}{cc}
\sqrt{1+\lambda} & \sqrt{\lambda} \\
\sqrt{\lambda} & \sqrt{1+\lambda}\end{array}\right]
\left[\begin{array}{cc}
\bm{v} & 0\\
 0 & \bm{v}^* \end{array}\right] ,
\end{equation}
where $\bm{u}, \bm{v}$ are arbitrary $N\times N$ unitary
matrices (each
contributing $N^2$ parameters) and $\bm{\lambda}$ is a real diagonal
matrix with $N$ arbitrary
non-negative elements.

Another useful representation was introduced by Pereyra
\cite{pereyra}
\begin{equation}
\bm{M}=\left[\begin{array}{cc}
e^{ih} & 0 \\
0 & e^{-ih^*}\end{array}\right]\left[\begin{array}{cc}
\sqrt{1+\eta\eta^{*}} & \eta\\
\eta^{*} & \sqrt{1+\eta^{*}\eta}\end{array}\right]  ,
\label{RepPereyra}
\end{equation}
where $h$ is an arbitrary $N\times N$ Hermitian matrix (thus contributing
$N^{2}$ parameters) and $\eta$ is an arbitrary $N\times N$ complex symmetric
matrix (with $N^{2}+N$ parameters).

\subsection{Multiplicativity property}
\label{M_multiplicativity}

Suppose we start with a system having a transfer matrix $\bm{M}_1$
and enlarge it by adding, on its right hand side, another system
having a transfer matrix $\bm{M}_2$;
the transfer matrix of the combined system is simply given by
\begin{equation}
\bm{M} = \bm{M}_2 \bm{M}_1 .
\end{equation}
The role of closed channels (evanescent modes)
is briefly discussed in the text, around Eq. (\ref{comb law ext M}).
This multiplicativity property is extremely useful and is
extensively used in the analytic study along the present work. However, from a numerical point of view, the successive multiplication of $\bm{M}$-matrices is unstable because the pseudounitary group, to which the
transfer matrices belong, is not compact \cite{mello-kumar}. This
property leads to numerical instabilities as the norm of the
transfer matrix elements can grow without limit. In contrast,
numerical methods based on the scattering matrix can be very
effective and stable \cite{JJS_evanesc}: if we have the scattering
matrices $\bm{S}_{1}$ and $\bm{S}_{2}$,  the scattering matrix $\bm{S}$
corresponding to the composition of the subsystems $(1)$ on the left
and $(2)$ on the right is given by
\begin{subequations}
\begin{eqnarray}
r & = & r_{1}+t'_{1}r_{2}Qt_{1}\label{nn.310}\\
t & = & t_{2}Qt_{1}\label{nn.330}\\
t' & = & (t)^\top\label{nn.320}\\
r' & = & t_{2}Qr'_{1}t'_{2}+r_{2}
\label{nn.340}
\end{eqnarray}
\label{SM_combine}
\end{subequations}
where
\begin{equation} Q\equiv\left(\bm{I}_N-r'_{1}r_{2}\right)^{-1}
\label{nn.350}
\end{equation}
and $\bm{I}_N$ denotes the $N\times N$ unit matrix.

\section{Evanescent Modes and the Effective Potential}
\label{veff}

In this appendix we define the effective potential for a delta
slice that was introduced in Eqs. (\ref{eps}) and
(\ref{vr N}).

Consider a problem admitting $N$ open channels and $N'$ closed ones.
We shall eventually be interested in the limit $N'\to \infty$.
The total number of channels will be denoted by $N_T = N + N'$.
It will be convenient to define projection operators $P$ and $Q$
(with $P+Q=I$) unto open and closed channels, respectively, i.e.,
\begin{subequations}
\begin{eqnarray}
P&=&\sum_{a=1}^N
\left| \chi _a \right\rangle
\left\langle \chi _a \right|, \\
Q&=&\sum_{a=N+1}^{N_T}
\left| \chi _a \right\rangle
\left\langle \chi _a \right| ,
\end{eqnarray}
\end{subequations}
where $\left| \chi _a \right\rangle$ represents the ``transverse" state defined in Eq. (\ref{chi a}).
The most general solution of the Schr\"odinger equation on either side of the scattering system contains:

i) incoming- and outgoing-wave amplitudes for all the open channels.
We denote by $\tilde{a}_P^{(1)}$, $\tilde{a}_P^{(2)}$ the $N$-component vectors of incoming open-channel amplitudes on the left and right of the system, respectively, while
$\tilde{b}_P^{(1)}$, $\tilde{b}_P^{(2)}$ denote the corresponding outgoing open-channel amplitudes.

ii) ``outgoing" closed-channel amplitudes,
denoted by the $N'$-component vectors
$\tilde{b}_Q^{(1)}$, $\tilde{b}_Q^{(2)}$, on the left and right of the system, respectively: these are the components that decrease exponentially at infinity.
The $N'$-component vectors $\tilde{a}_Q^{(1)}$, $\tilde{a}_Q^{(2)}$
represent the ``incoming" closed-channel amplitudes, i.e., the components that increase exponentially at infinity.
In order to have a normalizable (in the Dirac delta-function sense)
wave function, closed channels can only give an exponentially vanishing contribution at infinity, so that the components
$\tilde{a}_Q^{(1)}$, $\tilde{a}_Q^{(2)}$,
which we keep for convenience in the following equation, will eventually be set equal to zero.
We shall also use the notation $\widetilde{\alpha }_{PP}\equiv P \alpha P$,
etc.
The wave amplitudes on the two sides are then related by the
``extended transfer matrix" \cite{mello-kumar} as follows:
\begin{equation}
\left[
\begin{tabular}{c}
\begin{tabular}{c}
$\widetilde{b }_{P}^{(2)}$  \\
$\widetilde{b }_{Q}^{(2)}$
\end{tabular}
\\ \hline
\begin{tabular}{c}
$\widetilde{a }_{P}^{(2)}$ \\
$\widetilde{a }_{Q}^{(2)}$
\end{tabular}
\end{tabular}
\right]
=\left[
\begin{tabular}{c|c}
\begin{tabular}{cc}
$\widetilde{\alpha }_{PP}$ & $\widetilde{\alpha }_{PQ}$ \\
$\widetilde{\alpha }_{QP}$ & $\widetilde{\alpha }_{QQ}$%
\end{tabular}
&
\begin{tabular}{cc}
$\widetilde{\beta }_{PP}$ & $\widetilde{\beta }_{PQ}$ \\
$\widetilde{\beta }_{QP}$ & $\widetilde{\beta }_{QQ}$
\end{tabular}
\\ \hline
\begin{tabular}{cc}
$\widetilde{\gamma }_{PP}$ & $\widetilde{\gamma }_{PQ}$ \\
$\widetilde{\gamma }_{QP}$ & $\widetilde{\gamma }_{QQ}$
\end{tabular}
&
\begin{tabular}{cc}
$\widetilde{\delta }_{PP}$ & $\widetilde{\delta }_{PQ}$ \\
$\widetilde{\delta }_{QP}$ & $\widetilde{\delta }_{QQ}$
\end{tabular}
\end{tabular}
\right]
\left[
\begin{tabular}{c}
\begin{tabular}{c}
$\widetilde{a }_{P}^{(1)}$  \\
$\widetilde{a }_{Q}^{(1)}$
\end{tabular}
\\ \hline
\begin{tabular}{c}
$\widetilde{b }_{P}^{(1)}$ \\
$\widetilde{b }_{Q}^{(1)}$
\end{tabular}
\end{tabular}
\right] \; .
\label{M ext}
\end{equation}
Here we are using the notation of Ref. \cite{mello-kumar}, which was developed in terms of incoming- and outgoing- wave amplitudes, both for the $S$ matrix and for the $M$ matrix. This results in an asymmetry in the notation in the two vectors appearing in Eq. (\ref{M ext}).
Perhaps a more common notation expresses the $M$ matrix in terms of waves that travel to the right and to the left, giving a more symmetric definition.

The extended transfer matrix of Eq. (\ref{M ext}), which will be denoted by $\tilde{M}$,
contains four $N_T \times N_T$ matrix blocks.
When we set, as we already mentioned, the amplitudes
${a }_{Q}^{(1)}={a }_{Q}^{(2)}=0$
and consider, as given data, the $2N$ amplitudes
${a }_{P}^{(1)}$, ${b }_{P}^{(1)}$,
we obtain a set of $2(N+N')$ equations in the same number of unknowns:
${a }_{P}^{(2)}$, ${b }_{P}^{(2)}$, ${b }_{Q}^{(1)}$ ${b }_{Q}^{(2)}$.

The ``open-channel transfer matrix" of Eq. (\ref{M N 1}),
that relates the open-channel amplitudes on the two sides as
\begin{eqnarray}
\left[
\begin{array}{c}
{b}_{P}^{\left( 2\right) } \\
{a}_{P}^{\left( 2\right) }
\end{array}
\right] &=&\left[
\begin{array}{cc}
\alpha & \beta \\
\gamma & \delta
\end{array}
\right] \left[
\begin{array}{c}
{a}_{P}^{\left( 1\right) } \\
{b}_{P}^{\left( 1\right) }
\end{array}
\right] \; ,
\end{eqnarray}
can be obtained from the extended transfer matrix of Eq. (\ref{M ext}) by eliminating the closed-channel amplitudes ${b }_{Q}^{(1)}$ ${b }_{Q}^{(2)}$,
to obtain the four $N\times N$ blocks
\begin{subequations}
\begin{eqnarray}
\alpha
&=&\widetilde{\alpha }_{PP}
-\widetilde{\beta }_{PQ}\frac{1}{\widetilde{\delta }_{QQ}}
\widetilde{\gamma }_{QP},
\label{alfa}
\\
\beta &=&\widetilde{\beta }_{PP}
-\widetilde{\beta }_{PQ}\frac{1}{\widetilde{\delta }_{QQ}}
\widetilde{\delta }_{QP},
\label{beta}
\\
\gamma &=&\widetilde{\gamma }_{PP}
-\widetilde{\delta}_{PQ}\frac{1}{\widetilde{\delta }_{QQ}}
\widetilde{\gamma }_{QP},
\label{gama}
\\
\delta
&=&\widetilde{\delta }_{PP}
-\widetilde{\delta }_{PQ}\frac{1}{\widetilde{\delta }_{QQ}}
\widetilde{\delta }_{QP}.
\label{delta}
\end{eqnarray}
\label{M(ext_M)}
\end{subequations}

For a delta slice centered at $x=0$ and described by the potential of
Eq. (\ref{delta slice}), one finds the following extended transfer matrix:
\begin{equation}
\tilde{M}
=\left[
\begin{array}{cc}
\tilde{\alpha } &  \tilde{\beta } \\
\tilde{\beta }^{\ast}  &  \tilde{\alpha }^{\ast}
\end{array}
\right]
=\left[
\begin{array}{cc}
I_{N_T} + \frac{1}{2i}\frac{1}{\sqrt{K}} u \frac{1}{\sqrt{K}}
& \frac{1}{2i}\frac{1}{\sqrt{K}} u \frac{1}{\sqrt{K}}
 \\
- \frac{1}{2i}\frac{1}{\sqrt{K}} u \frac{1}{\sqrt{K}}
&I_{N_T} - \frac{1}{2i}\frac{1}{\sqrt{K}} u \frac{1}{\sqrt{K}}
\end{array}
\right] .
\label{ext_M 1} \\
\end{equation}
In this equation, $I_{N_T}$ denotes the $N_T$-dimensional unit matrix;
$u$ is the $N_T \times N_T$ matrix constructed from the matrix elements $u_{ab}$ of Eq. (\ref{delta slice 2});
$K$ is the diagonal $N_T \times N_T$ matrix
$K_{ab}=k_a \delta _{ab}$; for open channels ($a=1,\cdots, N$), $k_a$ is defined as the real positive square root of the r.h.s. of Eq. (\ref{ka});
for closed channels ($a=N+1, \cdots, N_T$), we define
$k_a = i \kappa _a$, $\kappa _a$ being real and positive.
It will be convenient to write
\begin{subequations}
\begin{eqnarray}
K_{PP} &\equiv& k_P   \\
K_{QQ} &\equiv& i \kappa _Q .
\label{K=(k,ikappa)}
\end{eqnarray}
\end{subequations}
Substituting the extended transfer matrix of Eq. (\ref{ext_M 1}) into
Eq. (\ref{M(ext_M)}) we find, for the blocks of the open-channel transfer matrix
\begin{subequations}
\begin{eqnarray}
\alpha &=&I_N + \frac{1}{2i}\frac{1}{\sqrt{k_{P}}}
\hat{u}_{PP}\frac{1}{\sqrt{k_{P}}},
\label{alfax=0} \\
\beta &=& \frac{1}{2i}\frac{1}{\sqrt{k_{P}}}
\hat{u}_{PP}\frac{1}{\sqrt{k_{P}}} ,
\label{betax=0}
\end{eqnarray}
\label{alpha_betax=0}
\end{subequations}
where $\hat{u}_{PP}$ is the ``effective potential"
referred to in the text, Eq. (\ref{vr N}), and defined as:
\begin{equation}
\hat{u}_{PP}
=u_{PP}
-u_{PQ}\frac{1}{\sqrt{2\kappa _Q}}
\frac{1}
{
I_{N'}
+ \frac{1}{\sqrt{2\kappa _Q}}u_{QQ}\frac{1}{\sqrt{2\kappa _Q}}
}
\frac{1}{\sqrt{2\kappa _Q}}u_{QP}.
\label{EffPot}
\end{equation}

For the present scattering problem associated with the potential of Eq.
(\ref{delta slice}), the reflection and transmission amplitudes to open channels are to be obtained
from Eq. (\ref{r,t(M)beta1}), where $\alpha $ and $\beta $ are given in Eq. (\ref{alpha_betax=0}) in terms of the effective potential of
Eq. (\ref{EffPot}).

As an exercise to understand the importance of including the effect of evanescent modes in the
construction of the open-channel transfer matrix for a given scatterer,
we shall apply the above formulation to verify that a 2D delta potential produces no scattering at all
(for the 3D version of this statement see Ref. \cite{roman}, p. 274, Problem 3-7).

Suppose that the potential $U(x,y)$ of Eq. (\ref{delta slice}) is given by
(notice that $u_0$, which we assume positive, is adimensional)
\begin{equation}
U(x,y) = u_0 \delta (x)\delta (y-y_0).
\label{2D_delta_pot}
\end{equation}
Then $u(y) = u_0 \delta (y-y_0)$ and the matrix elements $u_{ab}$ are now given by
\begin{equation}
u_{ab} = u_0 \; \chi _a (y_0) \; \chi _b (y_0) \equiv u_a u_b ,
\label{ua}
\end{equation}
where $\chi _a (y)$ are the transverse states defined in Eq. (\ref{chi a}).
Notice that $u_{ab}$ has the property of being ``separable" in the channel indices $a$ and $b$.
Defining $|u\rangle$ as the column vector
$\left[u_1, u_2, \cdots u_N, \cdots u_{N_T}\right]^T$,
we can write the matrix $\bm{u}=\{u_{ab} \}$ as
\begin{equation}
\bm{u} = |u\rangle \langle u |.
\label{matrix u}
\end{equation}
The projections
\begin{subequations}
\begin{eqnarray}
| u_P \rangle \equiv P| u \rangle
&=&\left[u_1, u _2, \cdots u_N \right]^T
\label{uP}\\
| u_Q \rangle \equiv Q| u \rangle
&=&\left[u_{N+1}, \cdots u_{N+N'} \right]^T
\label{uQ}
\end{eqnarray}
\label{uP,uQ}
\end{subequations}
allow writing $u_{PP}$, etc., needed in Eq. (\ref{EffPot}), as
\begin{subequations}
\begin{eqnarray}
u_{PP} &=& |u_P \rangle \langle u_P |, \\
u_{PQ} &=& |u_P \rangle \langle u_Q |, \\
u_{QP} &=& |u_Q \rangle \langle u_P|, \\
u_{QQ} &=& |u_Q \rangle \langle u_Q |.
\label{uPP,etc}
\end{eqnarray}
\label{uPP,etc}
\end{subequations}
We substitute the relations (\ref{uPP,etc}) in the expression (\ref{EffPot}) for the effective potential and obtain
\begin{equation}
\hat{u} = |u_P \rangle
\left[
1
-\langle v_Q |
\frac{1}{I_{N'} + |v_Q \rangle \langle v_Q |}
|v_Q \rangle
\right]
\langle u_P | ,
\label{EffPot 1}
\end{equation}
where we have defined
\begin{equation}
|v_Q \rangle = \frac{1}{\sqrt{2\kappa _Q}} |u_Q \rangle .
\label{vQ}
\end{equation}
The inverse appearing in Eq. (\ref{EffPot 1}) can be trivially calculated due to the separable nature of $|v_Q \rangle \langle v_Q |$, with the result
\begin{equation}
\frac{1}{I_{N'} + |v_Q \rangle \langle v_Q | }
= I_{N'} - \frac{|v_Q \rangle \langle v_Q |}{1+\langle v_Q |v_Q \rangle},
\label{1/{I+vv}}
\end{equation}
so that
\begin{equation}
\hat{u} = \frac{|u_P \rangle \langle u_P |}
{1+\langle v_Q |v_Q \rangle} .
\label{EffPot 2}
\end{equation}
The square of the norm of the vector $|v_Q \rangle$ is calculated from the definitions (\ref{vQ}), (\ref{uQ}) and (\ref{ua}), and the expression
(\ref{chi a}) for the transverse states to find
\begin{equation}
\langle v_Q |v_Q \rangle
= \frac{u_0}{W}
\sum_{a=N+1}^{N+N'}\frac{\sin^2(\frac{\pi a y_0}{W})}
{\sqrt{(\frac{\pi a}{W})^2-k^2}}.
\label{(vQ,vQ)}
\end{equation}
As an example, suppose that our 2D delta potential is located midway between the two boundaries of the waveguide, i.e., $y_0 = W/2$.
We then find
\begin{equation}
\langle v_Q |v_Q \rangle
= \frac{u_0}{W}
\sum_{a=N+1, {\rm odd}}^{N+N'}
\frac{1}{\sqrt{(\frac{\pi a}{W})^2-k^2}}.
\label{(vQ,vQ) 1}
\end{equation}
The tail of this sum, from $A\gg 1$, say, to $N+N'\gg A$ is
\begin{equation}
\langle v_Q |v_Q \rangle
\sim \frac{u_0}{W}
\sum_{a=A, {\rm odd}}^{N+N'}
\frac{W}{\pi a}
\sim \frac{u_0}{2 \pi } \ln \frac{N+N'}{A}.
\label{(vQ,vQ) 1}
\end{equation}
Thus, as $N'\to \infty$, $\langle v_Q |v_Q \rangle $ has a logarithmic divergence, and $\hat{u}$ of Eq. (\ref{EffPot 2}) vanishes in the limit.
As a result, the 2D delta potential inside our waveguide does not scatter at all.

In conclusion, should we replace, in Eq. (\ref{alpha_betax=0}), the effective
potential $\hat{u}$ by the ``bare" potential $u$, we would find that the 2D
delta potential produces non-zero scattering. The presence of an infinite
number of evanescent modes is essential to demonstrate that our 2D delta
potential produces no scattering.

\section
{The fourth-order term in the second-moment expansion, Eq. (\ref{vareps^2 4})}
\label{4th_order in second_moment}

We go back to the expression for the second moments of $\varepsilon $ for the BB, Eq. (\ref{vareps^2 series}).
The terms in the line (\ref{vareps^2 3}) vanish, being third order in the individual
$[\epsilon _r]_{ab}^{ij}$ and hence in the potentials $(\hat{v}_r)_{ab}$
[see Eq. (\ref{mu_(2t+1)})].
The fourth-order terms are given in the line (\ref{vareps^2 4}):
only the first of these three terms survives; in the other two there is no way to pair
the scatterer indices so as to get a non-vanishing result
(remember that the various $\epsilon _r$'s are statistically independent and have zero average).
For the nonvanishing term we have [see Eq. (\ref{eps})]
\begin{subequations}
\begin{eqnarray}
&&\left\langle  \left[ \varepsilon ^{(2)} \right]_{a b}^{i j}
\left[ \varepsilon ^{(2)} \right]_{c d}^{h l} \right\rangle _{\delta L}
=
\sum_
{
\substack{
r>s
\\
t>u
}
}
\Bigl<
\left[\bm{\epsilon}  _{r}\bm{\epsilon} _{s}\right]_{a b}^{i j}
\left[\bm{ \epsilon} _{t}\bm{\epsilon} _{u}\right]_{cd} ^{hl}
\Bigr>
\label{<eps2> BB 4 order a}
\\
&&\;\;\;\;\; =
\sum_
{
\substack{
r>s
\\
t>u
}
}
\sum_{\alpha '\beta '}
\Bigl<
(\hat{v}_r)_{a \alpha '} (\hat{v}_s)_{\alpha ' b} (\hat{v}_t)_{c \beta '} (\hat{v}_u)_{\beta 'd}
\Bigr>
\sum_{\lambda ' \mu '}\left[
(\vartheta _r)_{a \alpha '}^{i \lambda '}(\vartheta _s)_{\alpha ' b}^{\lambda ' j}
(\vartheta _t)_{c \beta '}^{h \mu '}(\vartheta _u)_{\beta ' d}^{\mu ' l}
\right]
\label{<eps2> BB 4 order b}
\\
&&\;\;\;\;\; =
\sum_{r>s}
\sum_{\alpha '\beta '}
\Bigl<(\hat{v}_r)_{a \alpha '} (\hat{v}_r)_{c \beta '}\Bigr>
\Bigl<(\hat{v}_s)_{\alpha ' b}  (\hat{v}_s)_{\beta ' d} \Bigr>
\sum_{\lambda ' \mu '}\left[
(\vartheta _r)_{a \alpha '}^{i \lambda '}(\vartheta _s)_{\alpha ' b}^{\lambda ' j}
(\vartheta _r)_{c \beta '}^{h \mu '}(\vartheta _s)_{\beta ' d}^{\mu ' l}
\right]
\\
&&\;\;\;\;\; =
\sum_{\alpha '\beta '}
\frac{\mu _2^{(v)}(a\alpha ',c\beta ')}{d} \;
\frac{\mu _2^{(v)}(\alpha ' b,\beta ' d)}{d}
\nonumber \\
&&\hspace{3cm}
\cdot \left\{
\sum_{\lambda '\mu '}
\sum_{r>s}
\left[
(\vartheta _r)_{a \alpha '}^{i \lambda '}(\vartheta _s)_{\alpha ' b}^{\lambda ' j}
(\vartheta _r)_{c \beta '}^{h \mu '}(\vartheta _s)_{\beta ' d}^{\mu ' l}
\right]
d^2
\right\} .
\end{eqnarray}
\label{<eps2> BB 4 order}
\end{subequations}
We now take the DWSL and find [see Eq. (\ref{<vv>/d})]
\begin{eqnarray}
\lim_{DWS}
\left\langle  \left[ \varepsilon ^{(2)} \right]_{a b}^{i j}
\left[ \varepsilon ^{(2)} \right]_{c d}^{h l} \right\rangle _{\delta L}
&=&\sum_{\alpha '\beta '}
\frac{C(a\alpha ', c\beta ')}
{\sqrt{\ell_{a\alpha '}(k) \ell_{c \beta '}(k)}}
\frac{C(\alpha ' b, \beta ' d)}
{\sqrt{\ell_{\alpha ' b}(k) \ell_{\beta ' d}(k)}}
\;\;\sum_{\lambda '\mu '}
\Delta_{a \alpha ',\alpha 'b,c \beta ',\beta 'd}^{i \lambda ',\lambda ' j,h \mu ',\mu ' l}
(k;{\cal R}(\delta L)),
\nonumber \\
\label{vareps(2)vareps(2) DWSL}
\end{eqnarray}
in analogy with Eq. (\ref{<vareps^2>}).
We have defined
\begin{subequations}
\begin{eqnarray}
&&\Delta_{a \alpha ',\alpha ' b,c \beta ',\beta ' d}
^{i \lambda ',\lambda ' j,h \mu ',\mu ' l}
(k;{\cal R}(\delta L))
=\int\int_{{\cal R}(\delta L)}
\vartheta _{a \alpha '}^{i \lambda '}(x)
\vartheta _{\alpha ' b}^{\lambda ' j} (x')
\vartheta _{c \beta '}^{h \mu '}(x)
\vartheta _{\beta ' d}^{\mu ' l}(x')
dx dx' ,
\label{Delta(k) 1 a}
\\
&& \;\;\;\;
= \frac{(-)^{i+h+\lambda ' + \mu '}}{iK_2}
\left[
\frac{\sin{\frac{K_1 + K_2}{2}\delta L}}{\frac{K_1 + K_2}{2}}
-e^{-i\frac{K_2}{2}\delta L}
\frac{\sin{\frac{K_1 \delta L}{2}}}{\frac{K_1}{2}}
\right] .
\label{Delta(k) 1 b}
\end{eqnarray}
Here, ${\cal R}(\delta L)$ denotes the region of integration $\{x>x'\}$,
i.e., half a square of size $\delta L$.
Eqs. (\ref{Delta(k) 1 a}) and (\ref{Delta(k) 1 b}) are analogous to the earlier definitions in
Eqs. (\ref{Delta(k)}) and (\ref{Delta}).
Eq. (\ref{Delta(k) 1 b}) is valid for $K_1 \neq 0$ and $K_2 \neq 0$.
The other possibilities are
\begin{eqnarray}
\Delta_{a \alpha ',\alpha ' b,c \beta ',\beta ' d}
^{i \lambda ',\lambda ' j,h \mu ',\mu ' l}
(k;{\cal R}(\delta L))
&=& (-)^{i+h+\lambda ' + \mu '}
\frac{(\delta L)^2}{2} \; ,
\hspace{1cm}K_1=K_2=0
\label{Delta(k) 1 c} \\
&=& (-)^{i+h+\lambda ' + \mu '} \frac{1}{iK_2}
\left[
\frac{\sin\frac{K_2\delta L}{2}}{\frac{K_2}{2}}
-e^{-i\frac{K_2\delta L}{2}} \delta L
\right] ,   \hspace{5mm}  K_1=0, \; K_2\neq 0
\label{Delta(k) 1 d}
\nonumber \\
\\
&=& (-)^{i+h+\lambda ' + \mu '} \frac{1}{iK_1}
\left[
e^{i\frac{K_1\delta L}{2}}\delta L
- \frac{\sin\frac{K_1\delta L}{2}}{\frac{K_1}{2}}
\right] ,    \hspace{5mm}  K_1\neq0, \; K_2 = 0 \; .
\nonumber \\
\label{Delta(k) 1 e}
\end{eqnarray}
\label{Delta(k) 1}
\end{subequations}
We have defined
\begin{subequations}
\begin{eqnarray}
K_1 &=& K_{a \alpha ', c \beta '}^{i \lambda ', h \mu '}
\\
K_2 &=& K_{\alpha ' b,\beta ' d}^{\lambda ' j,\mu ' l} \; .
\end{eqnarray}
\label{K1,K2}
\end{subequations}
We see from Eq. (\ref{Delta(k) 1 b})-(\ref{Delta(k) 1 e}),
or directly from the integral definition (\ref{Delta(k) 1 a}), that
in an expansion in powers of $\delta L$,
the leading term is quadratic in $\delta L$, i.e.,
\begin{equation}
\Delta_{a \alpha ',\alpha ' b,c \beta ',\beta ' d}
^{i \lambda ',\lambda ' j,h \mu ',\mu ' l}
(k;{\cal R}(\delta L))
= (-)^{i+h+\lambda '+ \mu '}
\frac{(\delta L)^2}{2} + \cdots ,
\label{Delta_leading 1}
\end{equation}
so that {\em the leading term, in a similar expansion,
of the fourth-order contribution to the second moments of
$\varepsilon $ in the DWSL, Eq. (\ref{vareps(2)vareps(2) DWSL}),
behaves as}
\begin{eqnarray}
\lim_{DWS}
\left\langle  \left[ \varepsilon ^{(2)} \right]_{a b}^{i j}
\left[ \varepsilon ^{(2)} \right]_{c d}^{h l} \right\rangle _{\delta L}
&=&
\sum_{\alpha '\beta '}
\frac{C(a\alpha ', c\beta ')}
{\sqrt{\ell_{a\alpha '}(k) \ell_{c \beta '}(k)}}
\frac{C(\alpha ' b, \beta ' d)}
{\sqrt{\ell_{\alpha ' b}(k) \ell_{\beta ' d}(k)}}
\nonumber \\
&&\times\sum_{\lambda '\mu '}
(-)^{i+h+\lambda ' + \mu '}
\left[\frac{(\delta L)^2}{2} + O(\delta L)^3 \right] .
\label{<vareps^2>leading 1}
\end{eqnarray}
This is the result mentioned at the end of
Sec. \ref{second moment BB propto v2}, right above Eq. (\ref{<vareps^2>leading}).

In the above analysis, the BB lies in the interval
$(-\delta L /2, \delta L /2)$.
If it is shifted to the interval $(L, L+\delta L)$,
Eq. ({\ref{vareps(2)vareps(2) DWSL}) is modified as
\begin{eqnarray}
\lim_{DWS}
\left\langle  \left[ \varepsilon ^{(2)} \right]_{a b}^{i j}
\left[ \varepsilon ^{(2)} \right]_{c d}^{h l} \right\rangle _{L, \delta L}
&=&\sum_{\alpha ' \beta '}
\frac{C(a\alpha ', c\beta ')}
{\sqrt{\ell_{a\alpha '}(k) \ell_{c \beta '}(k)}}
\frac{C(\alpha ' b, \beta ' d)}
{\sqrt{\ell_{\alpha ' b}(k) \ell_{\beta ' d}(k)}}
\nonumber \\
&&\hspace{1cm}\times\sum_{\lambda ' \mu '}
\Delta_{a \alpha ',\alpha ' b,c \beta ',\beta ' d}
^{i \lambda ',\lambda ' j,h \mu ',\mu ' l}
(k;{\cal R}(\delta L))e^{i(K_1+K_2)(L+\frac{\delta L}{2})},
\nonumber \\
\label{vareps(2)vareps(2) DWSL 1}
\end{eqnarray}
in analogy with Eq. (\ref{<vareps^2> L,dL}), while
Eq. (\ref{<vareps^2>leading 1}) becomes
\begin{eqnarray}
\lim_{DWS}
\left\langle  \left[ \varepsilon ^{(2)} \right]_{a b}^{i j}
\left[ \varepsilon ^{(2)} \right]_{c d}^{h l} \right\rangle _{L, \delta L}
&=&
\sum_{\alpha '\beta '}
\frac{C(a\alpha ', c\beta ')}
{\sqrt{\ell_{a\alpha '}(k) \ell_{c \beta '}(k)}}
\frac{C(\alpha ' b, \beta ' d)}
{\sqrt{\ell_{\alpha ' b}(k) \ell_{\beta ' d}(k)}}
\nonumber \\
&&\times\sum_{\lambda '\mu '}
(-)^{i+h+\lambda ' + \mu '}
\left[e^{i(K_1+K_2)L}\frac{(\delta L)^2}{2} + O(\delta L)^3 \right] .
\label{<vareps^2>leading 2}
\end{eqnarray}

\section{Analysis of the general term occurring in the calculation of the $p$-th moment of
$\varepsilon$ for the BB}
\label{pth moment varepsilon}

Eq. (\ref{vareps^2 series}) which gives the expansion of a second moment of $\varepsilon$
in terms of the $\varepsilon^{(\mu)}$'s, and hence to various orders in
the individual $\epsilon _r$'s,
can be generalized to an arbitrary $p$-th moment as
\begin{equation}
\left\langle
\varepsilon_{a_1 b_1}^{i_1 j_1}
\cdots
\varepsilon_{a_p b_p}^{i_p j_p}
\right\rangle _{\delta L}
= \sum_{\mu_1,\cdots, \mu_{p}}^{m}
\left\langle
\left[ \varepsilon ^{(\mu_1)} \right]_{a_1 b_1}^{i_1 j_1}
\cdots
\left[ \varepsilon ^{(\mu_{p})} \right]_{a_p b_p}^{i_p j_p}
\right\rangle _{\delta L}.
\label{vareps^p}
\end{equation}
In the analysis that follows 
{\em the BB will be centered at the origin}.
The term under the summation sign in this last equation is of order
$\mu_1 + \cdots +\mu_p$
in the individual
$[\epsilon _r]_{a b}^{i j}$'s, and hence of the same order in the potential
matrix elements $(\hat{v}_r)_{a b}$'s;
it survives only if it is of even order in these quantities, i.e., if
$\mu_1 + \cdots +\mu_p =2q$, say.

Using, for convenience, a simplified notation for the indices,
we express the term of order $2q$ under the summation sign in Eq. (\ref{vareps^p}) as
\begin{eqnarray}
&&
\left\langle
\left[ \varepsilon ^{(\mu_1)} \right]_{ab}^{ij}
\cdots
\left[ \varepsilon ^{(\mu_{p})} \right]_{ef}^{mn}
\right\rangle _{\delta L}
=
\sum_
{
\substack{
r_{1}>\cdots >r_{\mu _1} \\
\cdots \\
t_{1}> \cdots >t_{\mu _p}
}
}
\left\langle
\left[
\bm{\epsilon} _{r_1}\cdots \bm{\epsilon} _{r_{\mu _1}}
\right]_{ab}^{ij}
\cdots
\left[
\bm{\epsilon}_{t_1}\cdots \bm{\epsilon}_{t_{\mu _p}}
\right]_{ef}^{mn}
\right\rangle
\nonumber \\
&&\;\;\;\;\;\;=
\sum_
{
\substack{
r_{1}, \cdots,r_{\mu _1} \\
\cdots \\
t_{1}, \cdots, t_{\mu _p}
}
}
\;\;
\sum_
{
\substack{
\alpha_{1},\cdots,\alpha_{\mu _1 -1} \\
\cdots \\
\gamma_{1},\cdots,\gamma_{\mu _p -1}
}
}
\left\langle
(\hat{v}_{r_1})_{a \alpha _1}
\cdots
(\hat{v}_{r_{\mu _1}})_{\alpha _{\mu _1-1} b}
\cdots
(\hat{v}_{t_1})_{e \gamma _1}
\cdots
(\hat{v}_{t_{\mu _p}})_{\gamma _{\mu _p-1} f}
\right\rangle
\nonumber \\
&&\hspace{2cm}
\times \sum_
{
\substack{
\lambda _{1},\cdots,\lambda _{\mu _1 -1} \\
\cdots \\
\nu_{1},\cdots,\nu_{\mu _p -1}
}
}
[\vartheta _{r_1}]_{a \alpha _1}^{i \lambda _1}
\cdots
[\vartheta _{r_{\mu _1}}]_{\alpha _{\mu _1-1} b}^{\lambda _{\mu _1 -1}j}
\cdots
[\vartheta _{t_1}]_{e \gamma _1}^{m \nu _1}
\cdots
[\vartheta _{t_{\mu _p}}]_{\gamma _{\mu _p-1} f}^{\nu _{\mu _p -1}n}
\nonumber \\
&&\hspace{3cm} \times
h(r_1 - r_2) \cdots h(r_{\mu _1 -1} - r_{\mu _1})
\nonumber \\
&&\hspace{5cm}
\cdots
\nonumber \\
&&\hspace{3cm} \times
h(t_1 - t_2) \cdots h(t_{\mu _p -1} - t_{\mu _p})
\nonumber \\
&&\;\;\;\;\;\;
\equiv
\sum_
{
\substack{
\alpha_{1},\cdots,\alpha_{\mu _1 -1} \\
\cdots \\
\gamma_{1},\cdots,\gamma_{\mu _p -1}
}
}
\;\;
\sum_
{
\substack{
\lambda _{1},\cdots,\lambda _{\mu _1 -1} \\
\cdots \\
\nu_{1},\cdots,\nu_{\mu _p -1}
}
}
F_{a \alpha _1, \cdots , \alpha _{\mu_1 -1}b\; ; \cdots ; \;
e \gamma _1 , \cdots , \gamma _{\mu _p -1}f}
^{i \lambda _1, \cdots , \lambda _{\mu _1 -1}j; \; \cdots ; \;
m \nu _1 , \cdots , \nu _{\mu _p -1}n} \; .
\label{vareps^p 1}
\end{eqnarray}
We have introduced the step function $h(r-s)$ 
($=1$ for $r>s$ and $=0$ for $r\leq s$)
to implement the correct range of summation of the
scatterer indices.
The function $F$ defined in the last line can be read off from the equation itself; it has
$\mu_1 + \cdots +\mu_p =2q$  pairs of upper and lower indices and has the structure
\begin{eqnarray}
F_{a_1 b_1, \cdots, a_{2q} b_{2q}}^{i_1  j_1\cdots, i_{2q} j_{2q}}
&=&\sum _{r_1,\cdots,r_{2q}}
\Big\langle  (\hat{v}_{r_1})_{a_1b_1}
\cdots
(\hat{v}_{r_{2q}})_{a_{2q}b_{2q}}\Big\rangle
\nonumber \\
&&\hspace{2cm} \times
f_{a_1 b_1, \cdots, a_{2q} b_{2q}}^{i_1 j_1\cdots, i_{2q} j_{2q}}(r_1,\cdots,r_{2q}) ,
\label{F 0}
\end{eqnarray}
where the function $f_{a_1 b_1, \cdots, a_{2q} b_{2q}}^{i_1 j_1\cdots, i_{2q} j_{2q}}(r_1,\cdots,r_{2q})$
is given by:
\begin{eqnarray}
f_{a_1 b_1, \cdots, a_{2q} b_{2q}}^{i_1 j_1\cdots, i_{2q} j_{2q}}(r_1,\cdots,r_{2q})
&=&
[\vartheta _{r_1}]_{a_1 b_1}^{i_1 j_1}
\cdots
[\vartheta _{r_{2q}}]_{a_{2q} b_{2q}}^{i_{2q}j_{2q}}
\nonumber \\
&&\hspace{2cm} \times
{{\prod}'}_{i=1}^{2q-1}  h(r_i - r_{i+1}),
\label{f}
\end{eqnarray}
where the prime in the product sign means
$i \neq \mu_1, \mu_1 + \mu _2, \cdots, \mu_1 +\cdots + \mu _{p-1}$.
Two particular examples of the structure (\ref{F 0})
were already encountered earlier, in
Eqs. (\ref{var-eps^2 2b}) and (\ref{<eps2> BB 4 order b}) above.

The expectation value
appearing in Eq. (\ref{F 0}) can be written in terms of the cumulants of the various
blocks of $(\hat{v}_r)_{a b}$'s into which one can partition the product
$(\hat{v}_{r_1})_{a_1b_1} \cdots (\hat{v}_{r_{2q}})_{a_{2q}b_{2q}}$.
We first give a few examples, and then the general expression.

 i) $q=1$. One can write [see Eq. (\ref{<vv>})]
\begin{equation}
\Big\langle  (\hat{v}_{r_1})_{a_1 b_1}
(\hat{v}_{r_2})_{a_2 b_2}\Big\rangle
=\kappa_2^{(v)}(a_1 b_1, a_2 b_2)
\delta _{r_1 r_2} ,
\label{cum 2}
\end{equation}
where a second cumulant coincides with the corresponding second moment, i.e.,
\begin{equation}
\kappa_2^{(v)}(a_1 b_1, a_2 b_2)
=\mu_2^{(v)}(a_1 b_1, a_2 b_2),
\label{kappa_2}
\end{equation}
due to the vanishing of the first moments [Eq. (\ref{<u>})].

ii) For $q=2$ we have:
\begin{subequations}
\begin{eqnarray}
&&\Big\langle  (\hat{v}_{r_1})_{a_1 b_1} (\hat{v}_{r_2})_{a_2 b_2}
 (\hat{v}_{r_3})_{a_3 b_3} (\hat{v}_{r_4})_{a_4 b_4}\Big\rangle
 =\big[
\kappa _2^{(v)} (a_1 b_1,a_2 b_2)\kappa _2^{(v)} (a_3 b_3,a_4 b_4) \delta _{r_1 r_2} \delta _{r_3 r_4}
\nonumber \\
&& \hspace{.5cm} +
\kappa _2^{(v)} (a_1 b_1,a_3 b_3)\kappa _2^{(v)} (a_2 b_2,a_4 b_4) \delta _{r_1 r_3} \delta _{r_2 r_4}
+ \kappa _2^{(v)} (a_1 b_1,a_4 b_4)\kappa _2^{(v)} (a_2 b_2,a_3 b_3) \delta _{r_1 r_4} \delta _{r_2 r_3}
\big]
\label{q=2 pairs}
\nonumber \\
\\
&& \hspace{1cm} + \kappa _4^{(v)} (a_1 b_1,a_2 b_2, a_3 b_3,a_4 b_4) \delta _{r_1 r_2 r_3 r_4} \; ,
\label{cum 4}
\label{q=2 quartets}
\end{eqnarray}
\label{cum_exp_q=2}
\end{subequations}
where a fourth cumulant is defined in the usual way, i.e.,
\begin{eqnarray}
&&\kappa _4^{(v)} (a_1 b_1,a_2 b_2, a_3 b_3,a_4 b_4)
=\mu _4^{(v)} (a_1 b_1,a_2 b_2, a_3 b_3,a_4 b_4) \nonumber \\
&&\;\;\;\;\;
-\mu _2^{(v)} (a_1 b_1,a_2 b_2)\mu _2^{(v)} (a_3 b_3,a_4 b_4)
-\mu _2^{(v)} (a_1 b_1,a_3 b_3)\mu _2^{(v)} (a_2 b_2,a_4 b_4)
\nonumber \\
&&\hspace{3cm}-\mu _2^{(v)} (a_1 b_1,a_4 b_4)\mu _2^{(v)} (a_2 b_2,a_3 b_3).
\label{kappa_4}
\end{eqnarray}
In Eq. (\ref{cum_exp_q=2}), the two lines ending in (\ref{q=2 pairs})
contain all possible {\em pair contractions}, i.e., $3!! = 3$ terms altogether:
this partition of 4 elements can be represented by the Young diagram:
\begin{equation}
 \begin{array}{l}
\Box\Box \\
\Box\Box
\end{array}
\end{equation}
The last line, (\ref{q=2 quartets}), contains
the only possible {\em quartet} (the Kronecker delta with more than two indices is
defined to be nonzero only when all the indices are equal).
It can be represented by the Young diagram:
\begin{equation}
 \begin{array}{l}
\Box\Box \Box\Box
\end{array}
\end{equation}

iii) For $q=3$ we have:
\begin{subequations}
\begin{eqnarray}
&&\Big\langle  (\hat{v}_{r_1})_{a_1 b_1} (\hat{v}_{r_2})_{a_2 b_2}
 (\hat{v}_{r_3})_{a_3 b_3} (\hat{v}_{r_4})_{a_4 b_4}
 (\hat{v}_{r_5})_{a_5 b_5} (\hat{v}_{r_6})_{a_6 b_6}\Big\rangle
 \nonumber \\
&& \hspace{.5cm}
=\big[
\kappa _2 ^{(v)} (a_1 b_1,a_2 b_2)\kappa _2 ^{(v)} (a_3 b_3,a_4 b_4)
\kappa _2 ^{(v)} (a_5 b_5,a_6 b_6)
 \delta _{r_1 r_2} \delta _{r_3 r_4} \delta _{r_5 r_6}
\nonumber \\
&& \hspace{2cm} +
{\rm \; all \; possible \; combinations}
\big]
\label{q=2 222}
\\
&& \hspace{1cm}
+ \big[\kappa _4 ^{(v)} (a_1 b_1,a_2 b_2, a_3 b_3,a_4 b_4)
\kappa _2 ^{(v)} (a_5 b_5, a_6 b_6) \delta _{r_1 r_2 r_3 r_4 }
\delta _{r_5 r_6}
\nonumber \\
&& \hspace{2cm} +
{\rm \; all \; possible \; combinations}
\big]
\label{q=2 42}
\\
&& \hspace{1cm}
+ \kappa _6 ^{(v)} (a_1 b_1,a_2 b_2,a_3 b_3,a_4 b_4, a_5 b_5,a_6 b_6) \delta _{r_1 r_2 r_3 r_4 r_5 r_6} \; .
\label{q=2 6}
\end{eqnarray}
\end{subequations}
The two lines ending in (\ref{q=2 222}) of this last equation
contain all possible {\em pair contractions}, i.e., $5!!=15$ terms altogether:
this partition of 6 elements can be represented by the Young diagram:
\begin{equation}
 \begin{array}{l}
\Box\Box \\
\Box\Box \\
\Box\Box
\end{array}
\end{equation}
The two lines ending in (\ref{q=2 42}) contain all possible combinations of {\em one quartet plus one-pair contraction}, i.e.,
$
{\tiny
\left(
\begin{array}{c}
6 \\
2
\end{array}
\right)
}
=15$
terms altogether:
this partition of 6 elements can be represented by the Young diagram:
\begin{equation}
 \begin{array}{l}
\Box\Box \Box\Box\\
\Box\Box
\end{array}
\end{equation}
The last line, (\ref{q=2 6}), contains
the only possible {\em sextet}.
It can be represented by the Young diagram:
\begin{equation}
 \begin{array}{l}
\Box\Box \Box\Box\Box\Box
\end{array}
\end{equation}

It seems plausible that the particular examples given above can be
generalized to arbitrary $q$, so that we can write
$F_{a_1,\cdots,b_{2q}}^{i_1,\cdots,j_{2q}}$
of Eq. (\ref{F 0})
as
\begin{subequations}
\begin{eqnarray}
F_{a_1 b_1, \cdots, a_{2q} b_{2q}}^{i_1  j_1\cdots, i_{2q} j_{2q}}
&=&\sum _{r_1,\cdots,r_{2q}}
\Big\langle  (\hat{v}_{r_1})_{a_1b_1}
\cdots
(\hat{v}_{r_{2q}})_{a_{2q}b_{2q}}\Big\rangle
f_{a_1 b_1, \cdots, a_{2q} b_{2q}}^{i_1 j_1\cdots, i_{2q} j_{2q}}(r_1,\cdots,r_{2q})
\\
&=&
\sum _{r_1,\cdots,r_{2q}}
\left\{
\Big[
\kappa_2 ^{(v)} (a_1 b_1,a_2 b_2)
\cdots \kappa_2 ^{(v)} (a_{2q-1} b_{2q-1},a_{2q} b_{2q})
\delta _{r_1 r_2} \cdots \delta _{r_{2q-1} r_{2q}}
\right.
\nonumber \\
&& \hspace{2cm}
+ {\rm \; all \; possible \; combinations}
\Big]
\label{F 01}
\\
&& +
\Big[
\kappa_4 ^{(v)} (a_1 b_1,a_2 b_2,a_3 b_3,a_4 b_4)
\kappa_2 ^{(v)} (a_5 b_5,a_6 b_6)
\cdots
\kappa_2 ^{(v)} (a_{2q-1} b_{2q-1},a_{2q} b_{2q})
\nonumber \\
&& \hspace{2cm}
\times \delta _{r_1 r_2 r_3 r_4}\delta _{r_5 r_6}
\cdots \delta _{r_{2q-1} r_{2q}}
+ {\rm \; all \; possible \; combinations}
\Big]
\label{F 02}
\nonumber \\ \\
&& +
\cdots
\nonumber \\
&&
+ \kappa_{2q} ^{(v)} (a_1 b_1, a_2 b_2\cdots ,a_{2q} b_{2q})
\delta _{r_1 r_2 \cdots r_{2q}}
\Big\}
f_{a_1 b_1, \cdots, a_{2q} b_{2q}}^{i_1 j_1\cdots, i_{2q} j_{2q}}(r_1,\cdots,r_{2q})
 \; .
\label{F 03}
\end{eqnarray}
\label{F 0F}
\end{subequations}
Again, the partition of $2q$ elements contained inside each square bracket can be represented by a Young diagram.
The first square bracket ending in (\ref{F 01}) contains all possible pair contractions,
of which there are $(2q-1)!!$ altogether.

Eq. (\ref{F 0F}) shows that we can write
$F_{a_1,\cdots,b_{2q}}^{i_1,\cdots,j_{2q}}$
(omitting, for simplicity, the lower and upper indices, as well as the index $(v)$ in the cumulants) as
\begin{subequations}
\begin{eqnarray}
F
&=&
\Bigg[
\frac{\kappa_2 (a_1 b_1,a_2 b_2)}{d}
\cdots \frac{\kappa_2 (a_{2q-1} b_{2q-1},a_{2q} b_{2q})}{d}
\sum_{r_2 r_4 \cdots r_{2q}}
f\left(r_2, r_2, \dots,  r_{2q}, r_{2q} \right) d^q
\nonumber \\
&& \hspace{2cm} +
{\rm \; all \; possible \; combinations}
\Bigg]
\label{F 1}
\\
&& + d
\Bigg[
\frac{\kappa_4 (a_1 b_1,a_2 b_2,a_3 b_3,a_4 b_4)}{d^2}
\frac{\kappa_2 (a_5 b_5,a_6 b_6)}{d}
\cdots
\frac{\kappa_2 (a_{2q-1} b_{2q-1},a_{2q} b_{2q})}{d}
\nonumber \\
&& \hspace{2cm} \times
\sum_{r_4 r_6 \cdots r_{2q}}
f\left(r_4,r_4,r_4,r_4,r_6, r_6, \dots,  r_{2q}, r_{2q} \right) d^{q-1}
+ {\rm \; all \; possible \; combinations}
\Bigg]
\label{F 2}
\nonumber \\ \\
&& +\cdots
\nonumber \\ \nonumber \\
&& + d^{q-1}\frac{\kappa_{2q} (a_1 b_1,\cdots ,a_{2q} b_{2q})}{d^q}
\sum_{r_{2q}}f\left( r_{2q}, \cdots, r_{2q}  \right) d \; .
\label{F 3}
\end{eqnarray}
\label{F F}
\end{subequations}
The cumulants $\kappa_{2t}$ appearing in Eq. (\ref{F F}) are defined, for $t=1,2$, in Eqs. (\ref{kappa_2}) and (\ref{kappa_4}), respectively.

At this point we take the DWSL defined by Eqs. (\ref{dwsl}).
The various fractions $\kappa_{2t}/d^t$ appearing in Eq. (\ref{F F})
are finite because of the scaling assumed in Eq. (\ref{mu_(2t)}).
Also, the various summations in Eq. (\ref{F F}) tend to finite integrals, and all the terms with factors of $d$ ``left over", i.e., from (\ref{F 2}) up to (\ref{F 3}), vanish.
As a consequence, the cumulants $\kappa_4, \cdots \kappa_{2q}$,
{\em do not contribute in the DWSL}:
this is the {\em central-limit theorem} (CLT) that was discussed at the end of Sec. \ref{fokker-planck},
at the end of Sec. \ref{construc of BB in general regime}
and in Sec. \ref{D and diff eqn}.
The second cumulants $\kappa_2$ enter through the various mfp's, as we
see from Eq. (\ref{lab}).

In the DWSL we thus write Eq. (\ref{F 0}) as
\begin{eqnarray}
&&\lim_{DWS}
\sum _{r_1,\cdots,r_{2q}}
\Big\langle  (\hat{v}_{r_1})_{a_1b_1}
\cdots
(\hat{v}_{r_{2q}})_{a_{2q}b_{2q}}\Big\rangle
f_{a_1 b_1\cdots, a_{2q} b_{2q}}^{i_1 j_1,\cdots, i_{2q}j_{2q}}(r_1,\cdots,r_{2q})
\nonumber \\
&&\hspace{1cm}=
\frac{C(a_1 b_1,a_2 b_2) \cdots C(a_{2q-1} b_{2q-1},a_{2q} b_{2q})}
{\sqrt{l_{a_1 b_1} l_{a_2 b_2}\cdots l_{a_{2q-1} b_{2q-1}}l_{a_{2q} b_{2q}}}}
\Delta_{a_1 b_1, \cdots, a_{2q} b_{2q}}^{i_1 j_1,\cdots, i_{2q} j_{2q}}
(k;{\cal R}(\delta L); 12,34,\cdots,2q-1 \; 2q)
\nonumber \\ \nonumber \\
&&\hspace{2cm}+ {\rm \; all \; possible \; combinations} .
\label{F DWSL}
\end{eqnarray}
We have used Eq. (\ref{<vv>/d}) and we have defined
\begin{eqnarray}
&&\Delta_{a_1 b_1 \cdots, a_{2q} b_{2q}}^{i_1 j_1, \cdots, i_{2q} j_{2q}}
(k;{\cal R}(\delta L); 12,\cdots,2q-1 \; 2q)
\nonumber \\
&&= \int_{\delta L} \cdots \int _{\delta L}
f_{a_1 b_1,\cdots, a_{2q} b_{2q}}^{i_1 j_1, \cdots, i_{2q} j_{2q}}(x_2,x_2, \cdots ,x_{2q}, x_{2q})
dx_2 \cdots dx_{2q}
\nonumber \\
&&= \int \cdots \int _{{\cal R} \subset (\delta L)^q}
\vartheta _{a_1 b_1}^{i_1 j_1} (x_2)\vartheta _{a_2 b_2}^{i_2 j_2} (x_2)
\cdots
\vartheta _{a_{2q-1} b_{2q-1}}^{i_{2q-1} j_{2q-1}} (x_{2q})
\vartheta _{a_{2q} b_{2q}}^{i_{2q} j_{2q}} (x_{2q})
dx_2 \cdots dx_{2q} .
\label{Delta 2q}
\end{eqnarray}
Here, $\vartheta _{ab}^{jl}(x)$ is the continuous version of the function
$\left[ \vartheta_r \right]_{ab}^{jl}$ of Eq. (\ref{theta}).
The region of integration ${\cal R} \subset (\delta L)^q$
arises from the appropriate step functions, Eq. (\ref{f}), that implement the correct range of
summation of the scatterer indices, and from the type of pair contraction.
We have added, in a symbolic fashion,
in the argument of $\Delta $, the information
about the scatterer indices that have been contracted:
in the above cases, the contraction was
$r_1=r_2,r_3=r_4, \cdots, r_{2q-1}=r_{2q}$.
This information is redundant, but has beeen added for clarity.
Eq. (\ref{F DWSL}) [inserted in Eq. (\ref{vareps^p 1})]
and Eq. (\ref{Delta 2q})
generalize the earlier expressions (\ref{<vareps^2>}) and
(\ref{Delta(k)}).
One of the ``possible combinations", i.e., the one arising from the contraction $r_1=r_3, r_2=r_4$, that would be indicated symbolically as $13,24$,  generalizes Eqs.
(\ref{vareps(2)vareps(2) DWSL}) and
(\ref{Delta(k) 1}).
In an expansion of the integral (\ref{Delta 2q}) in powers of $\delta L$,
the leading term clearly behaves as
\begin{equation}
\Delta_{a_1 b_1 \cdots, a_{2q} b_{2q}}^{i_1 j_1, \cdots, i_{2q} j_{2q}}
(k;{\cal R}(\delta L))
\sim (\delta L)^q + \cdots .
\label{Delta=(delta L)^q}
\end{equation}

Consider now the particular case of an even moment of the BB
$\varepsilon $. For this purpose we set $p=2t$ in the above analysis, starting from Eq. (\ref{vareps^p}).
The lowest-order term in the expansion of Eq. (\ref{vareps^p})
corresponds to $\mu_1= \cdots =\mu_{2t} = 1$ and thus to $2q=2t$,
in the notation introduced right after Eq. (\ref{vareps^p})
(i.e., this term is of order $2t$ in the $\hat{v}_r$'s);
in the DWSL it is found, by setting $q=t$ in Eqs. (\ref{F DWSL}) and (\ref{Delta=(delta L)^q}), that its leading term in an expansion in powers of $\delta L$
behaves as $(\delta L)^t /\sqrt{l_{a_1 b_1} \cdots l_{a_{2t} b_{2t}}}$.
Higher-order terms in the expansion (\ref{vareps^p}) for the same moment are higher order in $\delta L$.
The contribution to a second moment obtained above,
Eq. (\ref{<vareps^2>leading}),
represents, for $t=1$, a particular case of this general result.

For an odd moment with $p=2t+1$, the first term in the expansion of Eq. (\ref{vareps^p}), i.e., the one with
$\mu_1= \cdots =\mu_{2t+1} = 1$,
vanishes, because it is of odd order in the $\hat{v}_r$'s.
The next-order terms in the expansion (\ref{vareps^p})
have one of the $\mu_i =2$ and all the other $\mu_i$'s equal to 1
[for instance: $\mu_1=2, \mu_2= \cdots =\mu_{2t+1} = 1$].
For these terms, $2q=2t+2$, so that from
Eqs. (\ref{F DWSL}) and (\ref{Delta=(delta L)^q})
we see that these terms are of order
$(\delta L)^{t+1} /\sqrt{l_{a_1 b_1} \cdots l_{a_{2t+2} b_{2t+2}}}$.

The conclusion of the last two paragraphs is not altered when we translate the BB to the interval $(L, L+\delta L)$.
We have thus proven, for the moments of $\varepsilon$, the behavior that was mentioned
at the end of Sec. \ref{construc of BB in general regime}.

\section{Some useful details for Sec. \ref{analytic}}
\label{derivation-analytic}

In the one-channel case, the quantity
$K_{ab,cd}^{ij,hl}$
of Eq. (\ref{K}) and the diffusion coefficient
$D_{ab,cd}^{ij,hl}(k,L)$
of Eq. (\ref{D(k,L)}), to be used in the diffusion equation
(\ref{diff eqn exp val sec3}), are given by:
\begin{subequations}
\begin{eqnarray}
K^{ij,hl}
&=&\left[ \left( -1\right)^{i}
+\left(-1\right) ^{j+1}
+\left( -1\right) ^{h}+\left( -1\right) ^{l+1}\right] k,
\label{K N=1}
\\
D^{ij,hl}\left( k,L\right)
&=&\frac{\left(-1\right) ^{i+h+1}}{2\ell }
e^{i K^{ij,hl}  L},
\label{D N=1}
\end{eqnarray}
\label{K,D N=1}
\end{subequations}
respectively. We have omitted the channel indices, which would take the value 1.

We can rewrite the pair of Eqs. (\ref{e-alfa/alfa*,alfa/beta}),
after multiplying the second one by $e^{ix_{0}s}$, as
\begin{subequations}
\begin{eqnarray}
\frac{1}{2}\frac{\partial A}{\partial s} &=&A+2b_{r},
\label{e-1} \\
\frac{\partial b_{r}}{\partial s}+x_{0}b_{i} &=&-A-2b_{r},
\label{e-2} \\
\frac{\partial b_{i}}{\partial s}-x_{0}b_{r} &=&0,
\label{e-3}
\end{eqnarray}
\label{e-1,2,3}
\end{subequations}
where 
\begin{subequations}
\begin{eqnarray}
A(s)
&=&2\left\langle \alpha \alpha ^{\ast }\right\rangle -1
\\
b(s)&=&b_r(s) + i b_i(s)
=\left\langle\alpha \beta \right\rangle_s e^{ix_0 s} .
\label{A,br,bi}
\end{eqnarray}
\end{subequations}

The quantities $p_{1}$, $p_{2}$ and $p_{3}$ appearing in Eq. (\ref{s_1,2,3})
are the roots of the third degree polynomial
$P\left( p\right) =p^{3}+x_{0}^{2}p-2x_{0}^{2}$
and are given by
\begin{subequations}
\begin{eqnarray}
p_{1} &=&u+v,  \label{raiz-1} \\
p_{2} &=&-\frac{1}{2}\left( u+v\right)
+i\frac{\sqrt{3}}{2}\left(u-v\right),
\label{raiz-2}
\\
p_{3} &=&-\frac{1}{2}\left( u+v\right)
-i\frac{\sqrt{3}}{2}\left(u-v\right) ,
\label{raiz-3}
\end{eqnarray}
\end{subequations}
with
\begin{equation}
u=\frac{x_{0}}{\sqrt{3}}
\left\{\left[
1+\left(\frac{3\sqrt{3}}{x_0}\right)^2 \right]^{1/2}
+ \frac{3\sqrt{3}}{x_0}
\right\}^{1/3},
\hspace{1cm}
v=-\frac{x_{0}}{\sqrt{3}}
\left\{\left[
1+\left(\frac{3\sqrt{3}}{x_0}\right)^2 \right]^{1/2}
- \frac{3\sqrt{3}}{x_0}
\right\}^{1/3} .
\end{equation}

When $x_{0}\gg 1$, we expand $u$ and $v$ as
\begin{subequations}
\begin{eqnarray}
u &=& \frac{x_{0}}{\sqrt{3}}
\left[ 1+ \frac{\sqrt{3}}{x_{0}} +\frac{3}{2x_{0}^{2}}
-\frac{4\sqrt{3}}{x_{0}^{3}} -\frac{105}{8x_{0}^{4}}
+\cdots \right]
\\
v &=& -\frac{x_{0}}{\sqrt{3}}
\left[ 1 - \frac{\sqrt{3}}{x_{0}} +\frac{3}{2x_{0}^{2}}
+\frac{4\sqrt{3}}{x_{0}^{3}} -\frac{105}{8x_{0}^{4}}
+\cdots \right] ,
\end{eqnarray}
\end{subequations}
and the roots are given approximately by
\begin{subequations}
\begin{eqnarray}
p_{1}&\simeq& 2 - \frac{8}{x_0^2} + O \left(\frac{1}{x_0^4}\right)
\\
p_{2}&\simeq& (-1+ix_{0}) +
\left(\frac{4}{x_0^2} + i \frac{3}{2x_0} \right)
+ O \left(\frac{1}{x_0^3}\right)
\\
p_{3}&=& p_2^*
\end{eqnarray}
\end{subequations}
We can thus write the exact solution (\ref{s_1,2,3}) as a power series in $1/x_0$
as
\begin{subequations}
\begin{eqnarray}
A(s)
&=&e^{2s}
+\frac{4}{x_0^2}
\left[
-(1+2s)e^{2s}+ e^{-s}\cos x_{0}s
\right]
+ O \left(\frac{1}{x_0^3}\right)
 \\
b_{r}(s) &=&
-\frac{1}{x_{0}}e^{-s}\sin x_{0}s
+\frac{2}{x_0^2}
\left[
\left(1-\frac{3s}{4}\right)e^{-s}\cos{x_0 s}-e^{2s}
\right]
+ O \left(\frac{1}{x_0^3}\right)
 \\
b_{i}(s)
&=&\frac{1}{x_{0}}
\label{A,b,1/x0}
\left[
-e^{2s} + e^{-s}\cos{x_0 s}
\right]
-\frac{3}{x_0^2}(s-1)e^{-s}\sin{x_0 s}
+ O \left(\frac{1}{x_0^3}\right) .
\end{eqnarray}
\label{e10}
\end{subequations}

\section{Detailed Structure of the Expectation Values
$\left\langle F(\bm{M}) \right\rangle _{L,k}$ in the SWLA, for the Case $N=1$.}
\label{swla from details of <>}

The aim of this appendix is to derive the diffusion equation in the
SWLA, Eq. (\ref{diff eqn exp val swla}), 
for the one channel case ($N=1$), assuming, for the expectation values
$\left\langle \cdots \right\rangle _{L,k}$, a more detailed structure than the one that was assumed in Eq. (\ref{<>0}).
We shall present such derivation in two ways:
i) starting from the BB (as in section \ref{SWLA}),
and ii) starting from the exact diffusion equation
(\ref{diff eqn N1}).

The justification for the structure that we shall suppose
comes from Sec. \ref{analytic}, in which we solved
Eq. (\ref{diff eqn N1}) for the quantities
$\left\langle \alpha \alpha ^{\ast }\right\rangle$ and
$\left\langle \alpha \beta \right\rangle$, giving the exact result of Eq. (\ref{s_1,2,3})
which, for large values of $k\ell$, was then expanded in inverse powers of this quantity, giving Eq. (\ref{<aa*> <ab> expansion}).
Inspired by this last result we shall suppose that for an arbitrary observable $F(\bm{M})$, the expectation value
$\left\langle F(\bm{M}) \right\rangle_{L,k}$
has the structure:
\begin{equation}
\left\langle F\left( \bm{M}\right) \right\rangle _{L,k}
=\left\langle F\left( \bm{M}\right) \right\rangle _{L}
^{\left( 0\right)}
+\sum_{n=1}^{\infty }\sum_{m=-\infty }^{\infty}
\frac{\left\langle F\left( \bm{M}\right)\right\rangle_{m,L}
^{\left(n\right) }}{\left( k\ell \right) ^{n}}e^{imkL},
\label{SWLA1}
\end{equation}
where the functions
$\left\langle F\left( \bm{M}\right) \right\rangle _{L}
^{\left( 0\right)}$, and
$\left\langle F\left( \bm{M}\right) \right\rangle _{m,L}
^{\left(n\right) }$
depend on the energy only through the mfp
which, just as in Sec. \ref{applications}, is taken as a fixed parameter denoted by $\ell_{11}=\ell$.
In the second term on the r.h.s of Eq. (\ref{SWLA1}) we have assumed, in principle, an infinite sum over $m$;
however, we shall show that this sum has in fact a finite number of
terms, as it occurs in Eq. (\ref{<aa*> <ab> expansion}).
Just as in Sec. \ref{analytic}, we have denoted the
longitudinal momentum, Eq. (\ref{ka}), by $k_{1}=k$.
For $k\ell \gg 1$, the quantities
$\left\langle F\left( \bm{M}\right)\right\rangle _{L}
^{\left( 0\right) }$ dominate the behavior of
$\left\langle F\left( \bm{M}\right) \right\rangle _{L,k}$ in Eq. (\ref{SWLA1}), precisely as we supposed in the ``ansatz" of Eq. (\ref{<>0}).

For the particular case of Eq.  (\ref{<aa*> <ab> expansion}) we make the following identifications:
\begin{eqnarray}
\left\langle \alpha \alpha ^{\ast }\right\rangle _{L}
^{\left( 0\right)}
&=&\frac{1}{2}\left( 1+e^{2L/\ell }\right),
\nonumber \\
\left\langle \alpha \alpha ^{\ast}\right\rangle _{m,L}
^{\left( 1\right) }
&=& 0, \;\;\; \forall m,
\nonumber \\
\left\langle \alpha \alpha ^{\ast }\right\rangle _{0,L}
^{\left(2\right) }
&=&-\frac{1}{2}\left(1+2\frac{L}{\ell }\right)e^{2L/\ell }, \;\;\;
\left\langle \alpha \alpha^{\ast}\right\rangle _{2,L}
^{\left( 2\right) }
=\left\langle \alpha \alpha ^{\ast }\right\rangle _{-2,L}
^{\left( 2\right)}
=\frac{1}{4}e^{-L/\ell },
\nonumber \\
&&\hspace{4cm}
\left\langle \alpha \alpha^{\ast}\right\rangle _{4,L}
^{\left( 2\right) }
=\left\langle \alpha \alpha ^{\ast }\right\rangle _{-4,L}
^{\left( 2\right)}
=0
\label{structure for N1}
\end{eqnarray}
and
\begin{eqnarray}
\left\langle \alpha \beta \right\rangle _{L}^{\left( 0\right) }
&=&0,
\nonumber \\
\left\langle \alpha \beta \right\rangle _{0,L}
^{\left( 1 \right) }
&=&\frac{i}{2}e^{-L/\ell },
\hspace{2.5cm}
\left\langle \alpha \beta \right\rangle_{2,L}
^{\left(1\right) }
=0,
\hspace{1.7cm}
\left\langle \alpha \beta \right\rangle_{-2,L}
^{\left(1\right) }
=-\frac{i}{2}e^{2L/\ell },
\nonumber \\
&&\hspace{4cm}
\left\langle \alpha \beta \right\rangle_{4,L}
^{\left(1\right) }
=0
\hspace{1.7cm}
\left\langle \alpha \beta \right\rangle_{-4,L}
^{\left(1\right) }
=0
\nonumber \\
\left\langle \alpha \beta \right\rangle _{0,L}^{\left( 2\right) } &=&\frac{1}{8}\left( 5-3\frac{L}{\ell }\right) e^{-L/\ell },
\quad
\left\langle \alpha \beta \right\rangle _{2,L}
^{\left( 2\right)}
=0
\hspace{1.7cm}
\left\langle \alpha \beta \right\rangle _{-2,L}
^{\left( 2\right)}
=-\frac{1}{2}e^{2L/\ell},
\nonumber \\
&&\hspace{4cm}
\left\langle \alpha \beta \right\rangle _{4,L}
^{\left(2\right) }
=0
\hspace{1.7cm}
\left\langle \alpha \beta \right\rangle _{-4,L}
^{\left(2\right) }
=-\frac{1}{8}e^{-L/\ell } \; .
\label{structure for N1 1}
\end{eqnarray}
Similarly, for the expectation values that appear on r.h.s of
Eqs. (\ref{diff eqn N1}) and (\ref{avF(L+dL) 1}) for one
channel, we suppose, in the SWLA, the structure
\begin{equation}
\left\langle \left( \cdot \cdot \cdot \right) ^{ijhl\lambda
\mu}\right\rangle _{L,k} =\left\langle \left(\cdots \right)
^{ijhl\lambda \mu }\right\rangle _{L}^{\left( 0\right)}
+\sum_{n=1}^{\infty }\sum_{m=-\infty } ^{\infty}\frac{\left\langle
\left( \cdots \right) ^{ijhl\lambda \mu }\right\rangle _{m,L}
^{\left( n\right)}}{\left( k\ell \right) ^{n}}e^{imkL} .
\label{SWLA2}
\end{equation}
[We have used the same abbreviation as in Eq. (\ref{<...> 2}).]
Again, the functions $\left\langle \left( \cdots \right)
^{ij,hl\lambda \mu }\right\rangle _{L}^{\left( 0\right)}$ and
$\left\langle \left( \cdot \cdot \cdot \right) ^{ijhl\lambda
\mu}\right\rangle _{m,L}^{\left( n\right)}$
depend on $k$ only thorugh the mfp
and, for $k\ell \gg 1$, the quantities $\left\langle \left( \cdots
\right) ^{ijhl\lambda \mu }\right\rangle _{L}^{\left( 0\right) }$
dominate the behavior of $\left\langle \left( \cdot \cdot \cdot
\right) ^{ij,hl\lambda \mu}\right\rangle _{L,k}$ in Eq.
(\ref{SWLA2}).

\subsection
{Derivation of Eq. (\ref{diff eqn exp val swla}) for $N=1$ Starting from the BB}
\label{swla from BB}

Introducing
Eqs. (\ref{SWLA1}), (\ref{SWLA2}) and the first term of (\ref{ee=e1e1+e2e2}) as given by (\ref{<vareps^2> L 1}}) into
Eq. (\ref{avF(L+dL) 1}) (for $N=1$), we obtain
\begin{eqnarray}
&&\left\langle F\left( \bm{M}\right) \right\rangle _{L+\delta
L}^{\left( 0\right) }+\sum_{n=1}^{\infty }\sum_{m=-\infty
}^{\infty }\frac{\left\langle F\left( \bm{M}\right) \right\rangle
_{m,L+\delta L}^{\left( n\right) }}{\left( k\ell
\right) ^{n}}e^{imk\left( L+\delta L\right) }
\nonumber \\
&=&\left[ \left\langle F\left( \bm{M}\right) \right\rangle _{L}^{\left(
0\right) }+\sum_{n=1}^{\infty }\sum_{m=-\infty }^{\infty
}\frac{\left\langle F\left(
\bm{M}\right) \right\rangle _{m,L}^{\left( n\right) }}{\left( k\ell \right) ^{n}}%
e^{imkL}\right]
\nonumber \\
&&+\left\{ \sum_{\substack{ ijhl\lambda \mu  \\ \left( K=0\right)
}}\frac{\left( -\right) ^{i+h+1}}{2\ell }\left[ \left\langle \left(
\cdot \cdot \cdot \right) ^{ijhl\lambda \mu }\right\rangle
_{L}^{\left( 0\right) }+\sum_{n=1}^{\infty }\sum_{m=-\infty
}^{\infty }\frac{\left\langle \left( \cdot \cdot \cdot \right)
^{ijhl\lambda \mu }\right\rangle _{m,L}^{\left( n\right) }}{\left(
k\ell \right) ^{n}}e^{imkL}\right] \delta L \right.
\nonumber \\
&&\hspace{1cm}+\sum_{\substack{ ijhl\lambda \mu  \\ \left( K\neq
0\right) }}\left(
-\right) ^{i+h+1}e^{iK\left( L+\frac{\delta L}{2}\right) }\frac{\sin \frac{K%
}{2}\delta L}{K\ell } \left[ \left\langle \left( \cdot \cdot \cdot
\right) ^{ijhl\lambda \mu }\right\rangle _{L}^{\left( 0\right) }
\right.
\nonumber \\
&&\hspace{4cm}+
\left.
\left.
\sum_{n=1}^{\infty }\sum_{m=-\infty }^{\infty }\frac{%
\left\langle \left( \cdot \cdot \cdot \right) ^{ijhl\lambda \mu
}\right\rangle _{m,L}^{\left( n\right) }}{\left( k\ell \right) ^{n}}%
e^{imkL}
\right]\right\}
\nonumber \\
&&\hspace{1cm}+\cdots .
\label{F5}
\end{eqnarray}
Here, $K$ is an abbreviation for $K^{ij,hl}$, as in the text.
The contribution of Eq. (\ref{<vareps^2> L 2}) is not indicated, since it either contains inverse powers of $k$ or is proportional to
$\left( \delta L\right) ^{2}$.
Moreover, we have used the expression (see Eq. (\ref{Delta}))
\begin{equation}
\Delta ^{ij,hl}\left( k, \delta L\right)
=\left( -\right)^{i+h+1}\frac{\sin \frac{K}{2}\delta L}
{\frac{K}{2}}.
\end{equation}
From Eq. (\ref{K N=1}) we see that the possible values of
$K$\ are $0$, $\pm 2k$ and $\pm 4k$: thus, if $K\neq 0$,
$\Rightarrow $\ $K\ell \propto $ $k\ell $.

We now take to the l.h.s of Eq. (\ref{F5}) all the quantities that do
not have inverse powers of $k\ell $, and take the rest to the
r.h.s, to obtain
\begin{eqnarray}
&&\left\langle F\left( \bm{M}\right) \right\rangle _{L+\delta
L}^{\left( 0\right) }-\left\langle F\left( \bm{M}\right) \right\rangle
_{L}^{\left( 0\right) }-\sum _{\substack{ ijhl,\lambda \mu
\nonumber \\
\left( K=0\right) }}\frac{\left( -\right) ^{i+h+1}}{2\ell
}\left\langle \left( \cdot \cdot \cdot \right)
^{ijhl\lambda \mu }\right\rangle _{L}^{\left( 0\right) }\delta L
\nonumber \\
&=&-\sum_{n=1}^{\infty }\sum_{m=-\infty }^{\infty }\frac{1}{\left(
k\ell \right) ^{n}}\left[ \left\langle F\left( \bm{M}\right)
\right\rangle _{m,L+\delta L}^{\left( n\right) }e^{imk\left(
L+\delta L\right) }-\left\langle F\left(
\bm{M}\right) \right\rangle _{m,L}^{\left( n\right) }e^{imkL}\right]
\nonumber \\
&&+\sum_{\substack{ ijhl\lambda \mu  \\ \left( K=0\right) }}%
\sum_{n=1}^{\infty }\sum_{m=-\infty }^{\infty }\frac{\left( -\right) ^{i+h+1}%
}{2\ell }\frac{\left\langle \left( \cdot \cdot \cdot \right)
^{ijhl\lambda
\mu }\right\rangle _{m,L}^{\left( n\right) }}{\left( k\ell \right) ^{n}}%
e^{imkL}\delta L
\nonumber \\
&&+\sum_{\substack{ ijhl\lambda \mu
\nonumber \\
\left( K\neq 0\right)
}}\left(
-\right) ^{i+h+1}e^{iK\left( L+\frac{\delta L}{2}\right) }\frac{\sin \frac{K%
}{2}\delta L}{K\ell } \left[ \left\langle \left( \cdot \cdot \cdot
\right) ^{ijhl\lambda \mu }\right\rangle _{L}^{\left( 0\right) }
\right.
\nonumber \\
&&\hspace{4cm}\left.+
\sum_{n=1}^{\infty }\sum_{m=-\infty }^{\infty }
\frac{\left\langle \left( \cdot \cdot \cdot \right) ^{ijhl\lambda
\mu }\right\rangle _{m,L}^{\left( n\right) }}{\left( k\ell \right)
^{n}} e^{imkL} \right]
\nonumber \\
&&+\cdots \; .
\end{eqnarray}
For $k\ell \gg 1$\ we neglect the r.h.s, so that we obtain:
\begin{equation}
\left\langle F\left( M\right) \right\rangle _{L+\delta L}^{\left(
0\right) }-\left\langle F\left( M\right) \right\rangle
_{L}^{\left( 0\right) }-\sum _{\substack{ ijhl\lambda \mu  \\
\left( K=0\right) }}\frac{\left( -\right) ^{i+h+1}}{2\ell
}\left\langle \left( \cdot \cdot \cdot \right) ^{ijhl\lambda \mu
}\right\rangle _{L}^{\left( 0\right) }\delta L+O\left( \delta
L\right) ^{2}\approx 0. \label{3.64(1)}
\end{equation}
This last equation is the analogous of
(\ref{avF(L+dL) SWLA a}) for the
case $N=1$; if we make the same expansion as in Eq. (\ref{Taylor <F>(L+dL)}),
we finally obtain Eq. (\ref{diff eqn exp val swla}) for $N=1$.

\subsection
{Derivation of Eq. (\ref{diff eqn exp val swla}) for $N=1$ Starting from the Exact Diffusion Equation.}
\label{swla from exact diff eqn}

We introduce Eqs. (\ref{SWLA1}) and (\ref{SWLA2}) into
Eq. (\ref{diff eqn N1}) to obtain
\begin{eqnarray}
&&\frac{\partial \left\langle F\left( \bm{M}\right) \right\rangle
_{L}^{\left(
0\right) }}{\partial L}+\sum_{n=1}^{\infty }\sum_{m=-\infty }^{\infty }\frac{%
1}{\left( k\ell \right) ^{n}}\frac{\partial \left\langle F\left(
\bm{M}\right)
\right\rangle _{m,L}^{\left( n\right) }}{\partial L}e^{imkL}
\nonumber \\
&&+\sum_{m=-\infty }^{\infty }\frac{ime^{imkL}}{\ell }\left[
\left\langle
F\left( \bm{M}\right) \right\rangle _{m,L}^{\left( 1\right) }
+\sum_{n=2}^{\infty }
\frac{\left\langle F\left( \bm{M}\right) \right\rangle _{m,L}
^{\left( n \right) }}{\left( k\ell \right) ^{n-1}}\right]
\nonumber \\
&=&\sum_{\substack{ ijhl\lambda \mu
\nonumber \\
\left( K=0\right)
}}\frac{\left( -\right) ^{i+h+1}}{2\ell }\left[ \left\langle \left(
\cdot \cdot \cdot \right) ^{ijhl\lambda \mu }\right\rangle
_{L}^{\left( 0\right) }+\sum_{n=1}^{\infty }\sum_{m=-\infty
}^{\infty }\frac{\left\langle \left( \cdot \cdot \cdot \right)
^{ijhl\lambda \mu }\right\rangle _{m,L}^{\left(
n\right) }}{\left( k\ell \right) ^{n}}e^{imkL}\right]
\nonumber \\
&&+\sum_{\substack{ ijhl\lambda \mu  \\ \left( K\neq 0\right) }}\frac{%
\left( -\right) ^{i+h+1}}{2\ell }e^{iKL}\left[ \left\langle \left(
\cdot \cdot \cdot \right) ^{ij,hl,\lambda \mu }\right\rangle
_{L}^{\left( 0\right) }+\sum_{n=1}^{\infty }\sum_{m=-\infty
}^{\infty }\frac{\left\langle \left( \cdot \cdot \cdot \right)
^{ijhl\lambda \mu }\right\rangle _{m,L}^{\left( n\right) }}{\left(
k\ell \right) ^{n}}e^{imkL}\right].
\end{eqnarray}
If we now take to the l.h.s all the quantities that do
not have inverse powers of $k\ell$, and all the rest to the
r.h.s, we obtain
\begin{eqnarray}
&&\left[ \frac{\partial \left\langle F\left( \bm{M}\right)
\right\rangle
_{L}^{\left( 0\right) }}{\partial L}-\sum_{\substack{ ijhl\lambda \mu
\nonumber \\
\left( K=0\right) }}\frac{\left( -\right) ^{i+h+1}}{2\ell
}\left\langle \left( \cdot \cdot \cdot \right) ^{ijhl\lambda \mu
}\right\rangle
_{L}^{\left( 0\right) }\right]
\nonumber \\
&&+\left[ \sum_{\substack{ m=-\infty  \\ m\neq 0}}^{\infty }\frac{im}{\ell }%
\left\langle F\left( \bm{M}\right) \right\rangle _{m,L}^{\left(
1\right)
}e^{imkL}-\sum_{\substack{ ijhl\lambda \mu  \\ \left( K\neq 0\right) }}%
\frac{\left( -\right) ^{i+h+1}}{2\ell }e^{iKL}\left\langle \left(
\cdot \cdot \cdot \right) ^{ijhl\lambda \mu }\right\rangle
_{L}^{\left( 0\right)
}\right]
\nonumber \\
&=&\sum_{n=1}^{\infty }\sum_{m=-\infty }^{\infty }\sum_{\substack{ %
ijhl\lambda \mu  \\ \left( K\neq 0,K=0\right) }}\frac{\left(
-\right) ^{i+h+1}}{2\ell }e^{iKL}\frac{\left\langle \left( \cdot
\cdot \cdot \right) ^{ijhl\lambda \mu }\right\rangle _{m,L}^{\left(
n\right) }}{\left( k\ell
\right) ^{n}}e^{imkL}  \label{(1/kl)1}
\nonumber \\
&&-\sum_{n=1}^{\infty }\sum_{m=-\infty }^{\infty
}\frac{e^{imkL}}{\left( k\ell \right) ^{n}}\left( \frac{\partial
\left\langle F\left( \bm{M}\right)
\right\rangle _{m,L}^{\left( n\right) }}{\partial L}+\frac{im}{\ell }%
\left\langle F\left( \bm{M}\right) \right\rangle _{m,L}^{\left(
n+1\right) }\right) .
\label{(1/kl)2}
\end{eqnarray}
For $k\ell \gg 1$ we neglect the r.h.s and obtain:
\begin{subequations}
\begin{eqnarray}
&&\left[ \frac{\partial \left\langle F\left( \bm{M}\right) \right\rangle
_{L}^{\left( 0\right) }}{\partial L}
-\sum_
{\substack{ ijhl\lambda \mu  \\ \left( K=0\right) }}
\frac{\left( -\right) ^{i+h+1}}{2\ell}\left\langle
(\cdots)^{ijhl,\lambda \mu } \right\rangle _{L}
^{\left( 0\right) }\right]
\label{swla -1 a}
\\
&&\hspace{2cm}+\left[ \sum_{\substack{ m=-\infty  \\ m\neq 0}}^{\infty }\frac{im}{\ell }%
\left\langle F\left( \bm{M}\right) \right\rangle _{m,L}^{\left(
1\right)
}e^{imkL}-\sum_{\substack{ ijhl\lambda \mu  \\ \left( K\neq 0\right) }}%
\frac{\left( -\right) ^{i+h+1}}{2\ell }e^{iKL}\left\langle \left(
\cdot \cdot \cdot \right) ^{ij,hl,\lambda \mu }\right\rangle
_{L}^{\left( 0\right) }\right] \approx 0 .
\label{swla -1 b}
\nonumber \\
\end{eqnarray}
\label{swla -1}
\end{subequations}
This last equation has two kinds of terms: the square bracket in
(\ref{swla -1 a}), which is independent of $k$, and the square bracket
in (\ref{swla -1 b}), which depends on $k$ through the complex exponentials.
Eq. (\ref{swla -1}) can only be consistent if each square bracket vanishes.
As a result, since, for $N=1$, $K$ takes on the values $0, \pm 2k, \pm 4k$,
the sum over $m$ appearing in the second bracket must have a finite number of terms: $m = \pm 2, \pm 4$.
For the soluble example of Eq. (\ref{<aa*> <ab> expansion}) we have verified explicitly that (\ref{swla -1 a}), (\ref{swla -1 b}) and,
in particular, the contribution of order $1/k\ell $ of the last two lines of Eq. (\ref{(1/kl)2}), are exactly zero.

Starting from Eq. (\ref{diff eqn N1}) we have thus obtained the diffusion equation for one channel in the SWLA, i.e.,
\begin{equation}
\frac{\partial \left\langle F\left( \bm{M}\right) 
\right\rangle_{L}^{\left( 0\right) }}
{\partial L}
=\sum_{\substack{ ijhl\lambda
\mu  \\ \left( K=0\right) }}\frac{\left( -\right) ^{i+h+1}}{2\ell}
\left\langle (\cdots)^{ijhl\lambda \mu} \right\rangle _{L}^{\left( 0\right)}
=\sum_{\substack{ ijhl\lambda \mu \\ K=0 }}\widetilde{D}^{ij,hl}\left\langle M^{j\lambda }M^{l\mu }%
\frac{\partial ^{2}F\left( \bm{M}\right) }{\partial M^{i\lambda }\partial
M^{h\mu }}\right\rangle _{L}^{\left( 0\right) }, \label{3.66N=1}
\end{equation}
which is Eq. (\ref{diff eqn exp val swla}) for $N=1$.

The diffusion coefficients $\widetilde{D}^{ij,hl}$ appearing in
Eq. (\ref{3.66N=1}) can obtained from Eq. (\ref{<D HEL>}) as:
\begin{subequations}
\begin{eqnarray}
\widetilde{D}^{11,11} &=&-\frac{1}{2\ell }, \\
\widetilde{D}^{11,22} &=&\frac{1}{2\ell }, \\
\widetilde{D}^{12,21} &=&\frac{1}{2\ell }, \\
\widetilde{D}^{11,12} &=&\widetilde{D}^{11,21}=0 .
\end{eqnarray}
\end{subequations}

\section{A Central-Limit Theorem for the Composition of Building Blocks}
\label{clt_bb}

In this Appendix we obtain the statistical properties of the transfer matrix
$\bm{M}_{T}$ for the whole waveguide of length $L$.
The waveguide is divided into $P$ BB's of length $\delta L$ each, so that $L=P\delta L$.
Since we can write the transfer matrix for the $p$-th BB as
$\bm{M}_{p}=I+\varepsilon _{p}$, the total transfer matrix $\bm{M}_{T}$ is given by
\begin{subequations}
\begin{eqnarray}
\bm{M}_{T} &=&\bm{M}_{P}\bm{M}_{P-1}\cdot \cdot \cdot \bm{M}_{2}\bm{M}_{1}=I+\delta \bm{M}_{T},
\\
\delta \bm{M}_{T}
&=&\sum_{p=1}^{P}\varepsilon _p
+ \sum_{p_1 > p_2}^{P}\varepsilon _{p_1} \varepsilon _{p_2}
+ \cdots
\equiv \sum _{\mu =1}^P \delta \bm{M}_{T} ^{(\mu )},
\end{eqnarray}
\end{subequations}
in analogy with Eq. (\ref{eps BB 7}).
Here, $p_{i}$ denotes the position $x_{p_{i}}$ of the center of
the $p_i$-th BB.
Suppose that we propose a model in which the BB's are statistically independent and that the various moments of the $\varepsilon $'s of the $p$-th BB
satisfy
\begin{subequations}
\begin{eqnarray}
\left\langle \left( \varepsilon _{p}\right)
_{a_{1}b_{1}}^{i_{1}j_{1}}\right\rangle
&=&0,
\\
\left\langle \left( \varepsilon _{p}\right) _{a_{1}b_{1}}^{i_{1}j_{1}}\left(
\varepsilon _{p}\right) _{a_{2}b_{2}}^{i_{2}j_{2}}\cdot \cdot \cdot \left(
\varepsilon _{p}\right) _{a_{2q}b_{2q}}^{i_{2q}j_{2q}}\right\rangle
&=&
\sum_{\alpha =0}^{\infty }f_{2q}^{\left( \alpha +q\right) }\left(
x_{p}\right) \left( \delta L\right) ^{\alpha +q},
\\
\left\langle \left( \varepsilon _{p}\right) _{a_{1}b_{1}}^{i_{1}j_{1}}\left(
\varepsilon _{p}\right) _{a_{2}b_{2}}^{i_{2}j_{2}}\cdot \cdot \cdot \left(
\varepsilon _{p}\right) _{a_{2q+1}b_{2q+1}}^{i_{2q+1}j_{2q+1}}\right\rangle
&=&
\sum_{\alpha =0}^{\infty }f_{2q+1}^{\left( \alpha + q + 1 \right) }\left(
x_{p}\right) \left( \delta L\right) ^{\alpha + q +1}  \; .
\end{eqnarray}
\end{subequations}
We shall make contact below with the physical model discussed in
Secs. \ref{BB q1d} and \ref{applications}.
We allow the functions
$f_{2q}^{\left( \alpha +q\right)}\left( x_{p}\right)$ and
$f_{2q+1}^{\left( \alpha + q +1\right) }\left(x_{p}\right)$
to depend on the positions $x_{p}$.
It is easy to show that in the mathematical limit
$P\rightarrow \infty $,
$\delta L\rightarrow 0$,
so that $P\delta L=L$, the statistical
distribution of the matrix $\delta \bm{M}_{T}$
{\em only depends on the functions}
$f_{2}^{\left( 1\right) }\left( x_{p}\right)$
appearing in the second moment
$\left\langle \left( \varepsilon _{p}\right) _{a_{1}b_{1}}^{i_{1}j_{1}}
\left(\varepsilon _{p}\right) _{a_{2}b_{2}}^{i_{2}j_{2}}\right\rangle$.
Therefore, the functions
$f_{2}^{\left( \alpha +1\right) }\left(x_{p}\right)$
appearing in the second moment
$\left\langle \left( \varepsilon _{p}\right) _{a_{1}b_{1}}^{i_{1}j_{1}}
\left( \varepsilon _{p}\right)_{a_{2}b_{2}}^{i_{2}j_{2}}\right\rangle$
with $\alpha \neq 0$ and the
functions
$f_{2q}^{\left( \alpha +q\right) }\left( x_{p}\right) $,
$f_{2q+1}^{\left( \alpha + q +1 \right) }\left( x_{p}\right)$
appearing in moments higher than the second do not play any role in the final statistics of
$\delta M_{T}$.
This result can be interpreted as a CLT for the composition of BB's.

The function
$f_{2}^{\left( 1\right) }\left(x_{p}\right)$
was considered above as an arbitrary function of $x_{p}$.
Now we identify it with the factor that is proportional to the
first power of $\delta L$ in the model discussed in the present paper:

i) For the exact description, Eq. (\ref{<eps.eps>_D}) gives
\begin{equation}
f_2^{(1)}(x_p)= 2 D_{a_1 b_1, a_2 b_2}^{i_1 j_1, i_2 j_2} (k, x_p),
\end{equation}

ii) For the SWLA, Eq. (\ref{<eps.eps>_D SWLA}) gives
\begin{equation}
f_2^{(1)}(x_p)= 2 \tilde{D}_{a_1 b_1, a_2 b_2}^{i_1 j_1, i_2 j_2} .
\end{equation}



\begin{thebibliography}{99}

\bibitem{Ishimaru78}
A. Ishimaru, {\em Waves Propagation and Scattering in Random Media} (Academic
  Press, New York, 1978).

\bibitem{Rytov89}
S.~M. Rytov, Y.~A. Kravtsov, and V.~I. Tatarskii, {\em Principles of
  Statistical Radiophysics} (Springer, Berlin, 1989).

\bibitem{Nieto-V90}
{\em Scattering in Volumes and Surfaces}, edited by M.
  {Ni\-eto-Ves\-pe\-ri\-nas} and J.~C. Dainty (North Holland,
Amsterdam, 1990).

\bibitem{Altshuler91}
{\em Mesoscopic Phenomena in solids}, edited by B.~L. {Al'tshuler}, P.~A. Lee,
  and R.~A. Webb (North Holland, Amsterdam, 1991).

\bibitem{Sheng95}
P. Sheng, {\em Introduction to Wave Scattering, Localization and Mesoscopic
  Phenomena} (Academic Press, New York, 1995).

\bibitem{carlo97}
C. W. J. Beenakker, Rev. Mod. Phys. {\bf 69}, 731 (1997).

\bibitem{Imrybook}
Y. Imry, {\em Introduction to Mesoscopic Physics} (Oxford Univ. Press, Oxford,
  1997).

\bibitem{Dattabook}
S. Datta, {\em Electronic Transport in Mesoscopic Systems} (Cambridge Univ.
  Press, Cambridge, 1997).

\bibitem{Alhassid}
Y. Alhassid, Rev. Mod. Phys. {\bf 72}, 895 (2000).

\bibitem{mello-kumar}
P. A. Mello and N. Kumar, {\em Quantum Transport in Mesoscopic Systems.
Complexity and Statistical Fluctuations}, Oxford University Press, 2004.

\bibitem{WRM}
A. Garc\'{\i}a-Mart\'{\i}n and J. J. S\'aenz, Waves in Random and Complex Media
{\bf 15}, 229 (2005).

\bibitem{mello_shapiro}
P. A. Mello and B. Shapiro,
Phys. Rev. {\bf B 37}, 5860 (1988).

\bibitem{gang_of_4}
E. Abrahams, P. W. Anderson, D. C. Licciardello and T. V. Ramakrishnan,
Phys. Rev. Lett, {\bf 42}, 673 (1979).

\bibitem{mello_rmp}
P. A. Mello,
J. Math. Phys. {\bf 27}, 2876 (1986).

\bibitem{mpk}
P. A. Mello, P. Pereyra and N. Kumar,
Ann. Phys. (N.Y.) {\bf 181}, 290 (1988).

\bibitem{dorokhov}
O.N. Dorokhov,
Pis'ma Zh. Eksp. Teor. Fiz. {\bf 36}, 259 (1982)
[JETP Lett. {\bf 36}, 318 (1982)].

\bibitem{oseledec}
V. I. Oseledec,
Trans. Mosc. Math. Soc. {\em 19}, 197 (1968).

\bibitem{UAM}
L. S. Froufe-P\'erez, P. Garc\'{\i}a Mochales, P. A. Serena, P. A. Mello and J.
J. S\'aenz, Phys. Rev. Lett {\bf 89}, 246403 (2002).

\bibitem{UAM2}
A. Garc\'{\i}a-Mart\'{\i}n and J. J. S\'aenz, Phys. Rev. Lett. {\bf 87}, 116603
(2001).

\bibitem{mello_tomsovic}
P. A. Mello and S. Tomsovic,
Phys. Rev. {\bf B 46}, 15963 (1992).

\bibitem{houches}
P. A. Mello,
in {\em Mesoscopic Quantum Physics}
(ed. E. Akkermans, G. Montambaux and J.-L Pichard).
Les Houches Summer School, Session LXI.
Elsevier, Amsterdam.

\bibitem{mello_physicaA}
P. A. Mello, M. Y\'epez, L. S. Froufe-P\'erez and J. J. S\'aenz,
Physica A, to be published.

\bibitem{dyson}
F. J. Dyson,
J. Math. Phys. {\bf 3}, 140 (1962).

\bibitem{JJS_evanesc}
J. A. Torres and J. J. S\'{a}enz, J. Phys. Soc. Jap. {\bf 73}, 2182 (2004).

\bibitem{chandra}
S. Chandrasekhar,
Rev. Mod. Phys. {\bf 15}, 1 (1943).
[Reprinted in {\em Selected papers on noise and stochastic processes}
(ed. N. Wax), p. 3. Dover Publications, New Yord, 1954].

\bibitem{Ziman} See for example, J.M. Ziman, {\em Electrons and
Phonons}, p. 269, Oxford University Press (2001).

\bibitem{pereyra}
P. Pereyra, J. Math. Phys. {\bf 36}, 1166 (1995).

\bibitem{born_wolf}
M. Born and E. Wolf,
{\em  Principles of Optics},
seventh edition, Cambridge University Press (1999),
pp. 116-120.

\bibitem{landauer}
R. Landauer,
Phil. Mag. {\bf 21}, 863 (1970).

\bibitem{luis_thesis}
L. S. Froufe-P\'erez,
Ph. D. Thesis, Universidad Aut\'onoma de Madrid, 2006.

\bibitem{PGM} P. Garc\'{\i}a-Mochales, P. A. Serena, N. Garc\'{\i}a and J. L. Costa-Kr\"amer, Phys. Rev. B {\bf 53}, 10268 (1996).

\bibitem{AGM_apl} A. Garc\'{\i}a-Mart\'{\i}n, J. A. Torres, J. J. S\'aenz, and M. Nieto-Vesperinas, Appl.
Phys. Lett. 71, 1912 (1997); Phys. Rev. Lett. 80, 41654168 (1998).

\bibitem{freilikher}
J. A. S\'anchez-Gil, V. Freilikher, I. V. Yurkevich, and A. A. Maradudin,
Phys. Rev. Lett. {\bf 80}, 948 (1998);
Phys. Rev. B {\bf 59}, 5915 (1999).

\bibitem{izrailev_1}
F. M. Izrailev, G. A. Luna-Acosta, J. A. M\'endez-Berm\'udez, and M. Rend\'on,
Phys. Stat. Sol. (c) {\bf 0}, 3032 (2003).

\bibitem{izrailev_2}
F. M. Izrailev, N. M. Makarov, and M. Rend\'on, Phys. Stat. Sol. (b) {\bf 242},
1224 (2005);
%
Phys. Rev. B {\bf 72}, 041403 (R) (2005).

\bibitem{rendon}
M. Rend\'on,
Ph. D. Thesis, Universidad Aut\'onoma de Puebla, 2006.

\bibitem{roman}
P. Roman, {\em Advanced Quantum Theory, an outline of the fundamental ideas},
Addison-Wesley, Reading, Mass., 1965.

\bibitem{beenakker_rejaei}
C. W. J. Beenakker and B. Rejaei,
Phys. Rev. Lett. {\bf 71}, 3689 (1993);
Phys. Rev. B {\bf 49}, 7499 (1994).




\end{thebibliography}
\end{document}